# A comprehensive overview of diffuse correlation spectroscopy: theoretical framework, recent advances in hardware, analysis, and applications

Quan Wang[1], Mingliang Pan[1], Lucas Kreiss[2], Saeed Samaei[3,6], Stefan A. Carp[4], Johannes D. Johansson[5], Yuanzhe Zhang[1], Melissa Wu[2], Roarke Horstmeyer[2], Mamadou Diop[3], and David Day-Uei Li[1]*

[1] University of Strathclyde, Department of Biomedical Engineering, Faculty of Engineering, Glasgow, United Kingdom
[2] Duke University, Department of Biomedical Engineering, Durham, North Carolina, United States
[3] Western University, Department of Medical and Biophysics, Schulich School of Medical & Dentistry, London, Ontario, Canada
[4] Harvard Medical School, Massachusetts General Hospital, Optics at Athinoula A. Martinos Center for Biomedical Imaging, Charlestown, Massachusetts, United States
[5] Linköping University, Department of Biomedical Engineering, Linköping, Sweden
[6] Imaging Program, Lawson Health Research Institute, London, Ontario, Canada

**Abstract**

Diffuse correlation spectroscopy (DCS) is a powerful tool for assessing microvascular hemodynamic in deep tissues. Recent advances in sensors, lasers, and deep learning have further boosted the development of new DCS methods. However, newcomers might feel overwhelmed, not only by the already complex DCS theoretical framework but also by the broad range of component options and system architectures. To facilitate new entry into this exciting field, we present a comprehensive review of DCS hardware architectures (continuous-wave, frequency-domain, and time-domain) and summarize corresponding theoretical models. Further, we discuss new applications of highly integrated silicon single-photon avalanche diode (SPAD) sensors in DCS, compare SPADs with existing sensors, and review other components (lasers, fibers, and correlators), as well as new data analysis tools, including deep learning. Potential applications in medical diagnosis are discussed, and an outlook for the future directions is provided, to offer effective guidance to embark on DCS research.

*David Day-Uei Li, E-mail: David.Li@strath.ac.uk

**Glossary**

| | |
|---|---|
| DCS | Diffuse correlation spectroscopy |
| DCT | Diffuse correlation tomography |
| SPAD | Single-photon avalanche diode |
| APD | Avalanche photon diode |
| PMT | Photomultiplier |
| SNSPD | Superconducting nanowire single-photon detector |
| DL | Deep learning |
| BF | Blood flow |
| BFi | Blood flow index |
| CBF | Cerebral blood flow |
| PET | Positron emission tomograph |
| SPECT | Single photon emission computed tomograph |
| XeCT | Xenon-enhanced computed tomography |
| MRI | Magnetic resonance imaging |
| DSC-MRI | Dynamic susceptibility contrast magnetic resonance imaging |
| LDF | Laser Doppler flowmetry |
| NIR | Near-infrared |
| NIRS | Near-infrared spectroscopy |
| DOS | Diffuse optical spectroscopy |
| CBV | cerebral blood volume |
| FCS | Fluorescence correlation spectroscopy |
| DLS | Dynamic light scattering |
| QELS | Quasi-elastic light |
| DWS | Diffusing wave spectroscopy |
| CHS | Coherent hemodynamics spectroscopy |
| RBC | Red blood cells |
| AI | Artificial intelligence |
| CMOS | Complementary metal-oxide-semiconductor |
| CW | Continuous wave |
| TD | Time domain |
| FD | Frequency domain |
| RTE | Radiative transfer equation |
| PDE | **Photon diffuse equation** (Section 2); **Photon detection efficiency** of detectors (Section 3) |
| CTE | Correlation transport equation |
| CDE | Correlation diffusion equation |
| MRI-ASL | MRI-based arterial spin labelling |
| RF | Radio-frequency |
| ANSI | American National Standards Institute |
| MPE | Maximal permissible exposure |
| PDE | Photon detection efficiency |
| LSCA | Laser Speckle Contrast Analysis |
| LSCI | Laser Speckle Contrast Imaging |
| DSCA | Diffuse speckle contrast analysis |
| DWS | Diffusing wave spectroscopy |
| DUS | Doppler ultrasound |
| PDT | Photodynamic therapy |
| TCD | Transcranial Doppler ultrasound |
| $pO_2$ | Oxygen partial pressure |
| $CMRO_2$ | Cerebral metabolic rate of oxygen |
| FPGA | Field Programmable Gate Arrays |

| | | |
|---|---|---|
| $D$ | Core diameter of multimode fiber | |
| $d$ | Speckle diameter | |
| SNR | Signal to noise ratio | |
| $g$ | Anisotropy factor | |
| $\mu_s$ | Scattering coefficient | |
| $\mu_s'$ | Reduced scattering coefficient | |
| $\langle \Delta r^2(\tau) \rangle$ | mean square displacement of moving scatterers | |
| $D_B$ | Effective diffusion coefficient for moving particles | |
| $V^2$ | mean square velocity | |
| $r_1$ | Distance between the detector and an approximated positive isotropic imaging source for a semi-infinite geometry | |
| $r_2$ | Distance between the detector and an approximated negative isotropic imaging source for a semi-infinite geometry | |
| $G1/g1$ | Unnormalized/normalized electric field autocorrelation function | |
| $G2/g2$ | Unnormalized/normalized intensity autocorrelation function | |
| $\lambda$ | Wavelength | |
| $k_0$ | Wavenumber in the medium | |
| $n$ | Refraction index | |
| $\alpha$ | Fraction of scattering events due to dynamic | |
| $\beta$ | Coherent factor | |
| $\tau$ | Correlation delay time | |
| $R_{eff}$ | Effective reflection coefficient | |
| $\rho$ | Distance between source and detection fibers | |
| $J_0$ | The zeros order Bessel function of the first kind | |
| $s_0$ | Point-like monochromatic light source | |
| $l_c$ | Coherence length | |
| $\Delta\lambda$ | The optical bandwidth | |
| $w$ | frequency corresponding to time in Fourier domain | |
| $q$ | The radial spatial frequency | |
| $p$ | Layer number of tissues | |
| $\omega$ | The source modulation frequency | |
| $T$ | The correlator bin time interval | |
| $T_{int}$ | Integration time (measurement duration) or the measurement time window | |
| $\tau_c$ | Decay constant | |
| $\langle M \rangle$ | Average number of photons within bin time T | |
| $I$ | Detected photon count | |
| $m$ | Bin index | |
| $s$ | Photon pathlength | |
| ToF | Time-of-flight | |
| $t$ | Photon time-of-flight | |
| NL | $N^{th}$-order | |
| SVR | Support vector regression | |
| EEG | electroencephalogram | |
| ECG | electrocardiogram | |
| 2DCNN | 2-dimentional convolution neural networks | |
| 3D | Three-dimensional | |
| LSTM | Long short-term memory | |
| RNN | Recurrent neural network | |
| $StO_2$ | cerebral tissue oxygenation | |

# 1. Introduction

Blood flow (BF) in a healthy person ensures stable delivery of oxygen and energy substrates (glucose and lactate) to and timely removal of metabolic waste products from organs[1]. Specifically, well-regulated cerebral blood flow (CBF) ensures healthy brain functions[2,3], brain metabolisms[4,5], and supports metabolic responses to external stimuli[6,7]. The average CBF for an adult human is around 50 ml/(100 g min)[8] and 10-30 ml/(100 g min) for a newborn[9]. Irregular CBF can cause brain damage through ischemic injury or stroke[10,11].

Effective real-time BF monitoring can aid in the diagnosis and management of broad range of medical conditions such as stroke, traumatic or hypoxic-ischemic encephalopathy (HIE)[12,13], neurological disorders, cardio-cerebral diseases, cancer treatment strategies, tissue perfusion in peripheral vascular diseases[14], brain health/functions[15], wound healing, sepsis and shock[16], skeletal muscle[17] injuries or tissue viability during surgeries.

Available real-time BF measurement tools are predominantly Doppler ultrasound based. However, Doppler ultrasonography requires a highly-skilled operator at the bedside and is operator-dependent[18]. Cerebral perfusion can be mapped using medical imaging scanners, including positron emission tomography (PET)[19], single photon emission computed tomography (SPECT)[20], xenon-enhanced computed tomography (XeCT)[21], dynamic susceptibility contrast magnetic resonance imaging (DSC-MRI)[22], and arterial spin labelling MRI (ASL-MRI)[23–25]. However, these techniques only provide 'snapshot' observations, are inappropriate for continuous monitoring, and typically require moving patients to imaging suites, which is unpractical for many patients. Additionally, a supine scan is necessary for MRI, PET and CT techniques. Further, PET, SPECT, and CT present additional risks of radiation exposure. Laser Doppler flowmetry (LDF)[26] is another perfusion technique, but it can only measure superficial tissue blood flow; thus, tissue samples must be thin to permit adequate sampling. Thus, there is a critical need to develop bedside techniques that are free from the limitations mentioned above and can noninvasively monitor microvascular BF in deep tissue at the bedside with a high sampling rate and at a low cost. For a thorough comparison of the modalities mentioned above, readers can refer to previous reviews[3,8,27].

In the late 1970s, Jöbsis observed a spectral window in the near-infrared (low optical absorption, $\mu_a$ and reduced scattering, $\mu_s'$, ~ 650-950 nm) wherein photons can penetrate deep/thick tissues up to several centimetres[28–30]. Subsequently, near-infrared spectroscopy (NIRS) or diffuse optical spectroscopy (DOS) was applied to study deep tissue hemodynamics, including cerebral oxygenation and cerebral blood volume (CBV), as early as the mid-1980s[31,32]. However, traditional NIRS primarily measures blood oxygenation saturation and hemoglobin concentrations, instead of tissue blood flow. Although NIRS can estimate tissue perfusion, this requires injecting exogenous contrast agents (such as indocyanine green)[33–35], limiting its applications in continuous monitoring.

Diffuse light correlation techniques, on the other hand, are rooted in the fundamental principles of dynamic light scattering (DLS). These methods, sometimes called 'quasi-elastic light (QELS) scattering' techniques[36–39], measure light intensity fluctuations scattered from samples to observe motions of sample constituents, e.g., Brownian motions of particles or macromolecules. Conventionally, DLS can provide detailed information about the dynamics of scattering media by using photon correlation techniques to analyze scattered light fluctuations[39]. However, QELS belongs to the single-scattering regime[36,40] and is unsuitable for turbid media in which the incident light is scattered multiple times. In 1987, Maret and Wolf[41] reported experimental measurements of the intensity autocorrelation function in the multiple-scattering regime and suggested a simple method for analyzing the measurements. One year later, Stephen derived a theoretical framework that extends QELS to the multiple-scattering regime[42].

Near-infrared diffuse correlation spectroscopy (DCS), also known as diffusing wave spectroscopy (DWS)[43,44] relates multi-scattered light's fluctuations to the underlying dynamics of scattering media. Despite the name, DCS is not a traditional spectroscopic technique[45–47] that uses multiple wavelengths; instead, it is a laser speckle method that analyzes light scattered over long distances through tissue. DCS, LDF[48,49], laser speckle contrast imaging (LSCI)[50], and diffuse speckle contrast analysis (DSCA)[51,52] are all based on laser speckles. Diffuse temporal correlation spectroscopy was first introduced by Boas and Yodh[53] and then formally named as 'DCS' by Cheung *et al.* in 2001[54]. It provides a theoretical framework that describes the underlying phenomenon using the popular diffuse approximation to the radiative transfer equation. Notably, a comprehensive diffuse correlation theory of diffuse speckle fields for predicting particle motions in highly scattered media was first introduced by Boas and Yodh in 1995[55,56]. The DCS theoretical model can be used to estimate deep-tissue microvascular blood flow index (BFi), which is a good surrogate for *in vivo* BF[57]. In the last two decades, DCS technologies have been further developed[56,58,59], validated, and employed for non-invasive BF measurements in deep tissue (up to ~ 2 centimeters[60,61]), such as skin, muscle[25,62–68], breast tumor[69–75] and the brain[11,13,76–82]. In 2001, the combination of DCS with NIRS/DOS was first introduced for cerebral monitoring in rats[83] and then in adult brains in 2004[84]. This combination allows for simultaneous monitoring of tissue BF and oxygenation.

A diagram of the history of DCS development is shown in Fig. 1(a). Fig. 1(b) displays the number of DCS publications over the past 20 years, with more than 400 publications to date (we only counted articles containing "DCS"). Fig. 1(c) presents DCS measurements from human brain tissue, organizing current studies by ρ (x-axis) and the blood flow sampling rate (y-axis). It highlights a trend towards employing parallel or multispeckle and interferometric DCS for higher sampling rates at a larger ρ. Additionally, it indicates the depth required to penetrate the scalp (refer to the top of Fig. 1(c)) and outlines different speed regimes needed to measure: 1) general changes, at a sampling rate of less than 1 Hz; 2) pulsatile blood flow, at 1-10 Hz; and 3) very rapid events, potentially detectable at rates greater than 10 Hz.

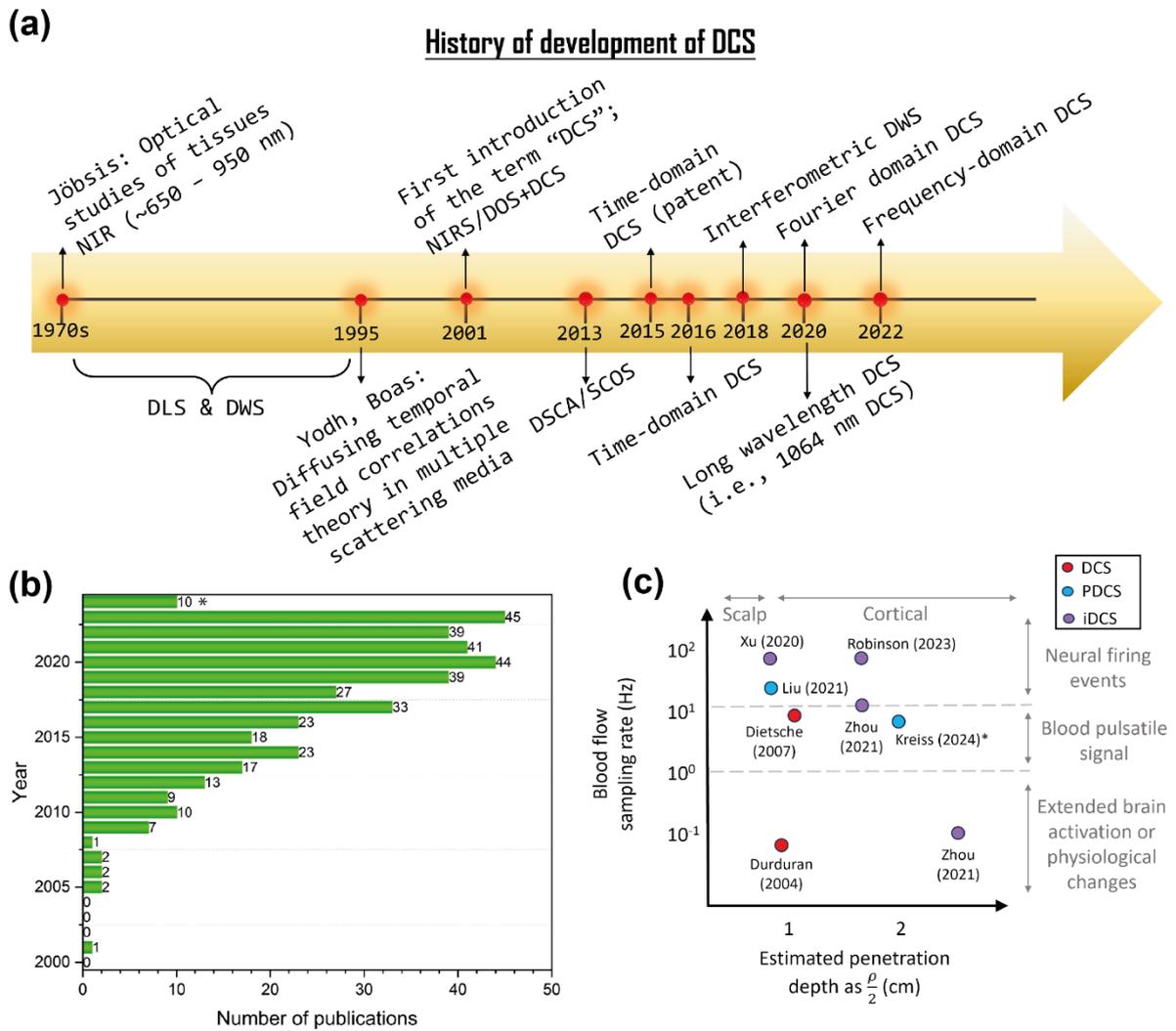

Fig. 1 (a) The roadmap of DCS historical development; (b) the number of published DCS papers based on PUBMED (*value for 2024 extrapolated as of the date of writing); (c) blood flow sampling rate vs measurement depths. PDCS: parallelized DCS, iDCS: interferometric DCS.

Fig. 2 illustrates the principle of DCS. Briefly, a long-coherence laser emits NIR light through an optical fiber to the tissue, Figure 2(a), and the recorded light intensity exhibits temporal fluctuations, Figure 2(b). These fluctuations are attributed to the motion of moving scatterers, such as red blood cells (RBC). To quantify the motion of RBC, a hardware or software correlator calculates the normalized intensity autocorrelation, $g_2(\tau)$ as shown in Figure 2(c). Typically, DCS systems are implemented in a reflection geometry, where a source and a detector are placed at a finite distance, $\rho$. Photons travelling from the source to the detector follow a "banana-shaped", stochastic scattering profile, as shown in Fig. 2(d), where the penetration depth of these DCS instruments is roughly between $\rho/3 \sim \rho/2$[85]. Fig. 2(c) and 2(e) show that the $g_2(\tau)$ curves decay faster with increased flow or $\rho$. The slope or the decay rate provides information about the optical properties and the motion of the scatters. The largest $\rho$ in the current state-of-the-art is 4 cm, corresponding to a depth of about 2 cm[60].

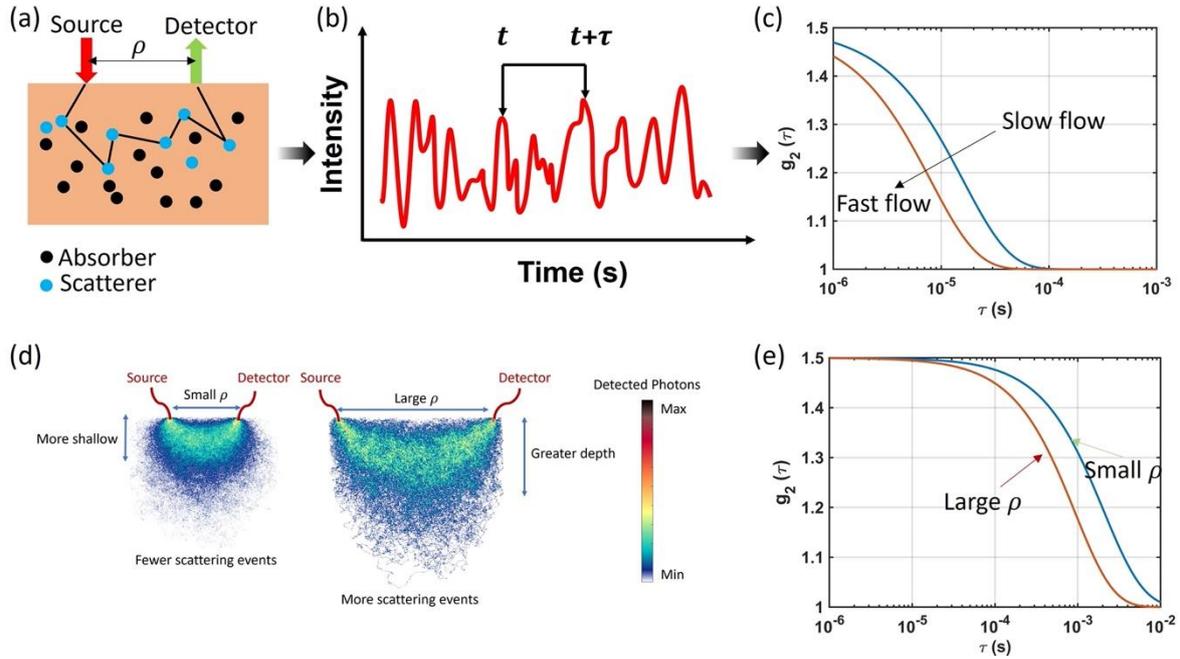

Fig. 2 The DCS principle for blood flow measurements. **(a)** The schematic of DCS measurements in the semi-infinite geometry. Highly coherent laser light is used to illuminate the sample via optical fibers. The source and detector fibers are placed on the tissue surface within a distance $\rho$; **(b)** the scattered light intensity fluctuates due to moving scatterers (e.g., red blood cells); **(c)** two intensity autocorrelation curves ($g_2(\tau)$) showing different flow rates. (d) Photons scattered from moving particles travel along "banana-shaped" paths between source and detection fibers; (e) Autocorrelation functions for different $\rho$.

Although there have been around 13 review DCS papers[3,31,76–80] in the last two decades, new approaches have emerged, including theoretical layered models, artificial intelligence (AI) methods for DCS analysis, and the use of novel sensors like highly integrated complementary metal-oxide-semiconductor (CMOS) single-photon avalanche diodes (SPAD) cameras. These aspects were not covered in previous reviews, which is why this review summarizes and systematically compares various analytical layered models, including continuous-wave (CW)-, time-domain (TD)-DCS, AI-enhanced DCS analysis methods, as well as the use of SPAD cameras in DCS. Furthermore, we also derived analytical models for the frequency domain (FD)-DCS, which was newly introduced in 2022[91]. The main contributions of this review include:

- We thoroughly derive and compare different layered analytical models used in CW-, TD-, and FD-DCS, highlighting their strengths and applications (Section 2).
- We discuss novel AI-enhanced DCS analysis strategies, addressing their effectiveness and potential (Section 4).
- Section 3.3 examines new applications of CMOS SPAD cameras and compares them with existing sensors used in DCS.
- Section 3.5 compares TD-DCS and CW-DCS systems and emphasizes the benefits of TD-DCS and its potential for future development.
- Discussion and outlooks are provided in Section 6.

This review aims to serve as a practical information resource for researchers and newcomers venturing into the field, offering a clearer understanding of the evolving DCS landscape and equipping them with the necessary knowledge to navigate it effectively.

## 2. Theory background

The propagation of light in highly scattering media such as biological tissues can be characterized by an absorption coefficient $\mu_a$ and a reduced scattering coefficient $\mu_s'$ using the radiative transfer equation (RTE)[89]. Similarly, to study the photon propagation under dynamic scatterers, the correlation transport equation (CTE)[3,87] is adopted to obtain the field (electrical) autocorrelation function $G_1(\tau)$ under general conditions of photon migration. The primary difference between the CTE and RTE lies in the fact that CTE describes the time-dependent specific intensity, reflecting an angular spectrum of the mutual coherence function. In the NIR spectral window, the unnormalized $G_1(\tau)$ can be expressed as, $G_1^T(r,\widehat{\Omega},\tau) = \langle E(r,\widehat{\Omega},r) \cdot E^*(r,\widehat{\Omega},t+\tau) \rangle$, where $\langle \cdots \rangle$ denotes a time average. $E(r,\widehat{\Omega},t)$ is the electric field at the position $r$ and time $t$ propagating in the $\widehat{\Omega}$ direction, inside the tissue that can be described by CTE[92–94] applicable for CW systems analogous to RTE:

$$\nabla \cdot G_1^T(r,\widehat{\Omega},\tau)\widehat{\Omega} + \mu_t G_1^T(r,\widehat{\Omega},\tau) = S(r,\widehat{\Omega}) + \mu_s \int G_1^T(r,\widehat{\Omega}',\tau) g_1^s(\widehat{\Omega},\widehat{\Omega}',\tau) f(\widehat{\Omega},\widehat{\Omega}') d\widehat{\Omega}', \qquad (1)$$

where $\mu_t = \mu_s + \mu_a$ is the transport coefficient. $S(r,\widehat{\Omega})$ is the source distribution; $g_1^s(\widehat{\Omega},\widehat{\Omega}',\tau)$ is the normalized field correlation function for single scattering; and $f(\widehat{\Omega},\widehat{\Omega}')$ is the normalized differential cross-section.

For a time dependent source, Eq. (1) becomes:

$$\nabla \cdot G_1^T(r,\widehat{\Omega},\tau,t)\widehat{\Omega} + \mu_t G_1^T(r,\widehat{\Omega},\tau,t) + \frac{1}{v}\frac{\partial}{\partial t} G_1^T(r,\widehat{\Omega},\tau,t) = S(r,\widehat{\Omega},t) + \mu_s \int G_1^T(r,\widehat{\Omega}',\tau,t) g_1^s(\widehat{\Omega},\widehat{\Omega}',\tau,t) f(\widehat{\Omega},\widehat{\Omega}') d\widehat{\Omega}', \qquad (2)$$

where $v$ is the light speed in the medium.

DCS BF measurements can be analyzed using the correlation diffusion equation (CDE)[3,58], derived from CTE using the standard diffusion approximation. The derivation procedure is summarized in Figure 3.

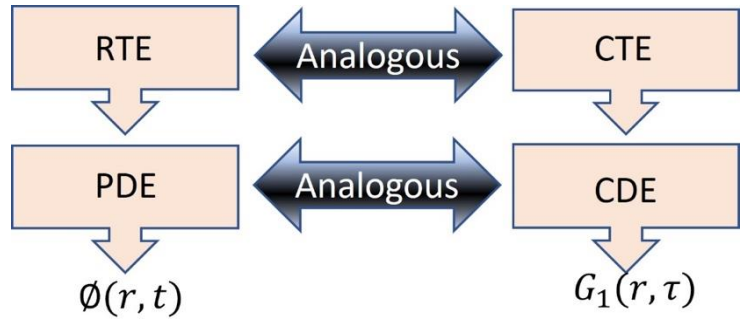

Fig. 3 Green's function for DCS derivation process

Furthermore, DCS instruments can be divided into three categories according to the light illumination strategy. Figure 4(a) depicts the simplest approach, which employs a CW laser. It is straightforward, and the instrumentation is relatively simple. The frequency-domain approach, Figure 4(b), utilizes an amplitude-modulated laser, with the modulation frequency set to the radio-frequency (RF) range (from tens to a thousand MHz). The time-domain (TD) approach, as shown in Figure 4(c), uses a short pulse laser and measures the delayed and temporally broadened output pulse. Time domain measurements have the most information content; however, they are more complex and expensive than the other two methods.

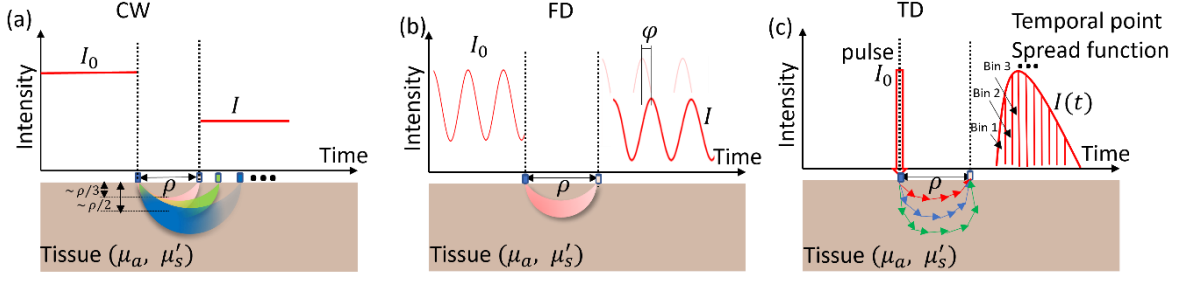

Fig 4. Three common illuminations schemes with the detected light intensity over time are depicted. (a) DCS data collected at different ρ to improve depth sensitivity for continuous wave (CW) light source. As a rule of thumb, the mean penetration depth in the reflection geometry is between one-third and one-half of the source-detector separation[95]. (b) intensity modulated (FD) light source, (c) time-resolved approaches.

The depth sensitivity of the DCS measurements can be improved using advanced techniques, such as TD- and FD-DCS. However, it may not be sufficient to minimize the superficial layer contamination. For this aim, different analytical models have been introduced to account for the contribution of the individual layers. These models typically include parameters of the optical system (e.g., the wavelength) and presumptions of optical tissue properties (e.g., $\mu_a$, $\mu_s'$, $n$) to fit mathematical models to the measurements. A summary of the analytical models commonly used in DCS analysis is shown in Figure 5.

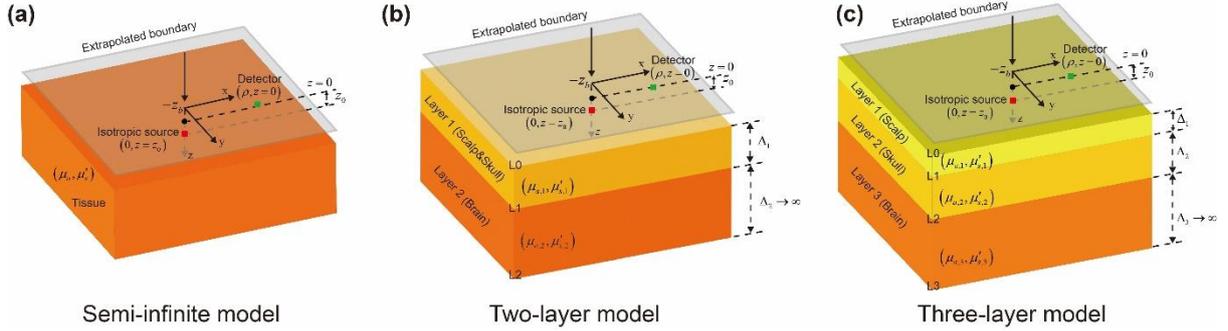

Fig. 5 Analytical models including the source and the detector for DCS (a) homogenous semi-infinite model, (b) two-layer analytical model, (c) three-layer analytical model. Here, $\mu_{a(n)}$ and $\mu'_{s(n)}$ are the absorption and reduced scattering coefficients in the $n$-th layer, respectively. $\Delta_i$ is the thickness of Layer $i$.

## 2.1 CW Semi-infinite homogenous (one layer) model

In traditional DCS systems, the tissue is commonly considered a homogenous semi-infinite medium, as shown in Figure 5(a). Under the standard diffusion approximation[96], we reduce Eq. (1) to CDE as:

$$\left(-\frac{D(r)}{v}\nabla^2 + \mu_a + \frac{1}{3}\alpha\mu_s' k_0^2 \langle \Delta r^2(\tau) \rangle\right) G_1(r,\tau) = S(r), \tag{3}$$

where $G_1(r,\tau) \equiv \langle \vec{E}(r,\tau) \cdot \vec{E}^*(r, t+\tau) \rangle$ is the electric field autocorrelation function. $D(r) = v/(3\mu_s')$ is the photon diffusion coefficient, $v$ is the speed of light in the medium, and $\mu_s' = \mu_s(1-g)$ is the reduced scattering coefficient, where $g \equiv \langle \cos\theta \rangle$ (ranging from -1 to 1) is the scattering anisotropy factor. $k_0$ is the wavenumber in the medium, $\alpha$ represents the probability that a light scattering event is with a moving scatterer (e.g., a flowing red blood cell), and $\langle \Delta r^2(\tau) \rangle$ represents the mean square displacement of moving scatterers, and is commonly described using two different models, including the Brownian motion and random ballistic models in biological tissues. For the Brownian motion,

$\langle \Delta r^2(\tau) \rangle = 6D_B\tau$[97], where $D_B$ is an 'effective' diffusion coefficient for moving particles. For random ballistic flow, $\langle \Delta r^2(\tau) \rangle = 6V^2\tau^2$, where $V^2$ is the mean square velocity of the scatterer in the vasculature.

In particular, for a semi-infinite, homogenous system with a point source $S(\vec{r}) = S_0\delta(r)$, $G_1(r,\tau)$ is the solution of Eq. (3), obtained using an image source approach following Kienle and Patterson[98] as,

$$G_1(\vec{r},\tau) = \frac{3\mu_s' S_0}{4\pi}\left[\frac{exp(-Kr_1)}{r_1} - \frac{exp(-Kr_2)}{r_2}\right], \quad (4)$$

where $K = \sqrt{3\mu_s'\mu_a + \alpha\mu_s'^2 k_0^2 \langle \Delta r^2(\tau)\rangle}$, $r_1$ and $r_2$ are the distances between the detector and the source/image source, respectively. $r_1 = \sqrt{\rho^2 + z_0^2}$ and $r_2 = \sqrt{\rho^2 + (z_0 + 2z_b)^2}$; $z_0 = 1/\mu_s'$ is the depth at which a collimated source on the tissue surface can be approximated as a point source; $z_b = 2(1 + R_{eff})/3\mu_s'(1 - R_{eff})$ and $R_{eff} = -1.440n^{-2} + 0.71n^{-1} + 0.668 + 0.0636n$ is the effective reflection coefficient, $n = \frac{n_{tissue}}{n_{air}} \approx 1.33$. Typically, $\alpha D_B$ is referred to as the blood flow index (BFi) in biological tissues[99]. In practice, the Brownian model can fit the observed correlation decay curves better over a wide range of tissue types, including rat [77,79,83,100], piglet [101,102], human brains[84,103–109], mouse tumours[110,111], human skeletal muscles[25,62–64,112], and human tumors[69,113–115].

## 2.2 CW two-layer model

We have stated above that the DCS theory is based on the correlation transport[55,92,93], approximated by CDE[94,96]. By assuming that light propagates in a homogenous medium, the simple solution of Eq. (3) has been widely used in the DCS community[116]. However, biological tissues[117] are usually layered encompassing unique physiological and optical properties[118,119]. Gagnon[120] *et al.* first proposed a two-layer analytical model, based on Kienle *et al.*'s model for reflectance spectroscopy with the two-layered geometry in Figure 5(b).

We assume that an infinitely thin beam shines the turbid two-layered medium. The first layer of the two-layer medium has a thickness $\Delta_1$, and the second layer is semi-infinite. The beam is scattered isotropically in the upper layer at a depth of $z = z_0$, where $z_0 = 1/(\mu_{a1} + \mu_{s1}')$. We also assume that the Brownian movement is independent in each layer, meaning that the particles can not move from one layer to another in the medium. The incident light is perpendicular to the surface of the turbid medium (on the x-y plane). Then Eq. (3) becomes:

$$\left(-D_1\nabla^2 + \mu_{a1} + \frac{1}{3}k_0^2\mu_{s1}'\langle \Delta r_1^2(\tau)\rangle\right)G_1^1(x,y,z,\tau) = S(x,y,z-z_0), \quad 0 \leq z \leq \Delta_1, \quad (5)$$

$$\left(-D_2\nabla^2 + \mu_{a2} + \frac{1}{3}k_0^2\mu_{s2}'\langle \Delta r_2^2(\tau)\rangle\right)G_1^2(x,y,z,\tau) = 0, \quad \Delta_1 \leq z, \quad (6)$$

where $D_i = 1/3(\mu_{a(i)} + \mu_{s(i)}')$ is the diffusion constant of Layer $i$. The mean-squared displacement $\langle \Delta r_i^2(\tau)\rangle = 6D_{B(i)}\tau$ for Layer $i$.

Although Kienle *et al.*'s derivations[121–123] are initially for diffuse reflectance spectroscopy (DRS), we re-derive them for DCS following the same procedure and obtain the solution of Eqs. (5) and (6) at z = 0 (Layer 1) in the Fourier domain by

$$\tilde{G}_1^1(q,z,\tau) = \frac{\sinh[\beth_1(z_b+z_0)]}{D_1\beth_1} \times \frac{D_1\beth_1\cosh[\beth_1(\Delta_1-z)]+D_2\beth_2\sinh[\beth_1(\Delta_1-z)]}{D_1\beth_1\cosh[\beth_1(\Delta_1+z_b)]+D_2\beth_2\sinh[\beth_1(\Delta_1+z_b)]} - \frac{\sinh[\beth_1(z_0-z)]}{D_1\beth_1}, \quad (7)$$

where $\beth_j^2 = (D_j q^2 + \mu_{aj} + 2c\mu_{sj}'k_0^2 D_{Bj})/D_j$, j =1 and 2, $q$ is the radial spatial frequency and

$$z_b = \frac{1+R_{eff}}{1-R_{eff}} 2D_1. \tag{8}$$

And $G_1^1(\boldsymbol{\rho}, z = 0, \tau)$ at $\boldsymbol{r} = \{\rho, z = 0\}$ on the medium surface is then obtained from the inverse spatial Fourier transform as,

$$G_1^1(\boldsymbol{\rho}, z = 0, \tau) = \frac{1}{2\pi} \int_0^\infty \tilde{G}_1^1(\boldsymbol{q}, z = 0, \tau) q J_0(q\rho) \, dq, \tag{9}$$

where $J_0$ stands for the zeroth order Bessel function of the first kind computed by the MATLAB function *besselj*.

## 2.3 CW three-layer model

Also, in the three-layer DCS model[103,124–126], $G_1(r, z, \tau)$ can be modelled similarly by CDE. A turbid medium consisting of 3 slabs was considered as shown in Figure 6(c). Each slab has a thickness $\Delta_p = L_p - L_{p-1}$, $p = 1, 2, 3$. To solve $G_1(r, z, \tau)$, Eq. (3) can be revised for the three-layer model as:

$$\left[\nabla^2 - \left(3\mu_a^{(p)} \mu_s'^{(n)} + 6k_0^2 \mu_s'^2 D_B^{(p)} \tau\right)\right] G_1(r, z, \tau) = -s_0 \delta(r - r'), \tag{10}$$

where $s_0$ is a point-like monochromatic light source located at $r' = \{\rho' = 0, z'\}$ inside Layer 1; $\rho$ represents the transverse coordinate. The field autocorrelation at the tissue surface, $G_1(r, z = 0, \tau)$, can be obtained by solving Eq. (10) in the Fourier domain with respect to $\rho$ as:

$$\hat{G}(\boldsymbol{q}, z, \tau) = \int d^2\rho \, G_1(r, \tau) \exp(i\boldsymbol{q} \cdot \boldsymbol{\rho}), \tag{11}$$

where $\boldsymbol{q}$ is the radial spatial frequency. Thus, in the Fourier domain Eq. (10) can be rewritten:

$$\left[\frac{\partial^2}{\partial z^2} - \kappa^2(\boldsymbol{q}, \tau)\right] \hat{G}(\boldsymbol{q}, z, \tau) = -s_0 \delta(z - z'), \tag{12}$$

where $\kappa_{(p)}^2(\boldsymbol{q}, \tau) = 3\mu_a^{(p)} \mu_s'^{(p)} + 6k_0^2 \mu_s'^2 D_B^{(p)} \tau + \boldsymbol{q}^2$.

We divided the top layer into two sublayers: Sub-layer 0 ($0 < z < z'$) identified by $p = 0$, and Sub-layer 1 ($z' < z < L_1$), identified by $p$ hereafter. The solution of Eq. (12) at Layer $p$ ($p = 1, 2, 3$) can be written as:

$$\hat{G}_p(\boldsymbol{q}, z, \tau) = A_p \exp(\kappa_{(p)} z) + B_p \exp(-\kappa_{(p)} z), \tag{13}$$

where $A_p$ and $B_p$ are constant factors for Layer $p$ determined by the boundary conditions:

$$\begin{cases} \hat{G}_0(\boldsymbol{q}, z, \tau) - z_0 \frac{\partial}{\partial z} \hat{G}_0(\boldsymbol{q}, z, \tau) = 0, & z = 0 \\ \hat{G}_0(\boldsymbol{q}, z, \tau) = \hat{G}_1(\boldsymbol{q}, z, \tau), & z = z' \\ \frac{\partial}{\partial z} \hat{G}_0(\boldsymbol{q}, z, \tau) = \frac{\partial}{\partial z} \hat{G}_1(\boldsymbol{q}, z, \tau) + 3\mu_s'^1, & z = z' \\ \hat{G}_p(\boldsymbol{q}, z, \tau) = \hat{G}_{p+1}(\boldsymbol{q}, z, \tau), & z = L_p, p = 1,2 \\ D_p \frac{\partial}{\partial z} \hat{G}_p(\boldsymbol{q}, z, \tau) = D_{p+1} \frac{\partial}{\partial z} \hat{G}_{p+1}(\boldsymbol{q}, z, \tau), & z = L_p, p = 1,2 \\ \hat{G}_3(\boldsymbol{q}, z, \tau) + z_3 \frac{\partial}{\partial z} \hat{G}_3(\boldsymbol{q}, z, \tau) = 0, & z = L_3 \end{cases} \tag{14}$$

where $z_0 \sim 1/\mu_s'^1$ and $z_3 \sim 1/\mu_s'^3$ are the extrapolation lengths taking into account internal reflections at external ($z = 0$ and $z = L_4$) boundaries.

Substituting Eq. (13) into Eq. (14), we can obtain $A_p$ and $B_p$ ($p = 1, 2, 3$). The Fourier transform $\hat{G}_0(\boldsymbol{q}, z, \tau)$ measured at $z = 0$ (the surface of the slab) is then obtained by substituting $A_0$ and $B_0$ into Eq. (13) under $\Delta_3 \to \infty$ to obtain:

$$\hat{G}_0(\boldsymbol{q}, z, \tau) = \frac{Num}{Denom}, \quad (15)$$

where $Num$ and $Denom$ when $p = 3$ and $\Delta_3 \to \infty$ are:

$$Num = 3\mu_s'^1 z_0 (\kappa_1 D_1 \cosh(\kappa_1(\Delta_1 - z'))(\kappa_2 D_2 \cosh(\kappa_2 \Delta_2) + \kappa_3 D_3 \sinh(\kappa_2 \Delta_2)) + \\ \kappa_2 D_2 (\kappa_3 D_3 \cosh(\kappa_2 \Delta_2) + \kappa_2 D_2 \sinh(\kappa_2 \Delta_2)) \sinh(\kappa_1 (\Delta_1 - z'))), \quad (16)$$

$$Denom = \kappa_2 D_2 \cosh(\kappa_2 D_2)\,(\kappa_1 (D_1 + \kappa_3 D_3 z_0) \cosh(\kappa_1 D_1) + (\kappa_3 D_3 + \\ \kappa_1^2 D_1 z_0) \sinh(\kappa_1 D_1)) + (\kappa_1 (\kappa_3 D_1 D_3 + \kappa_2^2 D_2^2 z_0) \cosh(\kappa_1 D_1) + (\kappa_2^2 D_2^2 + \\ \kappa_1^2 \kappa_3 D_1 D_3 z_0) \sinh(\kappa_1 D_1)) \sinh(\kappa_2 \Delta_2). \quad (17)$$

By performing the inverse Fourier transform of Eq. (15) with respect to $\boldsymbol{q}$, $\hat{G}_0(\boldsymbol{q}, z, \tau)$ can be obtained as:

$$G_0(r, \tau) = \frac{1}{(2\pi)^2} \int d^2 \boldsymbol{q}\, \hat{G}_0(\boldsymbol{q}, z = 0, \tau) \exp(-i\boldsymbol{q} \cdot \boldsymbol{\rho}) \\ = \frac{1}{2\pi} \int d\boldsymbol{q}\, \hat{G}_0(\boldsymbol{q}, z = 0, \tau) q J_0(\rho \boldsymbol{q}), \quad (18)$$

where $J_0$ denotes the first-kind zero-order Bessel function.

This three-layered solution has been tested with Monte Carlo simulations and used to analyze *in vivo* measurements [124,125,127].

Figure 6 (a), (b), and (c) show $g_1$ curves for semi-infinite, two-, and three-layer analytical models, respectively. Typically, in DCS data analysis, the measured $g_2$ is fitted with one of the models shown in Figure 6, using the Siegert relation $g_2(\tau) = 1 + \beta g_1^2(\tau)$. Usually, the homogenous semi-infinite analytical model is used in data analysis, assuming free diffusion for speckle decorrelation, giving rather poor agreement with experimental scenarios. This is because homogeneous fitting is more sensitive to the dynamic properties of the superficial layers. Compared with the semi-infinite model, two- and three-layered models can separate the signal between the superficial and brain layers. The layered models can mitigate the discrepancies between the one-layer model and realistic tissues. The accuracy of the three-layer analytical model has been investigated in previous studies[87,103,125]. Although multi-layered models provide a superior fit to measured data and are more accurate, they are susceptible to measurement noise, and much longer BFi estimation time is needed[126].

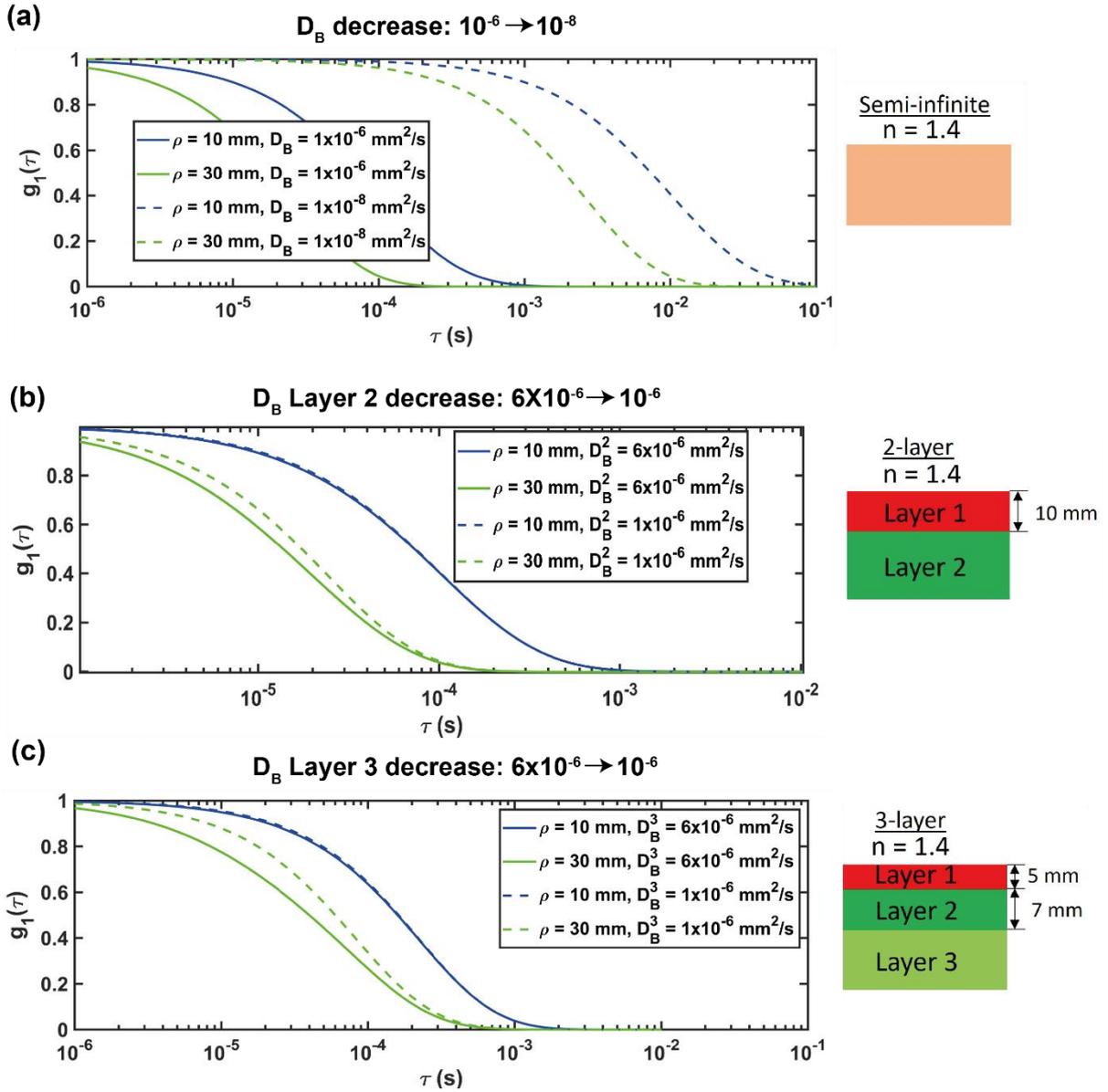

Fig. 6 (a) Representative $g_1(\tau)$ simulated from a sample with $\rho = 10\ mm$ (blue solid line) and $= 30\ mm$ (green solid line), varying $D_B$ from $1 \times 10^{-6} mm^2/s$ to $1 \times 10^{-8} mm^2/s$ (blue and green dot lines), $\mu_a = 0.013 mm^{-1}$, $\mu_s' = 0.86 mm^{-1}$, $\lambda = 785 nm$. (b) Representative $g_1(\tau)$ data simulated from a sample with $\rho = 10\ mm$ (blue solid line) and $= 30\ mm$ (green solid line), characterized with $\mu_a^{(1)} = 0.013 mm^{-1}$, $\mu_s'^{(1)} = 0.86 mm^{-1}$, $\Delta_1 = 10 mm$, $D_B^{(1)} = 1 \times 10^{-6} mm^2/s$, (Parameters for the top layer); $\mu_a^{(2)} = 0.018 mm^{-1}$, $\mu_s'^{(2)} = 1.11 mm^{-1}$, varying $D_B^{(2)}$ from $1 \times 10^{-6} mm^2/s$ to $1 \times 10^{-8} mm^2/s$ (Parameters for the bottom layer; blue and green dot lines); (c) Representative $g_1(\tau)$ data simulated from a sample with $\rho = 10\ mm$ (blue solid line) and $= 30\ mm$ (green solid line) characterized with $\mu_a^{(1)} = 0.013 mm^{-1}$, $\mu_s'^{(1)} = 0.86 mm^{-1}$, $D_B^{(1)} = 1 \times 10^{-8} mm^2/s$, $\Delta_2 = 5 mm$, (Parameters for the first layer); $\mu_a^{(2)} = 0.018 mm^{-1}$, $\mu_s'^{(2)} = 1.11 mm^{-1}$, $D_B^{(2)} = 1 \times 10^{-6} mm^2/s$, $\Delta_2 = 7 mm$ (Parameters for the second layer); $\mu_a^{(3)} = 0.03 mm^{-1}$, $\mu_s'^{(3)} = 1.19 mm^{-1}$, varying $D_B^{(3)}$ from $1 \times 10^{-6} mm^2/s$ to $1 \times 10^{-8} mm^2/s$ (Parameters for the third layer), the spatial frequency $q \in (0, 30] mm^{-1}$. All graphs are plotted using homemade software using MATLAB (Mathworks, Inc.).

## 2.4. TD semi-infinite (one layer) model

For TD-DCS systems, $G_1(\vec{r}, \tau, t)$ obeys the time-dependent correlation equation:

$$\left(-\frac{D(r)}{v}\nabla^2 + \mu_a + \frac{1}{3}\alpha\mu_s'k_0^2\langle\Delta r^2(\tau)\rangle + \frac{1}{v}\frac{\partial}{\partial t}\right)G_1(r,t,\tau) = S(r,t). \tag{19}$$

For a semi-infinite medium, it is straightforward to obtain the analytical solution of Eq. (19) under the boundary condition[128]. Thus $G_1(\rho, t, \tau)$ on the tissue surface ($z = 0$) is [129]:

$$G_1(\rho, t, \tau) = c\left(\frac{3\mu_s'}{4\pi ct}\right)^{\frac{3}{2}} \exp[-(\mu_a + 2\mu_s'D_B k_0^2\tau)ct] \exp\left(-\frac{3\mu_s'\rho^2}{4ct}\right) \times \left[\exp\left(-\frac{3\mu_s'z_0^2}{4ct}\right) - \exp\left(-\frac{3\mu_s'(z_0+2z_b)^2}{4ct}\right)\right]. \tag{20}$$

Thus, $g_1(\tau, s)$ for a photon pathlength $s$ can be written as:

$$g_1^{single}(\tau, s) = \frac{G_1(\rho, t, \tau)}{G_1(\rho, t, \tau=0)}$$

$$= \exp(-2\mu_s'D_B k_0^2 s\tau). \tag{21}$$

However, it is not easy to measure the pathlength of a photon in tissues. Therefore, the total scattered electric-field autocorrelation function $g_1(\tau, s)$ is obtained by incoherently summing the contributions over all $s$ [44,130]. Thus $g_1(\tau, s)$ is a weighted average over all possible pathlengths as:

$$g_1(\tau) = \int_0^\infty P(s)g_1^{single}(\tau, s)ds$$

$$= \int_0^\infty P(s)\exp(-2\mu_s'D_B k_0^2 s\tau)ds. \tag{22}$$

where $P(s)$ represents the probability that an incident photon travels a distance $s$ before emerging from the medium; it can be calculated as [131]:

$$P(s) = \frac{v}{(4\pi Ds/v)^{3/2}}\exp(-\mu_s s) \times \left[\exp\left(-\frac{r_1^2}{4Ds}\right) - \exp\left(-\frac{r_2^2}{4Ds}\right)\right], \tag{23}$$

where the variables are the same as in Eq. (4) and $s = vt$, with $t$ being the photon time-of-flight (ToF) and $v$ the speed of light in the medium.

By employing a sufficiently narrow time gate, Eq. (22) can be simplified, and the normalized time-gated $g_1(\tau)$ is modelled by a single exponential term:

$$g_1(\tau) = \exp(-2\mu_s'k_0^2 vtD_B), \tag{24}$$

Then $g_2(\tau)$ can be linked to $g_1(\tau)$ through the Siegert relation:

$$g_2(\tau) = 1 + \beta|g_1(\tau)|^2. \tag{25}$$

## 2.5. TD two-layer model

For the second layer model, Eq. (19) can be rewritten:

$$\left[\nabla^2 - \left(3\mu_a^{(p)}\mu_s'^{(p)} + 6k_0^2\mu_s'^2 D_B^{(p)}\tau\right) - \frac{3\mu_s'}{v}\frac{\partial}{\partial t}\right]G(r,\tau,t) = -3\mu_s'\delta(r-r'). \tag{26}$$

Similarly, we can derive the Fourier transform of $G(r, \tau, t)$ for the real space $(\rho, z)$, as well as time $t$, and then solve Eq. (26) in the Fourier space $(q, z, w)$.

$$\hat{G}(q, z, w, \tau) = \int dt \exp(iwt) \int d^2\rho\, G(\rho, z, t, \tau)\exp(iq\cdot\rho), \tag{27}$$

yielding

$$\left[\frac{\partial^2}{\partial z^2} - \left(3\mu_a^{(p)}\mu_s'^{(p)} + 6k_0^2\mu_s'^2 D_B^{(p)}\tau - 3\mu_s'^{(p)}\cdot\frac{iw}{c}\right) - q^2\right]\hat{G}(q, z, w, \tau) = -3\mu_s'\delta(z-z'). \tag{28}$$

The solution of Eq. (28) can be written as:

$$\hat{G}(q,z,w,\tau) = \gamma_p \exp(\Psi_p z) + \varphi_p \exp(-\Psi_p z), \tag{29}$$

where $\Psi_p = \sqrt{\left(3\mu_a^{(p)}\mu_s'^{(p)} + 6k_0^2 \mu_s'^2 D_B^{(p)} \tau - 3\mu_s'^{(p)} \cdot \frac{iw}{c}\right) + q^2}$, $\gamma_p$ and $\varphi_p$ are constant for Layer $p$ ($p = 1, 2$), determined by the boundary conditions:

$$\begin{cases}
\hat{G}_0(\mathbf{q},z,w,\tau) - z_0 \frac{\partial}{\partial z} \hat{G}_0(\mathbf{q},z,w,\tau) = 0, & z = 0 \\
\hat{G}_0(\mathbf{q},z,w,\tau) = \hat{G}_1(\mathbf{q},z,w,\tau), & z = z' \\
\frac{\partial}{\partial z}\hat{G}_0(\mathbf{q},z,w,\tau) = \frac{\partial}{\partial z}\hat{G}_1(\mathbf{q},z,w,\tau) + 3\mu_s'^1, & z = z' \\
\hat{G}_p(\mathbf{q},z,w,\tau) = \hat{G}_{p+1}(\mathbf{q},z,w,\tau), & z = L_p, p = 1,2 \\
D_p \frac{\partial}{\partial z}\hat{G}_p(\mathbf{q},z,w,\tau) = D_{p+1} \frac{\partial}{\partial z}\hat{G}_{p+1}(\mathbf{q},z,w,\tau), & z = L_p, p = 1,2 \\
\hat{G}_3(\mathbf{q},z,w,\tau) + z_3 \frac{\partial}{\partial z}\hat{G}_3(\mathbf{q},z,w,\tau) = 0, & z = L_3
\end{cases} \tag{30}$$

Thus, we can obtain the solution of Eq. (28):

$$\hat{G}_0(q, z=0, w, \tau) = \frac{3\mu_s' z_0 [\Psi_1 D_1 \cosh(\Psi_1(\Delta_1 - z_0)) + \Psi_2 D_2 \sin(\Psi_1(\Delta_1 - z_0))]}{\Psi_1(D_1 + \Psi_2 D_2 z_0)\cosh(\Psi_1 \Delta_1) + (\Psi_2 D_2 + \Psi_1^2 D_1 z_0)\sinh(\Psi_1 \Delta_1)}. \tag{31}$$

The inverse Fourier transform for $G(\rho, z, t, \tau)$ at $z = 0$ is:

$$G_0(\rho, z=0, t, \tau) = \frac{1}{2\pi} \int dw \exp(-iwt) \frac{1}{(2\pi)^2} \int d^2q \, \hat{G}_0(q, z=0, w, \tau) \exp(-i\mathbf{q} \cdot \boldsymbol{\rho}) =$$
$$\frac{1}{(2\pi)^2} \int dw \int dq \hat{G}_0(q, z=0, w, \tau) q J_0(\rho q) \exp(-iwt). \tag{32}$$

## 2.6. TD three-layer model

We start from Eq. (26), but derive similarly with Section 2.3 and $\Delta_3 \to \infty$, to obtain derive $G(\rho, z, t, \tau)$ for the three-layer model as the same with Eq. (32), where $\hat{G}_0(q, z=0, w, \tau) = \frac{Num}{Demo}$, where $Num$ and $Demo$ are shown below respectively,

$$Num = 3\mu_s' z_0 [\Psi_1 D_1 \cosh(\Psi_1(\Delta_1 - z'))(\Psi_2 D_2 \cosh(\Psi_2 D_2) + \Psi_3 D_3 \sinh(\Psi_2 D_2)) +$$
$$\Psi_2 D_2 (\Psi_3 D_3 \cosh(\Psi_2 D_2) + \Psi_2 D_2 \sinh(\Psi_2 D_2))\sinh(\Psi_1(\Delta_1 - z'))]. \tag{33}$$

$$Demo = \Psi_2 D_2 \cosh(\Psi_2 \Delta_2) [\Psi_1(D_1 + \Psi_3 D_3 z_0)\cosh(\Psi_1 \Delta_1) + (\Psi_3 D_3 + \Psi_1^2 D_1 z_0)\sinh(\Psi_1 \Delta_1)] +$$
$$[\Psi_1(\Psi_3 D_1 D_3 + \Psi_2^2 D_2^2 z_0)\cosh(\Psi_1 \Delta_1) + (\Psi_2^2 D_2^2 + \Psi_1^2 \Psi_3 D_1 D_3 z_0)\sinh(\Psi_1 \Delta_1)]\sinh(\Psi_2 \Delta_2) \quad .$$
(34)

$G_0(q, z=0, t, \tau)$ measured on the top of the surface ($z = 0$) of the slab is the inverse Fourier transform of $\hat{G}_0(q, z=0, w, \tau)$,

$$G_0(q, z=0, t, \tau) = \frac{1}{(2\pi)^2} \int dw \int dq \hat{G}_0(q, z=0, w, \tau) q J_0(\rho q) \exp(-iwt). \tag{35}$$

Figure 7 displays the numerical simulation $g_1$ for time-domain DCS from the semi-infinite, two-, and three-layer analytical models. Figure 7(a) is $g_1(\tau)$ for the early gate and late gate; Figure 7(b) is corresponding $g_2(\tau)$ for the early gate and late gate and Figure 7(c) is the $g_2(\tau)$ at different gate and lag time. Figure 7(d) is performed for $\rho = 10$ mm, two pathlengths are selected, t = 4.67×10$^{-10}$ s and t

= 9.34×10⁻¹⁰ s. Similarly, Figure 7(e) is performed for $\rho = 10$ mm, two pathlengths are selected, t = 4.67×10⁻¹⁰ s and t = 1.40×10⁻⁹ s.

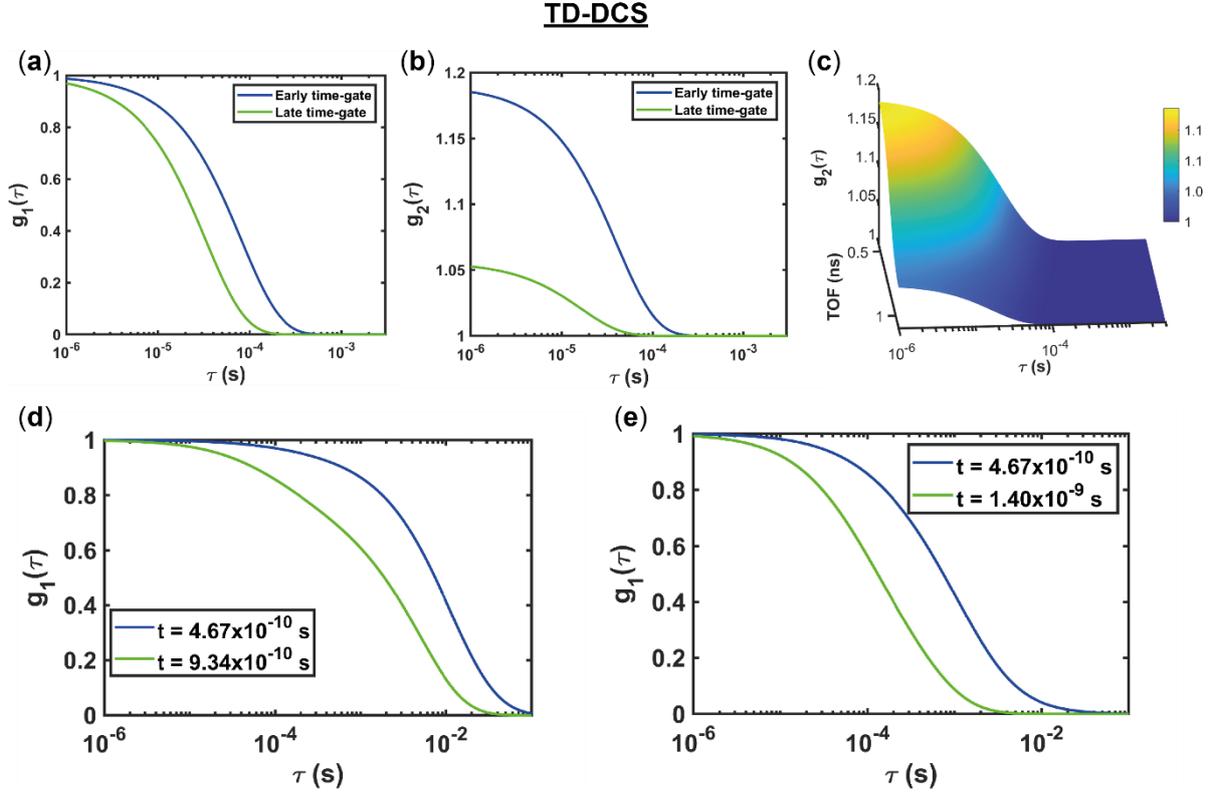

Fig. 7 (a) Simulated $g_1(\tau)$ with Eq. (24) $g_2(\tau)$ with (25), with $\rho = 10\ mm$, $D_B = 1.09 \times 10^{-8} mm^2/s$, $\mu_a = 0.013 mm^{-1}$, $\mu_s' = 0.86 mm^{-1}$, $\lambda = 785 nm$, $s = 135\ mm$ (ToF = 450 ps, data provided by Samaei[132]); (b) Simulated $g_1(\tau)$ from Eqs. (31) and (32) with $\mu_a^{(1)} = 0.013 mm^{-1}$, $\mu_s'^{(1)} = 0.86 mm^{-1}$, $\Delta_1 = 10 mm$, $D_B^{(1)} = 1 \times 10^{-6} mm^2/s$, $\mu_a^{(2)} = 0.018 mm^{-1}$, $\mu_s'^{(2)} = 1.11 mm^{-1}$, $D_B^{(2)} = 1 \times 10^{-6} mm^2/s$, $q \in (0, 30]$, $w \in (0\ 20]$Hz and $t = 4.67 \times 10^{-10} s$ and $t = 9.34 \times 10^{-10} s$. We adopted these parameters from Ref.[133] (c) $g_1(\tau)$ with $\mu_a^{(1)} = 0.013 mm^{-1}$, $\mu_s'^{(1)} = 0.86 mm^{-1}$, $D_B^{(1)} = 1 \times 10^{-6} mm^2/s$, $\Delta_1 = 2 mm$, $\mu_a^{(2)} = 0.018 mm^{-1}$, $\mu_s'^{(2)} = 1.11 mm^{-1}$, $D_B^{(2)} = 1 \times 10^{-7} mm^2/s$, $\mu_a^{(3)} = 0.03 mm^{-1}$, $\mu_s'^{(3)} = 1.19 mm^{-1}$, $D_B^{(3)} = 1 \times 10^{-6} mm^2/s$, $q \in (0, 30] mm^{-1}$, $w \in (0\ 20]$Hz, and $t = 4.67 \times 10^{-10} s$ and $t = 1.40 \times 10^{-9} s$. The settings are the same with Ref.[133].

## 2.7. Frequency domain semi-infinite model

We also obtain $G_1(\rho, \omega, \tau)$ when modulated illumination is used, $G_1(\rho, \omega, \tau)$ follows a slightly different CDE as:

$$\left[\nabla^2 - 3\mu_s'\left(\mu_a + 2\mu_s' k_0^2 D_B \tau - \frac{i\omega}{v}\right)\right] G_1(\rho, \omega, \tau) = -3\mu_s' s_0 e^{-i\omega t}, \quad (36)$$

where $\omega$ is the source modulation frequency and $s_0 e^{-i\omega t}$ is the modulated source term. For a semi-infinite homogeneous tissue, the solution of Eq. (36) is given by

$$G_1(\rho, \omega, \tau) = \frac{3\mu_s'}{4\pi} \left[\frac{\exp(-K_D(\omega,\tau)r_1)}{r_1} - \frac{\exp(-K_D(\omega,\tau)r_2)}{r_2}\right], \quad (37)$$

where $K_D(\omega, \tau) = \sqrt{3\mu_s'(\mu_a + 2\mu_s' k_0^2 D_B \tau - i\omega/v)}$ is the frequency-dependent wave vector. The other parameters are the same as before. Figure 8 shows $g_1(\tau)$ for the FD semi-infinite model. By fitting the measurement data from FD-DCS systems to Figure 8, we can extract optical properties ($\mu_a$ and $\mu_s'$) and blood flow simultaneously by multi-frequency measurements. In contrast, the traditional CW-DCS system is only used for blood flow measurements. Another merit is that the laser source for FD-DCS is much cheaper than CW-DCS and TD-DCS systems. There are two reasons: 1) FD-DCS removes the necessity for collocating the source and phase-sensitive detectors; 2) FD-DCS can be executed by simply substituting the source of a traditional DCS system with an intensity-modulated coherent laser.

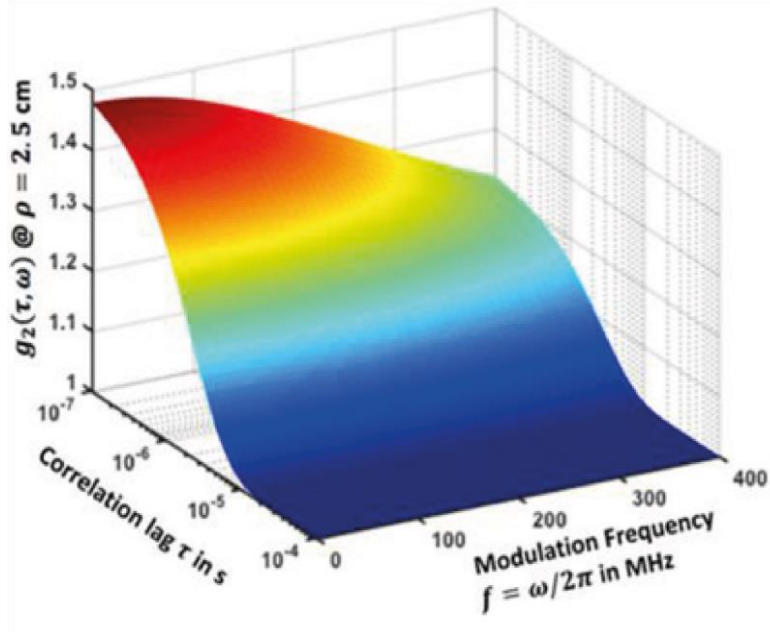

Fig. 8 Numerical simulated FD $g_1(\rho, \omega, \tau)$ at $\rho = 25mm$ with various modulation frequency. Image adopted from Ref.[91].

## 2.8. Noise model

In most simulation reports[134–137], a proper estimate of measurement noise is needed to reflect practical scenarios. A noise model suitable for photon correlation measurements was previously developed for a single scattering limit [138,139]. Later on, the noise model developed by Koppel[139] for fluorescence correlation spectroscopy (FCS) in the single scattering limit was introduced into DCS in 2006[79]. In DCS, the noise comes from photon counting statistics[138], and it has been derived[79] with the standard deviation of $(g_2(\tau) - 1)$, $\sigma(\tau)$ estimated as:

$$\sigma(\tau) = \sqrt{\frac{T}{T_{int}}} \left[ \beta^2 \frac{\left(1+e^{-T/\tau_c}\right)\left(1+e^{-\tau/\tau_c}\right)+2m(1-e^{-T/\tau_c})e^{-\tau/\tau_c}}{1-e^{-T/\tau_c}} + \langle M \rangle^{-2} \left(1 + \beta e^{-\tau/2\tau_c}\right) + 2\langle M \rangle^{-1} \beta (1 + e^{-\tau/\tau_c}) \right]^{1/2}, \qquad (38)$$

where $T$ is the frame exposure time (equal to the correlator bin time interval). $T_{int}$ is the integration time (measurement duration) or the measurement time window. $\tau_c$ is the speckle correlation time. $\langle M \rangle$ ($\langle M \rangle = IT$, where $I$ is the detected photon count rate) is the average number of photons within bin time $T$, $m$ is the bin index. To obtain $\tau_c$, $g_2(\tau)$ usually approximated with a single exponential function as $g_2(\tau) \approx 1 + \beta \exp(-\tau/\tau_c)$ under the Brownian motion model[79]. Once $\tau_c$ is obtained, we can obtain $\sigma(\tau)$. This noise model was then adopted by[140–143].

Fig. 9 shows noise (orange line) and noiseless (blue line) $g_2(\tau)$. The noise model predicted standard deviations for $g_2(\tau)$ at each $\tau$ was applied by randomly sampling a normal distribution, where the $T_{int} = 1s$ and 10 s and the delay time $1 \times 10^{-6} s \leq \tau \leq 1 \times 10^{-1} s$ (128 data points) was used. Considering realistic photon budgets, the photon count rate at 785 nm was assumed to be 8.05 kcps[135] at ρ of 30 mm. In Fig. 9, the DCS measurement noise decreases as $\tau$ increases.

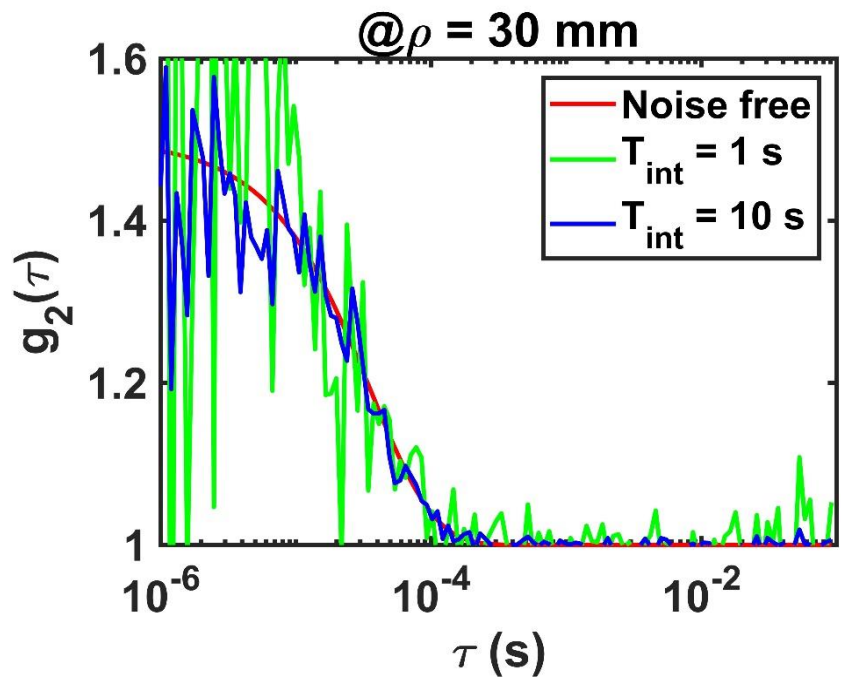

Fig. 9 simulated $g_2(\tau)$ curves with $\rho = 30mm$ on a homogeneous sample with $\mu_a = 0.01 mm^{-1}$, $\mu_s' = 1.2 mm^{-1}$, $\lambda = 785 nm$, $\beta = 0.5$, $T_{int} = 1 s$ (green line) and $T_{int} = 10 s$ (blue line), and $D_B = 2 \times 10^{-9} mm^2/s$, noise free (red solid line) and with Eq. (33) considered added assuming a 8.05 kcps at 785 nm[135].

## 3 Instrumentation

A DCS system consists of a laser source, source/detection fibers and sensors. Figure 10 shows representative systems for CW-, TD-, FD-, and Hybrid DCS systems. Figures 10(a) and (b) depict portable CW- and TD-DCS systems, respectively. The primary difference lies in the pulsed laser (VISIR-500) in the TD system. Figure 10(c) showcases the FD-DCS system, representing the latest DCS technology in the frequency domain. Lastly, Figure 10(d) presents a typical hybrid DCS system[144]. However, very few companies have initiated commercialization of DCS systems, including Hemophotonics, ISS Inc. (http://www.hemophotonics.com), and ISS Inc. (https://iss.com/biomedical/metaox).

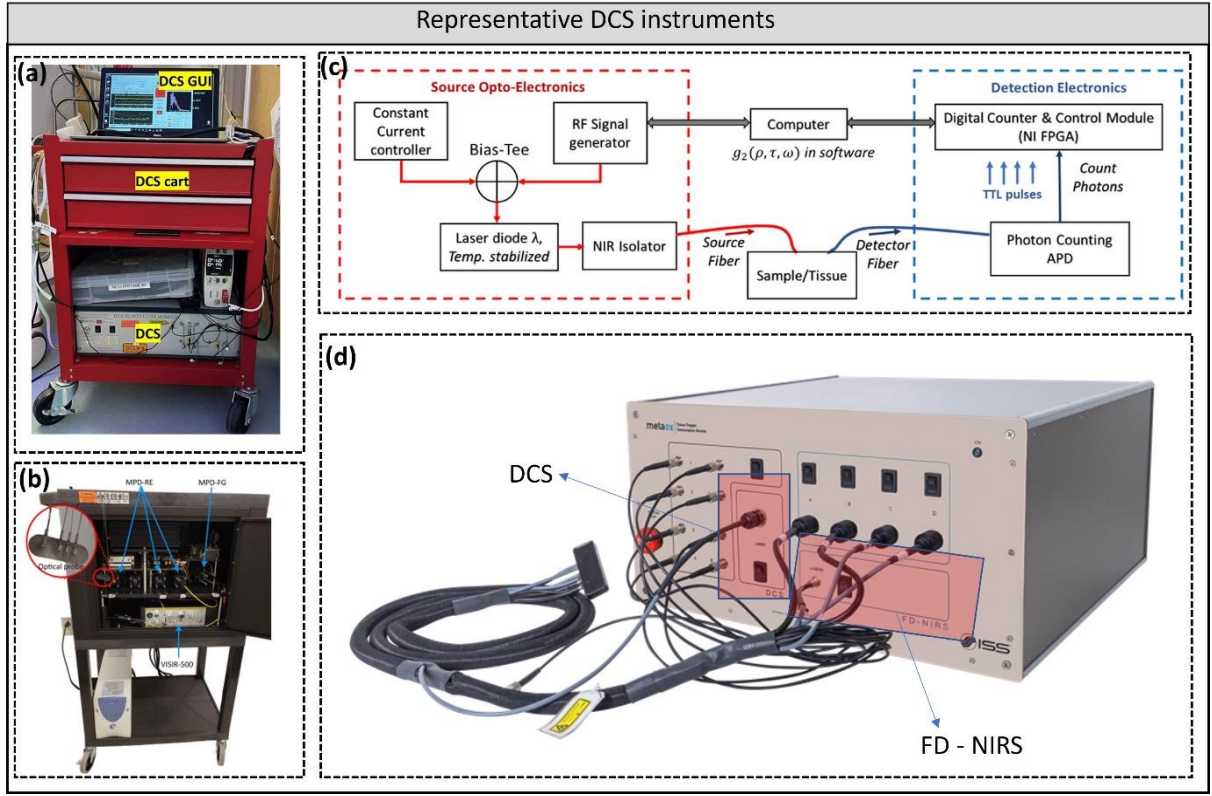

Fig. 10 (a) Sunwoo *et al.*'s CW-DCS system; the figure adopted from Ref.[145]; (b) Tamborini *et al*.'s TD-DCS system; the figure adopted from Ref.[146]; (d) Block diagram of Sadhu *et al.* FD-DCS system; the figure adopted from Ref.[147]; (d) Carp *et al.'s*[144] hybrid DCS system; the figure adopted from https://iss.com/biomedical/metaox.

## 3.1 Lasers

There are three types of laser used in DCS: CW, modulated, and pulsed lasers corresponding to CW-, FD-[147], and TD-DCS systems. As was mentioned above, the estimated BFi is derived from intensity fluctuations of the speckle pattern of back scattered light from the tissue surface, and the bright and dark patterns arise because photons emerging from the sample have travelled along different paths that interfere constructively and destructively at different detector positions[3,89,97]. Consequently, one of the main challenges is to select a laser with a long coherence length[97], $l_c$, designated by Eq. (39) assuming that the measured power spectral density has a Gaussian profile[148],

$$l_c = \frac{\lambda^2}{\Delta\lambda}, \tag{39}$$

where $\lambda$ is the central wavelength and $\Delta\lambda$ is the optical bandwidth. The diffusion theory and Monte Carlo simulations of light transport show that the minimum coherence length must be longer than the width of the photon path-length distribution[149], typically around $5\rho \sim 10\rho$ (e.g., 100 mm for $\rho = 10 mm$)[150]. For homodyne measurements, the coherence length needs to be substantially longer than the spread of pathlengths in tissue (which is within an order of $\rho$), and in heterodyne, care needs to be taken that the difference in length of the reference vs. sample arms, when summed with the expected pathlength variation, should also be substantially lower than the laser coherence length. Therefore, generally, the minimum coherence length is recommended as $l_{c,min} \gg 10\rho \sim 15\rho$, and since most practical DCS systems utilize $\rho \sim 3$ cm[3,135,151], the coherence length should be $35 \sim 50\ cm$, accounting for the variations of differential pathlength distances[152].

For clinical applications, the laser power should comply with the American National Standard for Safe Use of Lasers (ANSI)[153] limit for safe skin exposure with an proper irradiance. Spacers or prisms[150,154–

[156] are often between source fiber and sample to illuminate a larger area, which allows a higher laser power (more photons) while maintaining the same maximal permissible exposure (MPE) limit for intensity. Typically, lasers with wavelengths of 670 nm[157], 760nm[132], 785 nm[150,158], 850 nm[135,144,159], or 1064 nm[160] are employed. Although NIR wavelengths provide a higher number of photons for the same output power ($P = E/t = h\,c\,/\,\lambda$, E is photon energy), a higher MPE (more photons) and a deeper penetration depth, the photon detection efficiency (PDE) of most detectors is typically reduced for longer wavelengths. As a result, 785 nm and, more recently, 850 nm lasers are the most prevalent choice for most DCS techniques. This trade-off between the laser and the detector PDE is discussed in detail below.

Regarding TD-DCS, we can pinpoint the photons (either through gating or time-correlated single-photon counting[161]) that exhibit a similar path length in the tissue to provide depth-resolved information. This allows relaxing the requirement for a high coherence length compared with the scenario in which all the photon paths are considered. Moreover, the laser pulse width limits the maximum coherence length for a pulsed laser. Usually, a narrow laser pulse is preferable for precise depth-resolved measurements, however, a narrow pulse means a lower $l_c$, meaning a $g_2$ curve is closer to the noise floor. Therefore, there is a trade-off between $l_c$ and the pulse width[146]. In fact, $g_2's$ maximum amplitude depends on $l_c$, with β ranging from 0 for incoherence light to 1 for linearly polarized light (0.5 for unpolarized light) with $l_c$ longer than the longest photon path. Therefore, the main limitation of the broad use of TD-DCS is the availability of an ideal pulsed laser considering power settings, pulse width, coherence, stability, and robustness. To obtain a more in-depth investigation, readers can check Refs.[132,146,162]. In Table 1, we extend the conclusions made by Samaei *et al.*[132], Ozana, *et al.*[162] and Tamborini *et al.*[146] to show the relevant parameters of pulse lasers.

Table 1 Parameters of laser source used in TD-DCS, adopted from Samaei, *et al.*[132], Ozana, *et al.*[162] and Tamborini, *et al.*[163].

| Laser | Central wavelength (nm) | Temporal Coherence length (mm) | Spectrum bandwidth [nm] | Pulse width (ps) | Average output power (mw) |
|---|---|---|---|---|---|
| VIRIS-500 | 767 | 38 | N.A. | 550 | 50 |
| LDH-P-C-N-760 [132] | 760.4 | 6.1 | 0.095 | 106 | 12 |
| Ti: Sapphire [132] | 763.8 | 6.3 | 0.093 | 185 | 50 |
| VisIR-765-HP "STED"[163] | 765.7 | 38 | N.A. | 535 | ≤ 1500 |
| PicoQuant GMBH[162] | 1064 | 60 | N.A. | 600 | 100 |

### 3.2 Source and detection fibers

In DCS experiments, a pair of source and detection fibers are strategically placed on the tissue surface, with a separation of ρ (ranging from millimeters to centimeters). The laser emits long-coherence light through the source fiber into tissues, and the fiber collects the scattered light to a sensor. This distance ρ then defines the extent of the scattering paths of all detected photons, and thereby, the maximal measurement depth of DCS, as illustrated above in Fig. 2(d). The diagrams in Fig. 11(a), (b), and (c) illustrate three fibers with distinct modes, namely single-mode, few-mode, and multi-mode. Usually, a multi-mode fiber (core diameter D = 62.5, 200, 400, 600, 1000 $\mu m$)[115,162,164,165] is used for the source side. Here, it should be noted that a larger diameter fiber translates to a larger illumination area allowing a higher laser power (more photons) at the same MPE limit for intensity (see Section 3.1). For the detection, previously published DCS systems used single-mode (e.g., 5 $\mu m$)[17,70,115,164,166–174], few-mode[80,103], or multi-mode fibers[141,157,175,176]. Single-mode fibers are usually directly coupled to the respective detector. For parallelized DCS with SPAD arrays, multi-mode fibers are used for detection. In that case, the fiber is placed at a distance z to the detector, to match the speckle diameter (d) to the diameter of the detector's active area, according to Ref.[177]

$$d = \frac{\lambda z}{D}, \qquad (40)$$

Thus, adjusting the distance between fiber and detector (z) allows controlling the speckle size on the detector and therefore the number of measured speckles per pixel. Using single-mode fibers limits the measured light intensity because only the fundamental mode of light can be transported, limiting $\rho$'s dynamic range. Unlike conventional fibers, few-mode fibers allow not only the fundamental mode but also a few higher-order modes of light. Expanding the fiber diameter and numerical aperture (NA) in few-mode fibers to encompass multiple speckles enhances the detected signal intensity, consequently enhancing the signal-to-noise ratio (SNR). However, the multiple speckles detected by the few-mode fibers exhibit uncorrelated behaviour, and the decrease in $\beta$ largely counteracts the SNR enhancement. Finally, this flattens the autocorrelation function curve, potentially diminishing the sensitivity of DCS flow measurements[178,179]. To further increase the detected light intensity, multi-mode fibers with a larger core diameter have been used to accommodate larger sensor arrays (e.g., 5 × 5, 32 × 32, 192 × 128, 500 × 500 SPAD arrays). Usually, these SPAD arrays are set up in a way (Eq. (40)) that each pixel measures a single speckle on average. However, these detectors only have fill factors of 1-15% so there can be mismatches in the position.

Additionally, the PDE of SPAD arrays is often lower than single detectors, reducing $SNR$[141,157,175,180,181]. For more details on large SPAD arrays, see Section 3.3. He *et al.*[182] compared single-mode, few-mode, and multi-mode fibers on the detection side, and concluded that few-mode and multi-mode detection fibers can improve SNR compared with single-mode fibers, but it reduces $\beta$.

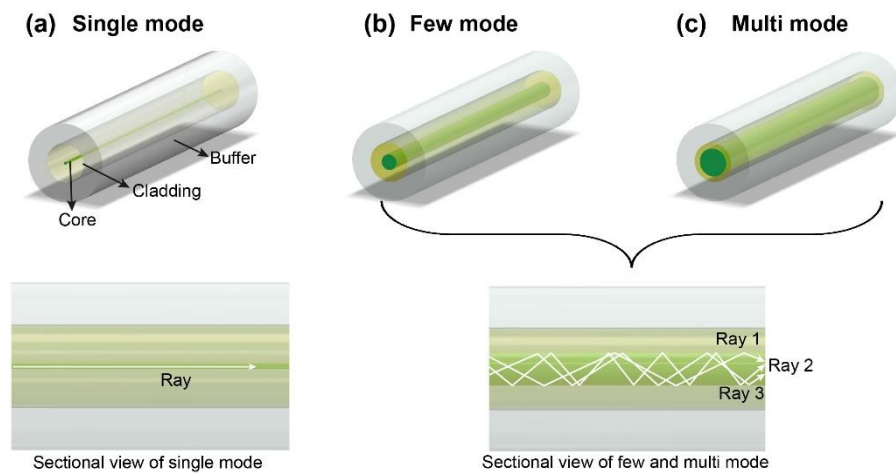

Fig. 11 Different optical fibers: (a) a single-mode fiber (SMF), (b) a few-mode fiber, (3) a multi-mode fiber.

### 3.3 Sensors

Detectors are pivotal in DCS systems for accurate BF measurements, with the advances being intricately connected to the adoption of new high-efficiency massively parallel detectors.

In early DCS systems, photomultipliers (PMTs) were commonly employed for detecting single photons[94,183]. However, PMTs are bulky, so early systems only contain a few channels. Additionally, driving these PMTs requires a high bias voltage, at least hundreds of volts, to start the electron multiplication process. These requirements pose challenges for developing compact and portable devices.

In the last two decades, avalanche photon diodes (e.g., APDs, such as the SPCM series, Excelitas, Canada)[137,167,182] were used nearly exclusively in DCS systems, replacing PMTs. APDs, known for their high sensitivity, leverage an internal avalanche multiplication effect for capturing single photons. These detectors offer several benefits compared with PMTs, including lower cost, simpler operations, and a smaller size. Although APDs offer high quantum efficiency, they are prone to higher dark current and noise in low-light conditions[184]. Additionally, these detectors are typically single-channel devices. In DCS, each speckle grain carries independent information about the dynamic scattering process. By averaging the autocorrelation signals from multiple speckles, we can enhance the SNR. However, advances in CMOS manufacturing technologies have enabled the integration of large SPAD arrays on a single chip, offering highly parallel single-photon detection.

Highly integrated CMOS SPAD arrays were boosted first by 3D/time-resolved fluorescence imaging applications[185–187], and later Richardson *et al.*'s low-noise SPAD structures[188] emerged from the EU6 MEGAFRAME project[189]. These SPAD arrays contain either time-correlated single-photon counting (TCSPC) or time-gating modules for time-of-flight or traditional photon counting measurements[190–196].

Using SPAD arrays in a multispeckle approach directly enhances SNR, with an enhancement of the square root of the number of independent speckle measurements. Using such new sensors in DCS experiments is straightforward without increasing the setup complexity. Dietsche *et al.*[105] verified this method by grouping 28 individual SPADs, enhancing SNR by $\sqrt{28}$. Johansson *et al.*[180] first developed a $5 \times 5$ SPAD DCS system to demonstrate an improved SNR on milk phantoms and *in vivo* blood occlusion tests, followed by $32 \times 32$[157,197–199], $192 \times 128$[181], and $500 \times 500$[60,175]. These systems significantly improve SNR by a factor of $\sqrt{N}$, where $N$ is the number of individual pixels. Figure 12 highlights the evolution of SPAD-based DCS systems (from APD to the state-of-the-art large SPAD arrays $500 \times 500$) with an enhanced SNR gain from 1 to ~500.

**PDCS – SNR gain of modern SPAD arrays**

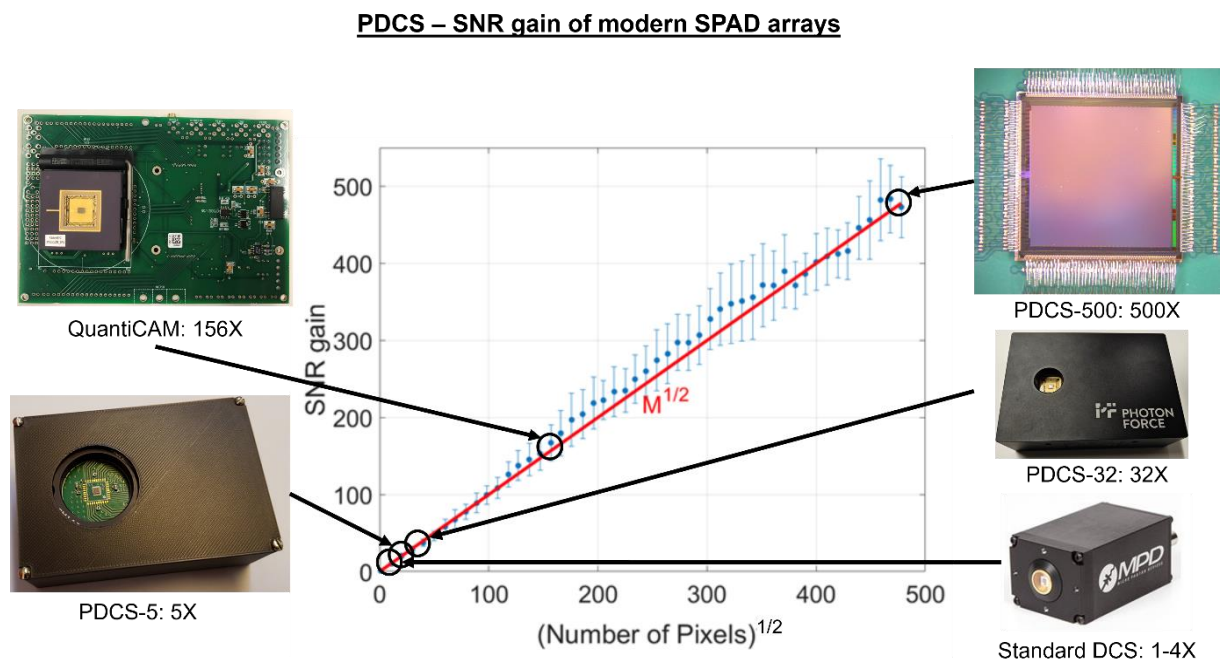

Fig. 12 SNR-vs-pixels plot adopted from Wayne *et al.*[175], with different SPAD sensors employed in DCS systems.

Besides SNR and PDE, the exposure time of SPAD arrays is another critical consideration, as it defines the interval between two adjacent time lags $\Delta\tau$ of the autocorrelation curves. Especially for fast decay rates (e.g., at large source-detector separations or for high flow rates), the relatively slow frame rate of large SPAD arrays (3 μs for $32 \times 32$[157,197,199] or 10 μs for $500 \times 500$[175]) can be a limiting factor in *in vivo* experiments. Another limitation of the SPAD arrays, though, is the difficulty in light coupling and the thinner active areas – thus an element of the SPAD array has a sensitivity lower than a dedicated SPAD. Nevertheless, the large number of elements allows one to exceed the performance of individual SPADs. Figure 13 shows the primary processing of a Parallelized DCS (PDCS) system[157].

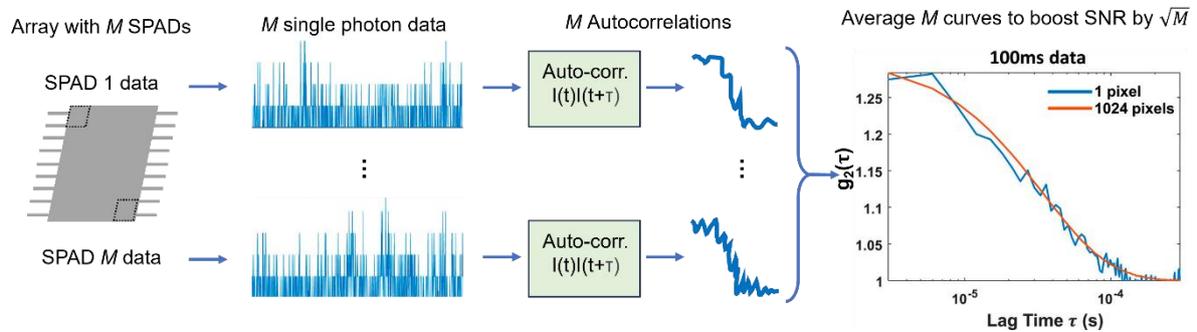

Fig. 13 An schematic layout of the SPAD array with representative raw data of temporal light intensity fluctuations from single pixels and the corresponding intensity autocorrelation curves. The blue and red lines in the rightmost figure represent the autocorrelation curves of a single pixel and the whole SPAD array (1024 pixels), respectively. Data and plots are adopted from Liu *et. al.*[157].

Commercial CMOS cameras are also used in DCS due to their larger array sizes, higher fill factors, and lower cost. However, they do not have single-photon sensitivity. To address this, Zhou *et al.*[200] employed a heterodyne detection method to enhance the signal; they also used MMFs to capture multiple speckle patterns, thereby increasing the throughput. They successfully conducted pulsatile blood flow measurements. Meanwhile, Liu *et al.*[201] integrated a CMOS detector into a wearable, fiber-free probe, enabling the testing of CBF in neonatal pigs. Of note, the heterodyne detection approach can also be applied in SPAD-based DCS systems, where it offers at least a doubling of SNR and reduced sensitivity to dark counts and environmental light[202].

Very recently, superconducting nanowire single-photon detectors (SNSPDs), a relatively new class of photo-detectors, have been used in TD-DCS systems[203,204]. SNSPD has many advantages, including a high PDE of >80% at longer wavelengths (e.g., 1064 nm), and a better timing resolution ($< 20$ ps)[205,206]. Nevertheless, SNSPD detectors come with a high cost, necessitating cryostats to maintain an operational temperature of 2 -3.1 K[206]. Moreover, their cooling time spans several hours, and they are noisy and emit a significant amount of heat, constraining their practical applicability in clinical settings. Table 2 summarizes the existing DCS systems with SPAD and representative non-SPAD sensors.

Table 2 shows the existing SPAD-DCS systems. Some SPAD are equipped with a TCSPC module, and TD-DCS systems can timetag detected photons to obtain their ToF, allowing distinguishing early and late arriving photons from fewer or more scattering events respectively, thereby enabling depth-resolved evaluation of BFi within tissues.

**Table 2** Existing DCS systems using SPAD array and other representative sensors

| Approaches | Detector | Laser, wavelength (nm) | $N_{pixel}$ | Applications | PDE | Fill factor | Frame rate (kHz/kfps) | $\rho$ (cm) | year | Ref. |
|---|---|---|---|---|---|---|---|---|---|---|
| CW | SPAD | 785 | $5 \times 5$ | Phantom, blood perfusion | 8% | 1.5% | 1000 | 2.5 | 2019 | [180] |
| CW | SPAD | 785 | $32 \times 32$ | Food, skin | 8% | 1.5% | 333 | 1.1 | 2020 | [141] |
| CW | SPAD | 670 | $32 \times 32$ | Phantom, in vivo | 16% | 1.5% | 333 | 2.1 | 2021 | [207] |
| CW | SPAD | 785 | $500 \times 500$ | Milk phantom, rotating diffuser | 15% | 10.6% | 92.2 | 3.3 | 2023 | [175] |
| CW | SPAD | 785 | $192 \times 128$ | rotating diffuser | 8% | 13% | 26 | N.A. | 2023 | [208] |
| CW | SPAD | 785 | $500 \times 500$ $128 \times 500$ | Human forearm and brain, in vivo | 15% | 10.6% | 100 for arm, 300 for brain | 4 | 2024 | [60] |
| iDCS | SPAD | 785 | $1 \times 1$ | Intralipid phantom | 61% | N.A. | N.A. | 3.6 | 2020 | [202] |
| LW-iDCS | InGaAs Linescan camera | 1064 | $2048 \times 1$ | Human brain, in vivo | N.A. | N.A. | 300 | 3.5 | 2023 | [209] |
| iDWS | CMOS | 852 | $512 \times 2$ | Human brain, in vivo | N.A. | N.A. | 333 | 2.5 | 2018 | [200] |
| fiDWS | Line-scan CMOS | 852 | $512 \times 2$ | Human brain, in vivo | >35% | N.A. | 333 | 4 | 2021 | [61] |
| πNIRS | CMOS | 785 | $1024 \times 1024$ | Forearm, forehead, human brain | 80% | N.A. | 16 | 2.5 | 2022 | [176] |
| TD | SNSPD | 785 | N.A. | Phantom, in vivo | 99% | N.A. | N.A | 1 | 2023 | [203] |

Notes: iDCS stands for interferometric diffuse correlation spectroscopy; iDWS is interferometric diffusing wave spectroscopy; fiDWS presents functional interferometric diffusing wave spectroscopy; πNIRS is abbreviation of parallel interferometric near-infrared spectroscopy, $\rho$ is source-detection separation; SNSPD stands for superconducting nanowire single-photon detectors; PDE is photon detection efficiency.

## 3.4 Correlators (incl. on-FPGA correlators)

To date, most DCS instruments employ commercial hardware correlators[11,66,83,84,111,112] to process detected signals and record the arrival of a Transistor-Transistor Logic (TTL) digital pulse for every photon from a photon counting detector. A commercial correlator[210], for example, uses the distribution of arrival times to quantify the temporal fluctuation of detected intensity. Traditionally, correlators embed a multi-$\tau$ processor[211–213] to compute the autocorrelation functions over a long delay period; this design was derived from early experiments in DLS[214] and diffusing wave spectroscopy (DSW)[105], primarily conducted on non-biological samples.

There are two kinds of hardware digital correlators, linear and multi-$\tau$ correlators, as shown in Fig. 14. Usually, the multi-$\tau$ framework is based on a logarithmic spacing spanning a massive lag-time range with a small number of channels without substantial sampling errors. Additionally, the multi-$\tau$ scheme significantly reduces the computational load compared with linear correlators. Although hardware correlators can operate at a faster sampling speed and offer real-time computing with a wide lag time dynamic range, they are relatively costly and not flexible since the fixed number of bits per channel results in a fixed lag time scale. Meanwhile, software correlators [215,216] (e.g., Fourier transform software correlators[217]) have also been developed. They show comparable performances with commercial hardware correlators and exhibit notable advantages in flexibility, cost-effectiveness, and seamless adaptability to evolving PC and data acquisition technologies. For most DCS applications with SPAD array, the autocorrelations are usually post-processed from raw data[175,180,197,207]. Table 3 shows the existing commercial correlators.

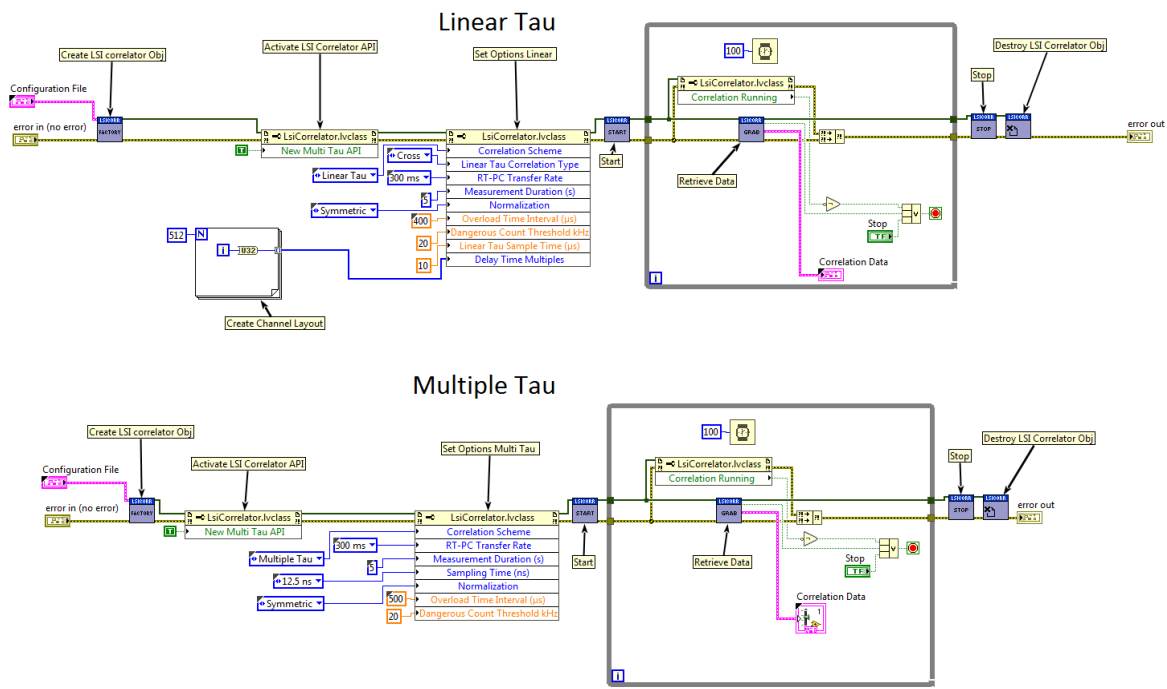

Fig. 14 Diagrams for linear- and multi-tau correlators provided by the CCO of LS Instruments, Dr Ian Block.

Table 3 Existing commercial correlator.

| Company | Correlator | Ref. |
|---|---|---|
| LSI Instruments | LSI Correlator | 218 |
| Becker & Hickl GmbH | SPC-QC-004 | 219 |
| ALV | ALV-5000/EPP | 220 |
| Photon Force, Ltd. | On-FPGA in PF's MF32 Sensor | 221 |

Note: Correlator.com provided good correlators to the DCS community.

### 3.5 Comparison between CW-, TD- and FD-DCS

Conventionally, enhancing depth sensitivity in CW-DCS measurements involves using a larger $\rho$. This allows detecting photons with longer pathlengths. An inherent drawback of this approach is the reduced detection of photons at a large $\rho$, reducing the SNR of $g_2$. Although Yodh et al.[130] have demonstrated pathlength-resolved DCS, their method required nonlinear optical gating and high laser powers, which are unsuitable for in vivo applications. Sutin et al.[129] first reported a novel time-domain (or pathlength-resolved) DCS on phantoms and a rat brain, showing the potential for clinical applications. Compared with CW-DCS, there are many advantages in TD-DCS:

Firstly, TD-DCS can measure the time point spread function (TPSF) of the tissue. Consequently, we can apply photon diffusion theories developed for time-domain near-infrared spectroscopy (TD-NIRS) to estimate tissue optical properties using the TPSF. Thus, we reduce errors in estimating dynamical properties, as we do not need to assume optical property values as traditional CW-DCS systems do[129].

Secondly, TD-DCS adds one further variable time, which can be exploited to select photons to increase the depth sensitivity[222]. Typically, the photons with a longer pathlength travelled deeper into the medium before reaching the detector. In contrast, those taking a shorter pathlength from source to detector reach only superficial tissue layers as shown in Fig. 4(c). Time-of-flight (ToF) measurements can achieve a higher depth resolution, as the ToF is proportional to the pathlength through the medium. Consequently, when computing the autocorrelation only with photons showing a ToF below a specific threshold, we can estimate the dynamic properties of the superficial layers, whereas a longer ToF allows for assessing deeper layers.

Thirdly, the pulsed laser utilized in the TD-DCS system can be integrated into the TD-NIRS setup[223]. This integration enables simultaneous measurements of NIRS and DCS, providing a comprehensive understanding of blood flow and hemodynamics variations. A temporal resolution of approximately one second and a favourable SNR in dynamic *in vivo* measurements was validated[224].

However, the primary obstacle preventing the broad adoption of TD-DCS is the need for an optimal pulsed laser (in power, pulse width, coherence, stability, and cost, around 6-fold more expensive than CW lasers). The effect of each of these factors has been evaluated in different studies, and various data processing strategies have been introduced to overcome the destructive influence of the instrument response function (IRF)[225] and the limited coherence length of the emitter. Moreover, Colombo et al.[226] demonstrated the contamination of non-moving scatters on the TPSF using a coherent pulsed laser utilized in the TD-DCS technique. Samaei et al.[132] have conducted the systematic discussion. Another drawback is that using narrow time gates to calculate the autocorrelation limits the SNR due to the scarcity of photons. Consequently, its applicability to *in vivo* experiments on human tissue is also restricted[224]. Although Ozana et al.[162] have designed a functional TD-DCS system that combines an optimized pulsed laser (a custom 1064 nm pulse-shaped, quasi transform-limited, amplified laser source), it is still costly, primarily due to the SNSPD.

Unlike CW-DCS, both TD- and FD-DCS can retrieve dynamic optical properties (e.g., BFi) and static optical properties (e.g., $\mu_a$ and $\mu_s'$), which are typically assumed in the conventional CW-DCS measurements. FD-DCS eliminates the requirement for collocated sources and phase-sensitive detectors, promising a portable and cost-effective system. Through data acquisition at a single $\rho$, FD-DCS effectively minimizes partial volume effects. This technology eliminates the need for extensive calibration in data analysis by acquiring flow and absorption from intensity-normalized data. FD-DCS shows high-speed acquisition, as flow and oxygenation information are inherently present in the dataset. Moreover, the implementation of FD-DCS is simplified by replacing a traditional DCS system's source with an intensity-modulated coherent laser. The detection mechanism remains unchanged, leading to reduced development time and cost.

Typically, to separate deep from superficial blood flow signals for CW-DCS, adding more detectors at different $\rho$ to obtain multiple-distance measurements is needed, as is shown in Fig. 4(a), which, however, increases the cost. Fig. 4(b) shows that FD-DCS can measure BFi, $\mu_a$ and $\mu_s'$ simultaneously. Table 4 summarizes representative existing TD-DCS systems. TD-DCS uses time-gating and TCSPC electronics to separate photons travelling the superficial layers from those that propagating deeper into the tissue, as illustrated in Fig. 4(c). General linear models (GLM) have been used for CW-DCS data from multiple source-detector separations ($\rho$) to regress out the effect of superficial flow. Therefore, large-$\rho$ DCS data is expressed as a linear combination of superficial blood flow (measured at a small $\rho$) and the desired deep blood flow[227], a method derived from fNIRS[228].

In contrast to fNIRS, which measures flow volume, DCS directly measures the BFi, which is related to flow speed. Since the flow speed differs significantly over different vessel diameters and tissue layers, the relation between superficial BFi and deep BFi is not linear. Therefore, new analysis tools that integrate additional data on vasculature structure are required to derive more accurate deep flow estimation from such multiple-distance DCS measurements.

Table 4 Representative existing time-domain DCS systems

| Year | Laser | Wavelength (nm) | Average power (mW) | Repetition rate (MHz) | Detection technique | IRF FWHM (ps) | Applications | Ref. |
|---|---|---|---|---|---|---|---|---|
| 2016 | DBR | 852 | 50 | 150 | Red-enhanced SPAD | 150 | Homogenous liquid phantom and small animal | 129 |
| 2017 | Ti: Sapphire | 785 | NA | 100 | SPAD | 100 | Two-layer liquid phantoms, forearm muscle, and adult human forehead | 224 |
| 2018 | VisIR STED, PicoQuant | 767 | 50 | NA | Red-enhanced SPAD | 500 | Homogenous liquid phantoms | 229 |
| 2018 | Ti:Sapphire | 785 | NA | 100 | Gated single-photon avalanche diode | 350 | Forearm muscle | 230 |
| 2019 | VisIR-500 | 767 | ≤ 1500 | ≤ 80 | SPAD | 550 | Homogenous liquid phantoms, forearm muscle, and adult human forehead | 163 |
| 2019 | Ti:Sapphire | 785 | NA | 100 | SPAD | 400 | Homogenous liquid phantoms | 225 |
| 2020 | Ti:Sapphire | 1000 | 30 | 100 | InGaAs PMT | NA | Homogenous liquid phantoms and forearm muscle | 231 |
| 2021 | LDH-P-C-760, Picquant | 760 | 12 | 80 | SPAD | 90 | Two-layer liquid phantoms, forearm muscle, and adult human forehead | 232 |
| 2022 | Custom-made two-stage fiber amplified pulsed laser | 1064 | 100 | 1-100 | SNSPD | 150-600 | Two-layer liquid phantoms and adult human forehead | 162 |
| 2023 | Ti:Sapphire | 785 | NA | 100 | SNSPD | 100-200 | Homogenous liquid phantoms and adult human forehead | 203 |

Note: SPAD stands for Single-Photon Avalanche Diode, and SNSPD stands for Superconducting Nanowire Single-Photon Detectors

## 4. Data processing

The accuracy and performance of multilayered analytical models have been extensively evaluated in prior literatures[120,124,125,127,232–235]. In addition to the analytical models described in Section 2, other data processing methods have been introduced to distinguish cerebral and extracerebral information. Baker *et al.*[236] introduced a pressure measurement paradigm combined with the modified Beer-Lambert law[237] and multi-distance measurement to reduce the extracerebral contamination from the signal associated with the deep layers. Furthermore, Samaei *et al.*[232] extended the bi-exponential model utilized in interferometric near-infrared spectroscopy (iNIRS)[238] to describe the TD-DCS signals influenced by scatterers moving at different speeds. They also conducted experimental validation using layered phantoms and *in vivo* experiments.

Traditionally, to extract BFi and β, we fit measured $g_2$ with Eqs. (4), (9), (18), (24), (32), and (35) in Section 2 by minimizing the cost function $\chi^2 = \sum_i [g_{2,analytical}(\rho, \tau_i) - g_{2,measured}(\rho, \tau_i)]^2$. Nonlinear least square fitting routines, e.g., Levenberg-Marquardt[103,136], fminsearchbnd[124] are usually used to quantify BFi. These approaches, however, are iterative, and sensitive to data noise. To address these constraints, the $N^{th}$-order (NL) algorithm[239,240], least-absolute minimization (L1 norm), and support vector regression (SVR) were introduced[143]. Yet, with the NL framework, the extraction of BFi is determined by the chosen linear regression approach[143]. Although L1 norm and SVR are novel for processing DCS data, they are sensitive to signal deviations[241]. For example, the computation time for BFi is 28.07 and 52.93 seconds[242] (using the Lenovo ThinkCentre M8600t desktop with a 3.4GHz CPU and 16GB memory) when employing L1 norm and SVR, respectively, still too slow for real-time applications.

In 1986, Dechter introduced "deep learning" (DL) to the machine learning community[243]. With rapid advances in computing technologies, DL has become a game-changer in many fields, including photonics[244], chemistry[245], biology[246], and medical diagnosis (such as electroencephalogram (EEG) and electrocardiogram (ECG)[247,248]), but is not yet broadly used in DCS. Recently, Zhang et al.[249] proposed the first recurrent neural network (RNN) regression model to DCS, followed by 2D convolution neural networks (2DCNN)[250], long short-term memory (LSTM)[251] and ConvGRU[252]. LSTM, as a typical RNN structure, has proven stable and robust for quantifying relative blood flow in phantom and *in vivo* experiments[251]. 2DCNN, on the other hand, tends to require massive training datasets for complex structures, demanding memory resources. ConvGRU, the newest deep learning method introduced to DCS, has also exhibited excellent performances in *BFi* extraction. Although the training of DL takes a long time, once it is done, DL is much faster than traditional fitting methods and more promising for real-time analysis and display. Fig. 14 and Table 5 summarize existing DL methods applied to DCS. It shows that DCS-NET's training is much faster than two-dimensional CNN, approximately 140-fold faster. Although the remaining models, RNN, LSTM and ConvGRU have fewer total layers, they are limited to a specific $\rho$[126]. Xu *et al.*[199] introduced a different DL approach and trained a deep neural network on DCS data of temporal speckle fluctuations from 12 fibers at different surface locations to reconstruct videos of flow dynamics 8 mm beneath a decorrelating tissue phantom. The reconstructed images had a millimetre-scale spatial resolution and a temporal resolution of 0.1-0.4 s.

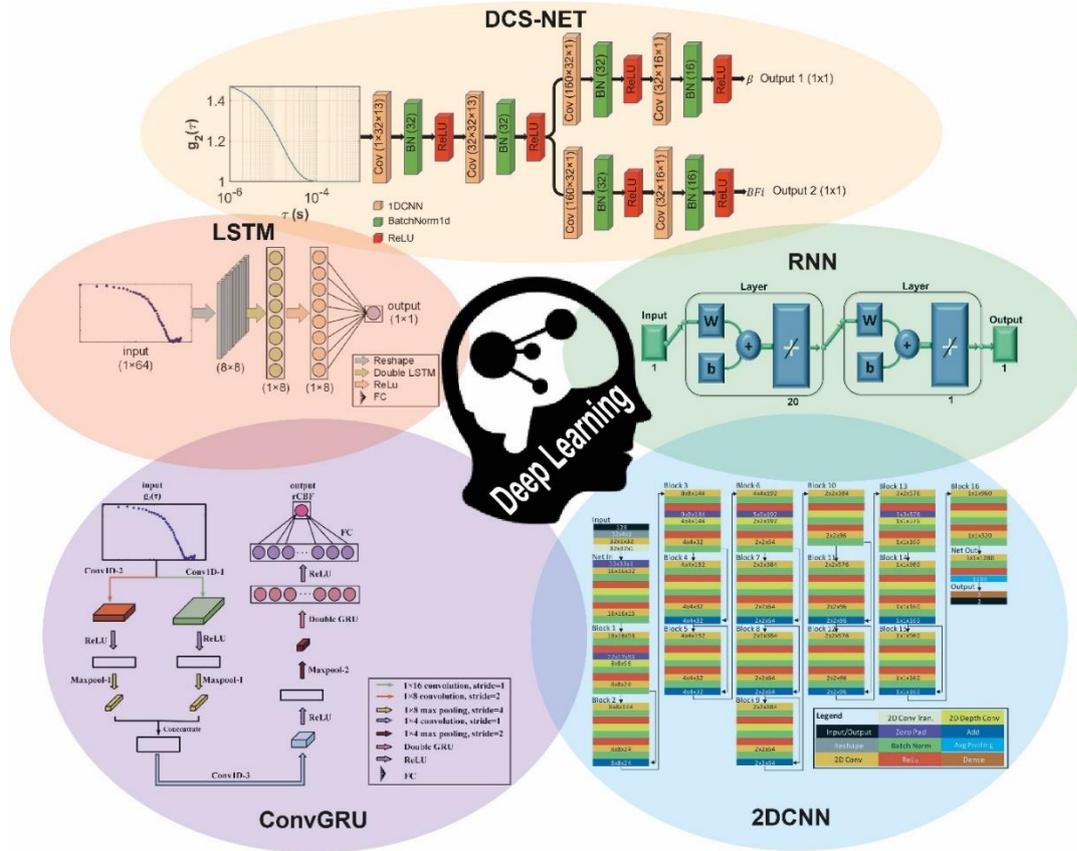

Fig. 14 The existing deep learning model applied in DCS, including RNN[253], 2DCNN[250], LSTM [254], ConvGRU[252] and DCS-NET. All of the graphs are re-printed from the published literatures.

**Table 5. Comparison of Existing AI methods for BFi estimation**

| Model | Training Parameters | Training time | Total layer | $\rho$ (mm) | Year |
|---|---|---|---|---|---|
| DCS-NET[126] | 25506 | ~ 13 (minute) | 18 | 5 to 30 | 2024 |
| RNN[249] | 174080 | N/A | 20 | 25 | 2019 |
| CNN(2D)[250] | 75552 | ~ 30.5 (hour) | 161 | 27.5 | 2020 |
| LSTM[251] | 1161 | N/A | 2 | 15 | 2021 |
| ConvGRU[252] | 11557 | N/A | 10 | 20 | 2022 |
| LSTM[174] | N/A | N/A | 5 | 30 | 2023 |

Note: the training parameters of RNN and CNN(2D) are not given in the literature; we calculate them according to the structure shown in the literature.

## 5. Applications

DCS has a broad range of applications. The integrated DCS systems with near-infrared spectroscopy (NIRS)[85], Doppler ultrasound, time-resolved near-infrared technique (TR-NIR)[255,256], and frequency-domain NIRS[90] are powerful for collecting abundant information[90]. This integration yields valuable insights into tissue oxygenation, blood oxygen metabolism, and hemodynamics[233,257]. Consequently, DCS has been used for neuromonitoring and has the potential for monitoring tissue and skeletal muscle blood flow, tumor diagnosis and therapy, and neonate cardio-cerebrovascular health evaluation.

The hypothesis proposed by Roy and Sherrington suggests that the increase in CBF is attributed to increased neuronal metabolic activity[258]. There exists a strong correlation between changes in CBF and psychological conditions. The tight coupling between neuronal activity and cerebral perfusion has been demonstrated in many articles[259–262]. As a result, studying regional CBF allows for observing local neuronal activities, performing diagnoses, and developing treatment procedures. Many articles have been published in the last decade. Interested readers can also refer to these review articles[3,85,88,263].

This section categorizes DCS applications into three categories: animals, human pediatrics, and human adults. The structure of this section includes: first, we provide a comprehensive overview of DCS applications in animals. We list application scenarios and preclinical trials. Next, we delve into DCS applications in neonates, focusing on perinatal care, cardio-cerebral diseases in neonates, neonatal brain development, and children's brain health. Finally, we explore DCS applications in adults, categorizing them into four sections: neuroscience, cardio-cerebrovascular diseases, skeletal muscle, and exercise physiology, as well as tumor diagnosis and therapy.

### 5.1 Animals

DCS has been applied to animals since the end of the 1990s to estimate the burn depth in pigs[264], as shown in Fig. 15(b). In 2001, Cheung *et al.* proposed a hybrid instrument integrating DCS with NIRS was applied to probe rat vascular hemodynamics[265]. Carp *et al.* used DCS to examine CBF during hypercapnia-induced cerebrovascular perturbation, with MRI-ASL as the standard measuring reference[100]. Furthermore, Menon *et al.* conducted the first DCS application in tumor monitoring[266]. They assessed tumor oxygenation in athymic nude mice (aged 6-8 weeks) bearing hypervascular human melanoma xenografts, achieved through vascular endothelial growth factor (VEGF) transfection. They combined DCS with Doppler ultrasound (DUS) to investigate microvessel density (MVD), BF, blood volume (BV), blood oxygen saturation, tissue oxygen partial pressure ($pO_2$), and oxygen consumption rate.

Moreover, DCS is pivotal in monitoring tumor blood flow changes in animal studies related to photodynamic therapy (PDT). Marrero *et al.*[267], Yu *et al.*[268], and Busch *et al.*[269] have employed DCS to monitor BF in tumors before, during, and after PDT. Sunar *et al.*[111] also used DCS to assess anti-vascular and ionizing radiation therapies. Farzam *et al.*[169] observed a dropped BFi in the high oxygen saturation tumor region using DCS and DOS after anti-vascular chemotherapy. These preclinical investigations have paved the way for human cancer research and clinical applications. Table 5.1 shows the DCS applications in animals.

Ischemia monitoring assesses potential damage to the brain or the secondary brain injury and paraparesis. Experiments have been conducted to study the perturbation of hemodynamics and cerebral blood metabolism induced by ischemia brain injury in rats[270], piglets[271] and sheep[272], see Fig. 15. Notably, Diop *et al.* developed a method integrating (TR-NIR) and DCS to quantify the absolute cerebral metabolic rate of oxygen ($CMRO_2$)[271,273].

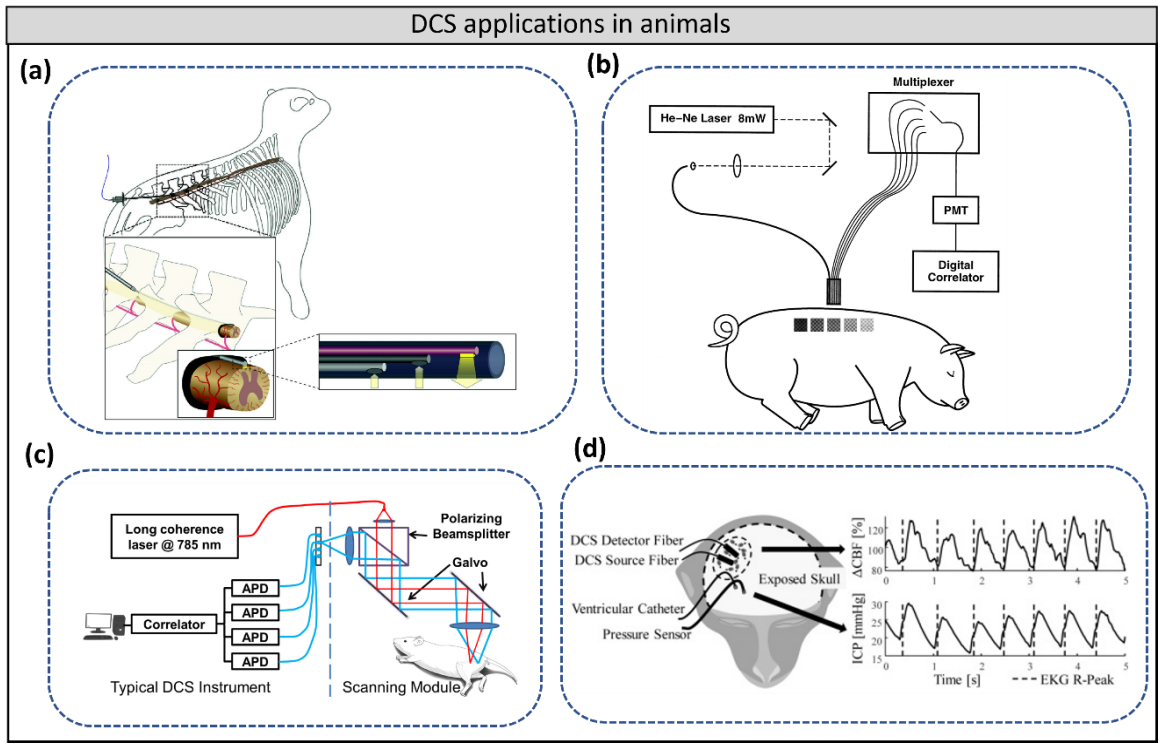

Fig. 15 (a) Detailed schematics of measurements on sheep, featuring the instrument and its thin fiber optic probe, images adopted from Ref.[272]; (b) The setup for pig experiments. The shaded areas on the pig indicate burns of various depths. Figures were reproduced from Ref[264]; (c) The non-contact scanning system set-up for mice. (black lines: outline of the bones; red lines: outline of the graft). Figures were reproduced from Ref.[274]; (d) Experiment setup with the placement of optical fibers and pressure sensor as well as the catheter on the exposed skull of monkey. The traces at the right show an example of changes in cerebral blood flow (ΔCBF) and ICP. Figures were reproduced from Ref.[275].

Table 5.1 Classification of DCS application on animals.

| Application | Subject | Overview | System | Reference |
|---|---|---|---|---|
| Hypercapnia | Rodent | rCBF and blood oxygen information | DCS & NIRS | 265 |
| | Rodent | DCS: rCBF; MRI: validation | DCS & MRI (ASL) | 100 |
| Tumor diagnosis/therapy | Rodent | Tumor oxygen status monitoring | DCS & DUS | 266 |
| | Rodent | Tumor blood flow monitoring before, during and after PDT | DCS | 267–269 |
| | Rodent | Anti-vascular therapy and ionizing radiation of malignant mouse melanoma tumor models | Contrast-enhanced DUS & DCS | 111 |
| | Rodent | Hemodynamic response to anti-vascular chemotherapy | DOS & DCS | 169 |
| Ischemia | Rodent | rCMRO2 | DOT & DPDW & DCS | 270 |
| | Piglet | Absolute CMRO2 and CBF | DCS & TR-NIRS | 271,273 |
| | Sheep | Spinal cord ischemia | DCS & DOS | 272 |
| Neurovascular coupling | Rodent | Effects of the secondary and late cortico-cortical transmission in neurovascular coupling | EEG & DOI & DCS | 276 |
| Head injury | Piglet | Hemodynamic changes monitoring after a head injury | DRS & DCS | 101,102 |
| Intracranial pressure | Monkey | Cerebral blood flow monitoring as an indicator of intracranial pressure | DCS | 275 |
| Diffuse optical correlation tomography | Rodent | Measuring blood flow contrast | DCT | 178,274,277 |

To further investigate vessel hemodynamics, diffuse correlation tomography (DCT) has been developed by measuring blood flow perturbation, contributed by optical heterogeneities, to provide blood flow contrast imaging of the region of interest[178,274,277]. DCT is a safe and cost-effective imaging technique, providing 3D imaging rather than just point measurements, real-time monitoring, and functional information on hemodynamics. DCT can complement other imaging modalities, such as MRI, CT, or PET scans, by providing additional functional and physiological information.

## 5.2 Pediatrics

The cortex of newborns is more easily detectable as the scalp and skull are much thinner in newborns and more light reaches the cerebral tissue than in adults. Thus, neonates are an attractive population for bedside DCS measurements, as discussed below. Generally, DCS is often combined with NIRS, which can measure human blood metabolism[278,279] or transcranial Doppler ultrasound (TCD). The synergy between these approaches enables comprehensive measurements of microvascular blood flow and oxygen metabolism in neonatal human subjects[85].

### 5.2.1 Perinatal care

Babies born before 37 weeks of pregnancy are premature, and preterm birth is the leading cause of neonatal mortality[280]. According to the World Health Organization (WHO) 2023 report, there are around 13.4 million premature babies worldwide[281]. Perinatal health refers to health from 22 completed weeks of pregnancy until seven completed days after birth. Premature babies are more likely to suffer from brain injuries such as HIE, stroke, and periventricular leukomalacia, related to neurological deficits[282]. To study brain hemodynamics and blood oxygen metabolism of premature neonates, Roche-Labarbe *et al*. developed a hybrid instrument combining DCS for measuring CBF and quantitative FD-NIRS for assessing cerebral tissue oxygenation ($StO_2$) and CBV. The results indicate that the CBF-CBV correlation is unstable in premature neonates[283]. In addition, Germinal matrix-intraventricular hemorrhage (GM-IVH) in premature neonates can be monitored by measuring CBF and $CMRO_2$ to identify the vulnerability of potential brain damage in newborns[284]. Buckley *et al*. used DCS for continually monitoring CBF in the middle cerebral arteries of low birthweight premature infants during a postural manipulation. They discovered a significant correlation between TCD and DCS measurements[81]. CBF monitoring during the first three days after birth was conducted to assess the risk of brain injury due to CBF instabilities in preterm infants[285]. DCS holds a promising potential for preterm human infants' brain health care.

### 5.2.2 Neonate cardio-cerebral diseases

DCS is also a promising tool for the monitoring of congenital heart defects in newborns. Durduran *et al*. used a hybrid NIRS-DCS instrument to study the changes in oxyhemoglobin, deoxyhemoglobin, total hemoglobin concentrations, $CMRO_2$, and CBF during hypercapnia. The validation of CBF and $CMRO_2$ was conducted using MRI-ASL, and the results showed a good agreement with DCS measurements ($R = 0.7$, $p = 0.01$)[286]. Buckley *et al*.[287] and Shaw *et al*.[288] measured changes in cerebral hemodynamics and oxygen metabolism during cardiac surgeries using DCS and DOS to evaluate the risk of surgery duration and surgical procedures, respectively. In addition, therapeutic hypothermia (TH) for neonatal HIE has also been studied using hybrid FD-NIRS and DCS[289,290]. TH is the standard of care for moderate to severe HIE in newborns[290]. Sutin *et al*. revealed the effects presented by therapeutic hypothermia (TH) on cerebral hemodynamics and blood oxygen metabolism by measuring CBF and $CMRO_2$. Researchers pointed out that $CMRO_2$ is a good indicator of TH evaluation and can be measured repeatedly at the point of care[290].

### 5.2.3 Neonates brain development

Hemodynamics and $CMRO_2$ are potential indicators of neonates' brain health and development[291–293]. DCS combined with FD-NIRS has also been used to monitor newborns' brain development, which revealed the differences of CBF in cortical regions and $CMRO_2$ in the frontal areas between male and female babies with the right-left brain functional asymmetry[294]. Besides, Dumont *et al*. used DCS to monitor activities in the somatosensory cortex of premature neonates to evaluate brain neurodevelopment[295].

### 5.2.4 Children brain health evaluation

Busch *et al*.[296] observed CBF attenuation in the brains of children (aged 6-16 years) diagnosed with obstructive sleep apnoea syndrome (OSAS) and hypercapnia using DCS. Besides, Nourhashemi *et al*.[297] combined EEG, NIRS, and DCS to simultaneously capture changes in electrical and optical dynamics in children (aged 6-10 years) affected by absence epilepsy. The outcomes revealed a consistent correlation among EEG, NIRS, and DCS, suggesting that DCS holds promise in detecting hemodynamic changes of pediatric brain disorders. Moreover, DCS has been employed for real-time CBF measurements during chronic transfusion therapy for children with sickle cell diseases[155,173,298]. Figure 15 shows representative applications of DCS in neonates.

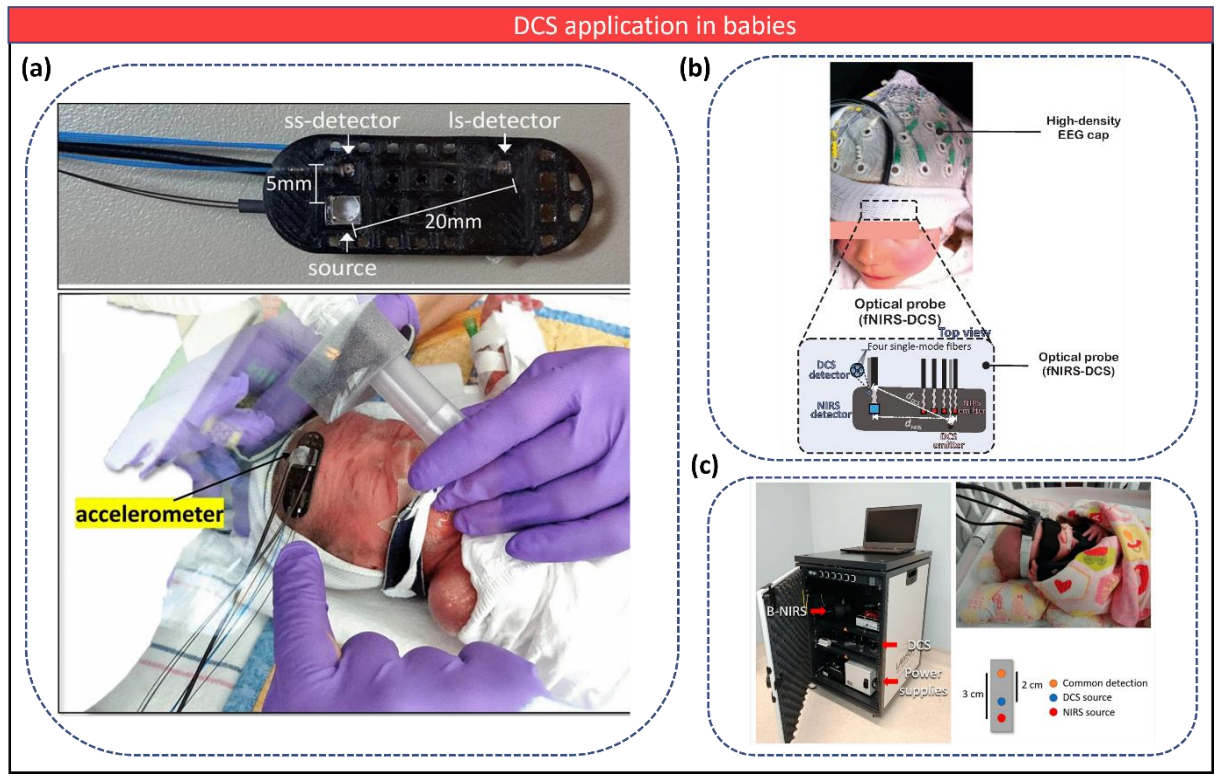

Fig. 15 (a) DCS sensor was attached to the infant's head for blood flow monitoring, figures adopted from Ref.[145] (b) The high-density EEG cap and optical probe (NIRS-DCS) and schematic representation of the location of the EEG and optical probes on a child's head. The figure was reproduced from Nourhashemi *et al.*[297]. (c) The hybrid DCS system for neonatal blood flow monitoring, figures reproduced from Ref.[299].

## 5.3 Adults

In this section, we focus on DCS applications in human adults and divide them into four sections: neuroscience study, cardio-cerebrovascular diseases, skeletal muscle and exercise physiology study, and tumor diagnosis and therapy evaluation. Figure 16 shows the use of DCS in adults.

### 5.3.1 Neuroscience study

Measuring CBF facilitates investigating neurovascular coupling, brain injuries, stroke, and neurological disorders. Neurovascular coupling denotes the connection between regional neural activity and subsequent alterations in CBF. The extent and spatial positioning of blood flow fluctuations are intricately connected to shifts in neural activity through a sophisticated sequence of coordinated processes involving neurons, glial cells, and vascular elements[300]. DCS can quantify changes in human cerebral blood flow in response to various stimuli, including but not limited to sensorimotor cortex activation[301], visual cortex activation[104,106], Broca's area activation[302], transcranial magnetic stimulation (TMS)[303], and vasoactive stimuli[304]. These studies presented noninvasive and straightforward means of monitoring cognitive neuronal activity in human brains. Older adults with mild cognitive impairment exhibit significantly higher CBF increments during motor and dual-task activities, whereas their counterparts display normal cognitive functions[305]. Another investigation highlighted the consistency of CBF with the posture changes within a healthy population (aged 20 to 78 years). Zavriyev *et al.* examined the role of DCS during hypothermic circulatory arrests (HCA) therapy among older people (mean age 61.8 ± 19.4 years)[154]. These findings offer good references for future research on age-related alterations in CBF[109]. In addition, DCS has been effectively applied for assessing cerebral hemodynamics under hypotension[306], obstructive sleep apnea[296], and adult comatose[307]. However, most state-of-the-art DCS setups are relatively limited for measuring blood flow in deeper cerebral tissue since the most common source-detector separations only enable measurements at ~1-1.5cm depth, which barely penetrates the non-cerebral tissues of the scalp and skull.

### 5.3.2 Cardio-cerebrovascular diseases

Several studies have assessed human artery diseases and treatment outcomes. Carotid endarterectomy (CEA), for instance, has been associated with hypoperfusion syndrome in the internal carotid artery (ICA), leading to potential cerebral ischemia. Evaluating and monitoring cerebral hemodynamics during and after CEA emerges as a critical measure to assess associated risks. Shang *et al.* conducted a comparative analysis between DCS and EEG, revealing that DCS-measured CBF exhibited more prompt responses to ICA clamping than EEG measurements[308]. Furthermore, Kaya *et al.* integrated DCS with NIRS to demonstrate the feasibility of real-time cerebral hemodynamics and oxygen metabolism monitoring during CEA procedures[309]. Mesquita *et al.* also established a physiological connection between CBF and oxygenation in patients with peripheral artery disease[310]. CBF during the cardiac cycle has been acquired using DCS before and during ventricular arrhythmia in adults[311]. DCS has also been used for monitoring CBF[312,313] and critical closing pressure (CrCP)[314] of ischemic stroke patients, intrathecal nicardipine treatment after subarachnoid hemorrhage[315], and thrombolysis therapy evaluation in ischemic stroke[316]. Notably, in the neurocritical care unit, DCS coupled with NIRS is a good bedside monitoring tool for individualized CBF management and manipulation of head-of-bed treatment for patients with critical brain injuries[151,317].

### 5.3.3 Skeletal muscle and exercise physiology

DCS has found applications in investigating human skeletal muscle physiological states, offering a valuable approach for assessing tissue vascular diseases and enhancing clinicians' understanding. For instance, Yu *et al.* compared muscle blood flow and oxygenation between healthy individuals and those with peripheral arterial disease during cuff occlusion and plantar flexion exercise[318]. Subsequently, they integrated MRI-ASL with DCS to monitor BFi during cuff inflation and deflation[319]. Shang *et al.* characterized muscle blood flow, oxygenation, and metabolism in women with fibromyalgia during leg fatiguing exercise and arm arterial cuff occlusion[320]. Matsuda *et al.* evaluated local skeletal muscle blood flow during manipulative therapy (MT), suggesting that MT can enhance blood flow with minimal effects on systemic circulatory function[321].

Nevertheless, conventional technologies such as DUS, electromyography (EMG), and MRI encounter challenges when measuring physiological signals due to motion-induced artifacts, leading to inaccurate blood flow measurements. DCS offers more reliable measurements against experimental variations[322]. However, it is noteworthy that muscle fiber motion artefacts may still result in overestimating the change in BFi. Sang *et al.* have proposed methods to extract accurate blood flow measurements, including the co-registration of a dynamometer[64]. Alternative techniques, such as hardware-integrated gating[17,323] and a random walk correction model with FD-NIRS[68], have been introduced to address fiber motion artefacts in DCS measurements.

### 5.3.4 Tumor diagnosis and therapy evaluation

DCS has been employed in the diagnosis of human breast cancer, prostate, and neck tumors. Durduran *et al.* conducted an initial comparative analysis of blood flow disparities between tumor and normal tissues in the human breast. The investigation revealed a noteworthy increase in blood flow within tumor tissues[324]. This observation paves the way for noninvasive tumor diagnosis. Choe *et al.* used DCS in human breast cancer diagnosis[74]. Besides, the findings align with the results reported by Durduran *et al.*, which underscored the increased blood flow within tumor regions. Besides, noncontact DCT has been adopted for three-dimensional (3-D) visualizing of blood flow distribution in human breast tumors, showing that DCS is a promising technique for localizing human tumors[325].

Yu *et al*. combined DCS with NIRS to measure BF and oxygenation in human prostate cancer[114] and head/neck tumors[113] to assess treatment. Also, DCS has been used to evaluate the photosensitizer 2-1[hexyloxyethyl]-2-devinylpyropheophorbide-a (HPPH)- mediated PDT (HPPH-PDT). They showed that HPPH-PDT could induce a significant drug photobleaching with a reduction of blood flow and blood oxygenation[326]. In addition, DCS can evaluate chemotherapy[327,328] or radiation delivery[329] in human tumors.

However, more patient statistics are needed when DCS is applied to human tumor diagnosis and therapy. Currently, most earlier prediction studies involved 7 to 11 patients[330], segmented into two or three response groups. This is additionally complicated by different definitions of responding and non-responding groups utilized by each research team. Longitudinal studies in larger patient populations for a more extended monitoring period are needed for more precise clinical application references. Besides, more precise DCS theoretical models according to application scenarios are also needed[90,331].

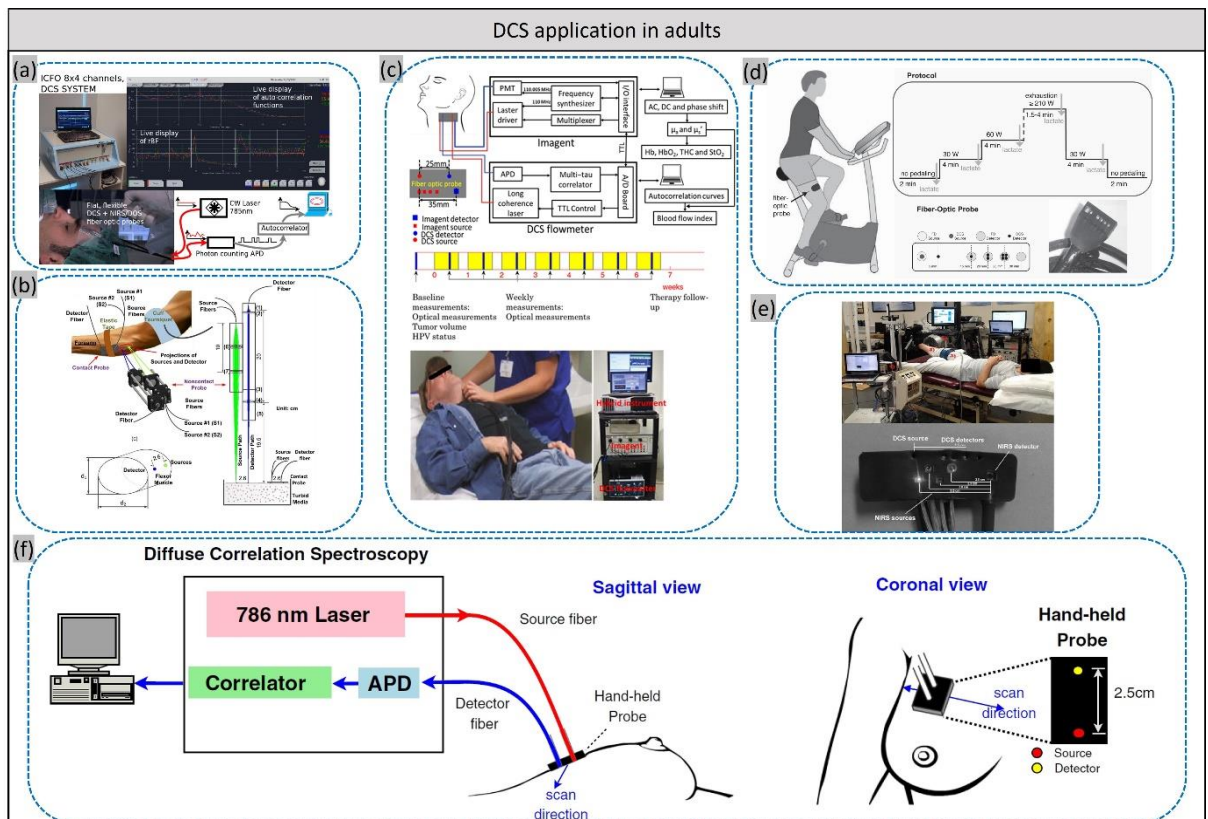

Fig. 16 (a) Hybrid DCS system applied to the human forehead, image reprinted from Ref.[332]; (b) Experimental configuration with a contactless probe, figures adopted from Ref.[333]; (c) Schematic of hybrid instrument, hybrid Imagent/DCS instrument for simultaneous measurement of tumor oxygenation and blood flow during chemoradiation therapy, images adopted from Ref.[334]; (d) Drawing of a subject cycling on a stationary bicycle with a multi-distance FDNIRS-DCS probe attached to the right superficial rectus femoris. The figure was adopted from Ref.[154]; (e) Hybrid DCS/NIRS device for muscle measurement. Figures were adopted from Ref.[323]; (f) Diagram of DCS working on a breast, figures adopted from Ref.[325].

In addition to the applications listed above (Figures 14, 15, and 16), DCS has also been used for critical care[335], anesthesiology[336], and thyroid blood flow measurements[337]. DCS is a relatively new and evolving technology, and its applications continue to expand as new studies emerge and sensor technologies advance. DCS's non-invasive and portable nature makes it particularly attractive for studying dynamic physiological processes *in vivo*. Theory models have evolved from semi-infinite to multi-layer models to obtain more precise measurements and expanded from CW-DCS to TD- and FD-DCS. Besides, DCT can visualize blood flow contrast deep in tissues, making it more understandable for blood-related disease diagnosis and therapy. However, DCT shows a limited SNR and requires a long data processing time; therefore, it is still not applied to clinical applications. Thus, efforts are necessary to propel the development of DCS further, and we expect that DCS will offer increasingly reliable BFi measurements and find expanded applications in the future.

## 6. Discussion and outlook

Non-invasive DCS techniques have great potential for early diagnosis, prognosis, and a broad range of clinical conditions. Although DCS is simple and cost-effective, human applications still face challenges. Increasing DCS's SNR is crucial for effective probing through thick near-surface tissue layers, especially at larger source-detector separations. A solution for increasing SNR is simply increasing the amount of light delivered to tissues under the maximum permissible exposure (MPE) limited by safety standards (ANSI safety limit[338]) or using high photon detection efficiency sensors that collect more scattered photons. Additionally, with new CMOS manufacturing techniques, the improvement in SNR has been shown in multi-speckle DCS systems using SPAD arrays with $5 \times 5^{180}$, $32 \times 32^{197,199,207}$, $192 \times 128^{181}$, or $500 \times 500^{175}$ pixels. The latest parallelized DCS system with a SPAD array of 500×500 pixels has already been demonstrated to boost the SNR by 500, compared to a single SPAD pixel of the same device. In 2020, a SPAD camera with 1024×1000 pixels was demonstrated[195], although its relatively low frame rate of 24 kfps still prevented a practical use in DCS. We believe this ongoing development of larger and faster SPAD technologies[195,339] will continue to boost the SNR of DCS, thereby allowing feasible measurements at longer source-detection separation and effectively enabling the measurement of deeper blood flow.

Another method that has a similar goal is the interferometric approach based on a Mach-Zehnder interferometer. Over the past five years, the interferometric detection for diffusely scattered light in biological tissues has been investigated[61,200,202,238,340–345]. There are many advantages, including:

1) offering comparable or superior functionality to photon counting but at a significantly lower cost per pixel[61,202,342,346];

2) altering the temporal coherence of light proves to be an effective and adaptable method for attaining Time-of-flight (ToF) resolution or discrimination within an interferometric arrangement, eliminating uncertainties for precise signal interpretation[238,343,344,346];

3) holding significant promise for analyzing blood flow fluctuations, whereas conventional DCS is hindered by its expensive nature and limited throughput[61,105];

4) insensitive to ambient light, which is a considerable benefit for practical use cases. Recently, Robinson *et al.*[209] proposed long wavelength (1064 nm), interferometric DCS (LW-iDCS), which outperforms the long wavelength DCS (LW-DCS) based on SNSPD[347] in terms of SNR and implementation cost. However, the drawback of this approach is its relatively complex setup with a reference arm and higher stability requirements for the platform accommodating the setup.

One of the substantial advantages of the TD-DCS technique, as described in Section 3.5, is its capability to reduce the superficial layer contamination by selecting photons propagated into the deep tissues. Although TD-DCS measurements are typically conducted at a short ρ, due to the limited coherence length of the currently available emitters, this feature overcomes the influence of short ρ measurements and provides a higher depth sensitivity than CW-DCS methods. Therefore, TD-DCS requires a pulsed lasers and a TCSPC (or time-gating) module, which increases cost. To reduce the cost, Moka *et al.*[147]

proposed FD-DCS. A faster acquisition speed can be achieved using FD-DCS as BF and oxygenation information is implicit in the collected data. This can be a good solution for some traditional DOS and DCS systems. Moreover, implementing FD-DCS is simplified using an intensity-modulated coherence laser, which can be cost-effective.

Indeed, large arrays comprising thousands of SPADs equipped either with in-pixel Time-to-Digital Converters (TDCs)[348–350] or with a set of TDCs shared across various pixels[351,352] are being developed in cost-effective CMOS process. Despite recent advances in SPAD technologies, state-of-the-art TD-DCS has not yet been implemented using TCSPC techniques based on TDC techniques[353–358]. There is no doubt that large SPAD arrays with embedded TCSPC can be a parallelizable solution for next-generation TD-DCS, with a potential breakthrough in the SNR of the measurements and the depth-encoding. We expect this kind of TD-DCS system to be released in the coming years.

Combining DCS and DRS[66,112,265] for concurrent BF and oxygenation measurements is also a trend. Quantifying blood oxygenation, metabolism, and tissue BF is essential for the diagnosis and therapeutic assessments of vascular/cellular diseases[359–364]. However, most relevant instruments assess tissue hemodynamics and metabolism by employing optical probes in direct contact with tissue surfaces. Contact measurements pose notable challenges, such as an elevated risk of infection in ulcerous tissues and potential deformation of delicate tissues (e.g., breasts and muscles) due to probe-tissue contact. This deformation can lead to distortions in the measured tissue properties. Thus, noncontact probes have been designed for deep tissues[165,365,366].

Regarding data processing, traditional nonlinear fitting methods are usually based on analytical models (homogenous semi-infinite one-[367], two-[234,368], and three- layer[103,124,125,133] models). However, they are computationally demanding and less accurate as the SNR decreases, especially for multi-layer fitting[126]. Deep learning methods have been proposed in DCS analysis since 2019[253], including 2DCNN[250], LSTM[174,254], ConvGRU[369], and DCS-NET[126]. New AI techniques will be introduced to DCS applications soon.

Over the past 25 years, we have witnessed the emergence of DCS to quantify BF dynamics of deep tissues more accurately with a higher SNR. We expect low-cost, user-friendly DCS technologies will be introduced and applied soon. Combining NIRS with DCS will provide a better solution for critical bottlenecks in neuroscience and clinical applications.

**Funding**

This work has been funded by the Engineering and Physical Science Research Council (Grant No. EP/T00097X/1): the Quantum Technology Hub in Quantum Imaging (QuantiC) and the University of Strathclyde.

**Acknowledgments**

We thank Professor David A. Boas from the Department of Biomedical Engineering at Boston University for his invaluable insights and contributions. We also thank Professor Robert Henderson from the School of Engineering at the University of Edinburgh for supporting us with SPAD sensors. A special thanks to Dr Wenhui Liu from the Department of Biomedical Engineering at Durham, North Carolina, United States.

**Appendix 1: optical-based blood flow monitoring modalities**

The tree diagram in Figure 17 shows the optical-based blood flow monitoring modalities, including LSCI[50], LDF[26,48], DCS, speckle contrast optical spectroscopy (SCOS), and DSCA[51,52], all sharing the advantage of non-invasive measurement of blood flow using non-ionizing radiation. Goodman developed the fundamental principles linking temporal statistics of fluctuations in laser speckle patterns in the 1960s[370]. In the 1970s, the study of time-varying speckles, induced by motion, emerged as a focal point for research. LSCI is an excellent BF imaging technique that transforms a featureless laser speckle image of a tissue surface into a high-contrast BF image, but it is only suitable for shallow-depth tissue. DSCA methodology has drawn heavily from concepts in LSCI, focusing primarily on measuring average values rather than imaging BF. Consequently, advances in LSCI can be readily implemented in DSCA with minimal difficulty. After two original papers were published by Ren *et al.*[51,52,371], DSCA has been extensively studied theoretically[372–374] and experimentally[375,376]. DSCA can be categorized into spatial DSCA and temporal DSCA depending on how the statistics are applied when calculating speckle contrast. LDF relies on measuring the Doppler shift caused by moving red blood cells to the illuminating coherent light. Since its introduction to the commercial market in the early 1980s, it has maintained a modest yet consistent and progressively expanding presence[377,378]. SCOS, also known as DSCA, was initially introduced in Valdes[379] *et al.*'s study. In the research conducted by Kim *et al.*[380], they confirmed that SCOS outperforms DCS, delivering over a 10-fold improvement in SNR at a comparable cost. Notably, fiber-based SCOS offers a viable avenue for functional neuroimaging in cognitive neuroscience and health science domains. Unlike techniques like LDF and LSCI, designed for superficial tissue measurement, DCS is a deep-tissue blood flow monitoring modality. Initially employing a continuous wave laser source, known as CW-DCS, various approaches have since been developed, including heterodyne/interferometric, multi-speckle, time-domain, long-wavelength, and Fourier-domain methods. Readers are encouraged to consult the review papers from Carp *et al.*[381], and James *et al.*[159] to delve into the detailed comparisons among these approaches.

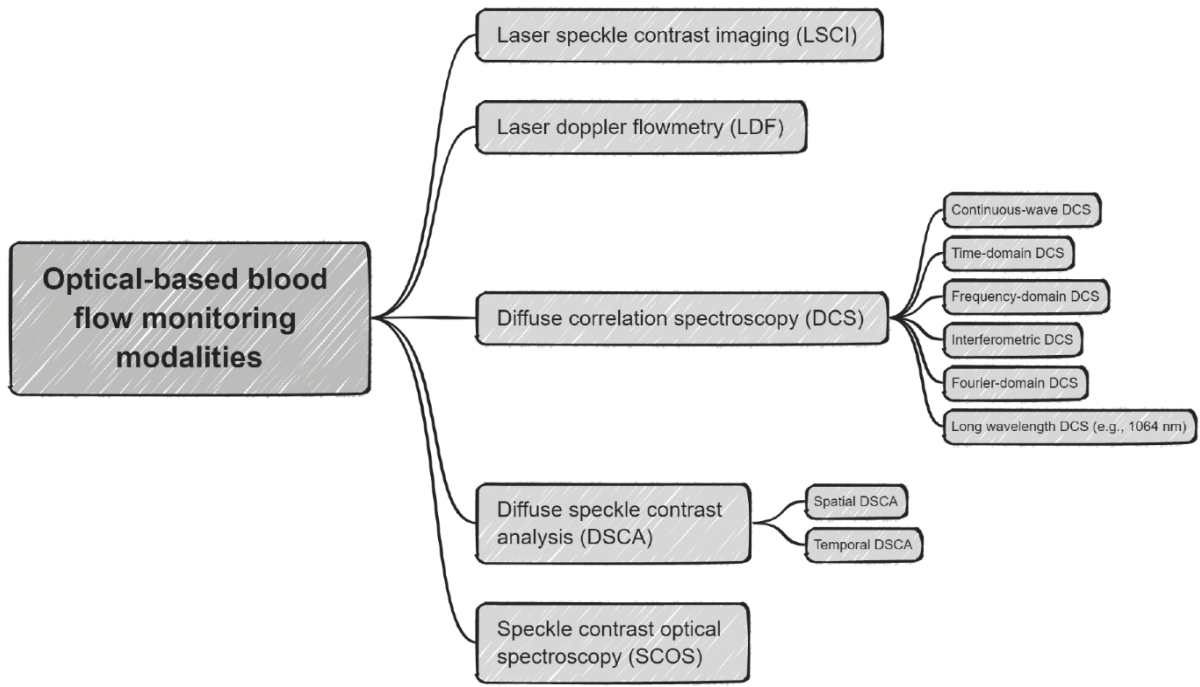

Figure 17. Optics-based blood flow monitoring modalities, including laser speckle contrast imaging (LSCI), laser doppler flowmetry (LDF), diffuse correlation spectroscopy (DCS), diffuse speckle contrast analysis (DSCA)/speckle contrast optical spectroscopy (SCOS).

**Appendix 2: other novel representative DCS instruments**

We present additional novel representative DCS instrument diagrams not previously shown. Fig. 18(a) depicts the optimized functional TD-DCS system[162], which integrates a custom 1064 nm pulse-shaped, quasi transform-limited, amplified laser source with a high-resolution time-tagging system and SNSPD sensors. Fig. 18(b) illustrates the setup of functional interferometric diffuse wave[61], with the interferometer detection path depicted in horizontal and vertical views. Fig. 18(c) showcases a Fourier domain implementation of the off-axis heterodyne parallel speckle detection instrument[340]. Fig. 18(d) is the schematic of the fiber-based SCOS setup and the corresponding data analysis pipeline[380].

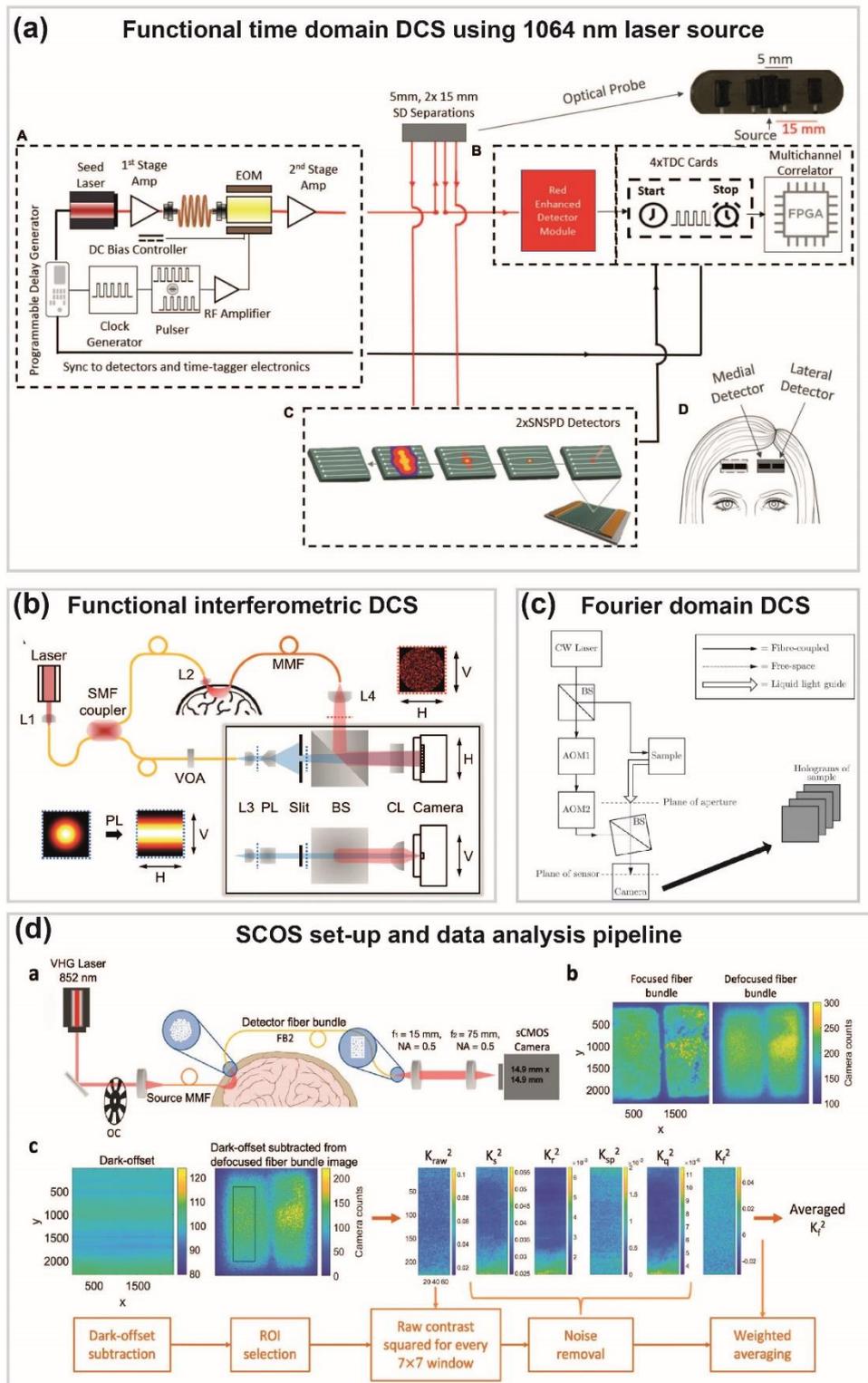

Figure 18. (a) Schematic diagram of functional TD-DCS at 1064 nm; the figure adopted from Ref.[162]. (b) Schematic of functional interferometric DWS; the figure adopted from Ref.[61]. (c) Schematic of the off-axis heterodyne parallel speckle detection (the Fourier-domain approach); the figure adopted from Ref.[340]. (d) The schematic of the fiber-based SCOS set-up and the corresponding data analysis pipeline[380].

# Appendix 3: correlation between acquisition time and spatial resolution across various BFi measurement modalities

Fig. 19 illustrates the correlation between acquisition time and spatial resolution across various BFi measurement modalities. Although DCS does not have a high spatial resolution, it outperforms others regarding acquisition speed.

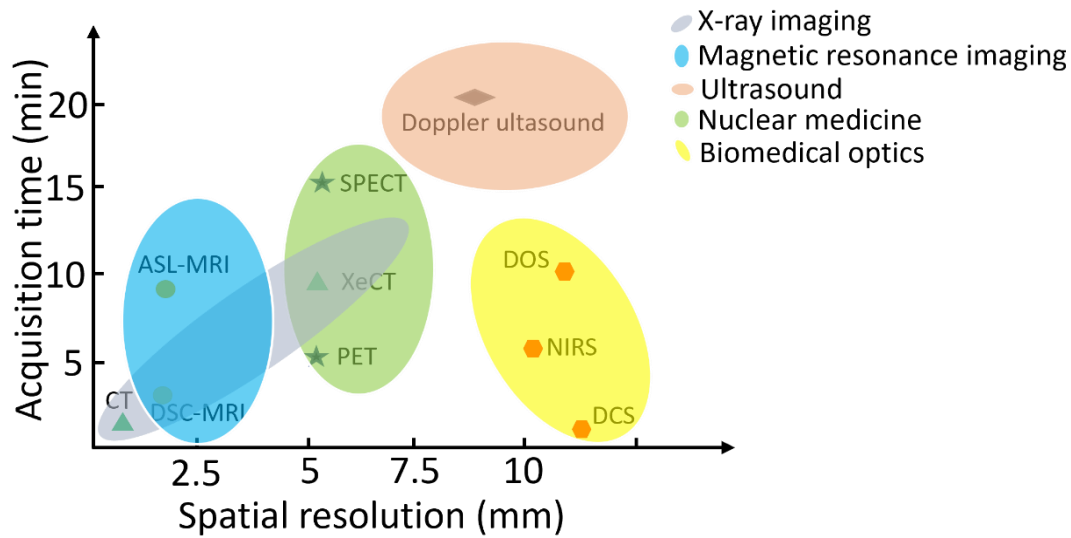

Fig. 19 Available techniques for measuring CBF in terms of the spatial resolution and acquisition time. (Note: SPECT: single photon emission computed tomography; PET: positron emission tomography; CT: computed tomography; DSC-MRI: dynamic susceptibility contrasts magnetic resonance imaging; ASL: arterial spin labeling; TCD: transcranial Doppler; TDF: thermal diffusion flowmetry; LDF: laser Doppler flowmetry; DCS: diffuse correlation spectroscopy; NIRS: near-infrared spectroscopy; CHS: coherent hemodynamics spectroscopy).

# Appendix 4: DCS simulation tools

The Monte Carlo (MC) method for simulating light propagation through tissue is a benchmark technique[382], extensively discussed in Zhu and Liu's review paper[383]. MC has been widely used in the NIRS and DCS communities. To aid researchers in performing and documenting more intricate experimental analyses, various analysis platforms and specialized software tools[384,385] have been created. The exhaustive NIRS/DCS MC software tools are listed in another review paper[88]. Here, we only list the tools commonly used in DCS, as shown in Table 2.8. MCML, developed by Jacques et al.[386], is a steady-state MC tool for analyzing multi-layered turbid media using an infinitely narrow photon beam as the light source. Operating in a 3D environment, it provides outputs including the radial position, angular dependence of local reflectance and transmittance, and the internal distribution of energy deposition and fluence rate within the multilayered medium. The program can be easily modified. Alternative software packages like Monte Carlo eXtreme (MCX)[387] or mesh-based Monte Carlo (MMC)[388], developed by Fang and his colleagues, can simulate arbitrary optode placements on diverse, intricate tissue models with heterogeneity. MCX and MMC can record the path lengths and momentum transfer from the detected photons to obtain the electric field autocorrelation function[126,367]. ScatterBrains, developed by Wu et al.[389], is an open database of human head models with companion optode locations of interest and a toolkit designed for generating specifications to execute MC simulations of light propagation, including the code to create input files compatible with MMC. Additionally, an illustration of post-processing techniques for DCS is provided.

Table 2.8 Existing software tools related to DCS

| Name | Language | Website |
|---|---|---|
| MCML[390] | Standalone | https://omlc.org/software/mc/ |
| MMC[388] | Standalone/Matlab | http://mcx.space/#mmc |
| MCX[387] | Standalone/Matlab | http://mcx.space/ |
| scatterBrains[389] | Matlab | https://github.com/wumelissa/scatterBrains |

# References

1. A. Zauner et al., "Brain Oxygenation and Energy Metabolism: Part I—Biological Function and Pathophysiology," Neurosurgery **51**(2), 289 (2002).
2. K. Uludağ et al., "Coupling of cerebral blood flow and oxygen consumption during physiological activation and deactivation measured with fMRI," NeuroImage **23**(1), 148–155 (2004) [doi:10.1016/j.neuroimage.2004.05.013].
3. T. Durduran and A. G. Yodh, "Diffuse correlation spectroscopy for non-invasive, micro-vascular cerebral blood flow measurement," NeuroImage **85**, 51–63 (2014) [doi:10.1016/j.neuroimage.2013.06.017].
4. M. G. Kaiser and M. J. During, "Combining laser doppler flowmetry with microdialysis: a novel approach to investigate the coupling of regional cerebral blood flow to neuronal activity," Journal of Neuroscience Methods **60**(1–2), 165–173 (1995) [doi:10.1016/0165-0270(95)00008-I].
5. A. Devor et al., "Frontiers in Optical Imaging of Cerebral Blood Flow and Metabolism," J Cereb Blood Flow Metab **32**(7), 1259–1276 (2012) [doi:10.1038/jcbfm.2011.195].
6. C. Cheung et al., "*In vivo* cerebrovascular measurement combining diffuse near-infrared absorption and correlation spectroscopies," Phys. Med. Biol. **46**(8), 2053–2065 (2001) [doi:10.1088/0031-9155/46/8/302].
7. V. Quaresima, S. Bisconti, and M. Ferrari, "A brief review on the use of functional near-infrared spectroscopy (fNIRS) for language imaging studies in human newborns and adults," Brain and Language **121**(2), 79–89 (2012) [doi:10.1016/j.bandl.2011.03.009].
8. S. Fantini et al., "Cerebral blood flow and autoregulation: current measurement techniques and prospects for noninvasive optical methods," Neurophoton **3**(3), 031411 (2016) [doi:10.1117/1.NPh.3.3.031411].
9. C. J. Rhee et al., "Neonatal cerebrovascular autoregulation," Pediatr Res **84**(5), 602–610 (2018) [doi:10.1038/s41390-018-0141-6].
10. B. C. Campbell et al., "Failure of Collateral Blood Flow is Associated with Infarct Growth in Ischemic Stroke," J Cereb Blood Flow Metab **33**(8), 1168–1172 (2013) [doi:10.1038/jcbfm.2013.77].
11. T. Durduran et al., "Transcranial optical monitoring of cerebrovascular hemodynamics in acute stroke patients," Opt. Express **17**(5), 3884 (2009) [doi:10.1364/OE.17.003884].
12. W. Weigl et al., "Application of optical methods in the monitoring of traumatic brain injury: A review," J Cereb Blood Flow Metab **36**(11), 1825–1843 (2016) [doi:10.1177/0271678X16667953].
13. T. Durduran et al., "Diffuse optical measurement of blood flow, blood oxygenation, and metabolism in a human brain during sensorimotor cortex activation," Opt. Lett. **29**(15), 1766 (2004) [doi:10.1364/OL.29.001766].
14. K. F. Ma et al., "A systematic review of diagnostic techniques to determine tissue perfusion in patients with peripheral arterial disease," Expert Review of Medical Devices **16**(8), 697–710 (2019) [doi:10.1080/17434440.2019.1644166].
15. A. Duncan et al., "Measurement of Cranial Optical Path Length as a Function of Age Using Phase Resolved Near Infrared Spectroscopy," Pediatr Res **39**(5), 889–894, Nature Publishing Group (1996) [doi:10.1203/00006450-199605000-00025].
16. W. Becker et al., "Advanced time-correlated single photon counting techniques for spectroscopy and imaging in biomedical systems," presented at Lasers and Applications in Science and Engineering, 1 June 2004, San Jose, Ca, 104 [doi:10.1117/12.529143].
17. K. Gurley, Y. Shang, and G. Yu, "Noninvasive optical quantification of absolute blood flow, blood oxygenation, and oxygen consumption rate in exercising skeletal muscle," J. Biomed. Opt **17**(7), 0750101 (2012) [doi:10.1117/1.JBO.17.7.075010].
18. R. Tupprasoot and B. J. Blaise, "Continuous cerebral blood flow monitoring: What should we do with these extra numbers?," BJA Open **7**, 100148 (2023) [doi:10.1016/j.bjao.2023.100148].

19. J. J. Vaquero and P. Kinahan, "Positron Emission Tomography: Current Challenges and Opportunities for Technological Advances in Clinical and Preclinical Imaging Systems," Annu. Rev. Biomed. Eng. **17**(1), 385–414 (2015) [doi:10.1146/annurev-bioeng-071114-040723].
20. M. Ljungberg and P. H. Pretorius, "SPECT/CT: an update on technological developments and clinical applications," BJR **91**(1081), 20160402 (2018) [doi:10.1259/bjr.20160402].
21. H. Yonas, R. R. Pindzola, and D. W. Johnson, "Xenon/Computed Tomography Cerebral Blood Flow and its use in Clinical Management," Neurosurgery Clinics of North America **7**(4), 605–616 (1996) [doi:10.1016/S1042-3680(18)30349-8].
22. K. K. Kwong et al., "Dynamic magnetic resonance imaging of human brain activity during primary sensory stimulation.," Proc. Natl. Acad. Sci. U.S.A. **89**(12), 5675–5679 (1992) [doi:10.1073/pnas.89.12.5675].
23. E. L. Barbier, L. Lamalle, and M. Décorps, "Methodology of brain perfusion imaging: Methodology of Brain Perfusion Imaging," J. Magn. Reson. Imaging **13**(4), 496–520 (2001) [doi:10.1002/jmri.1073].
24. T. Durduran et al., "Optical measurement of cerebral hemodynamics and oxygen metabolism in neonates with congenital heart defects," J. Biomed. Opt. **15**(3), 037004 (2010) [doi:10.1117/1.3425884].
25. G. Yu et al., "Validation of diffuse correlation spectroscopy for muscle blood flow with concurrent arterial spin labeled perfusion MRI," Opt. Express **15**(3), 1064 (2007) [doi:10.1364/OE.15.001064].
26. A. P. Shepherd and P. Å. Öberg, *Laser-Doppler Blood Flowmetry*, Springer Science & Business Media (2013).
27. M. Wintermark et al., "Comparative Overview of Brain Perfusion Imaging Techniques," Stroke **36**(9) (2005) [doi:10.1161/01.STR.0000177884.72657.8b].
28. F. F. Jöbsis, "Noninvasive, Infrared Monitoring of Cerebral and Myocardial Oxygen Sufficiency and Circulatory Parameters," Science **198**(4323), 1264–1267 (1977) [doi:10.1126/science.929199].
29. F. F. Jöbsis-vanderVliet and P. D. Jöbsis, "Biochemical and Physiological Basis of Medical Near-Infrared Spectroscopy," J. Biomed. Opt. **4**(4), 397 (1999) [doi:10.1117/1.429953].
30. F. F. Jöbsis-vanderVliet, "Discovery of the Near-Infrared Window into the Body and the Early Development of Near-Infrared Spectroscopy," J. Biomed. Opt. **4**(4), 392 (1999) [doi:10.1117/1.429952].
31. D. A. Benaron et al., "Noninvasive Methods for Estimating In Vivo Oxygenation," Clin Pediatr (Phila) **31**(5), 258–273 (1992) [doi:10.1177/000992289203100501].
32. C. G. Pope and MurielF. Stevens, "Preliminary Communication," The Lancet **262**(6797), 1190 (1953) [doi:10.1016/S0140-6736(53)90736-3].
33. H. Habazettl et al., "Near-infrared spectroscopy and indocyanine green derived blood flow index for noninvasive measurement of muscle perfusion during exercise," J Appl Physiol **108** (2010).
34. D. J. Hawrysz and E. M. Sevick-Muraca, "Developments Toward Diagnostic Breast Cancer Imaging Using Near-Infrared Optical Measurements and Fluorescent Contrast Agents1," Neoplasia **2**(5), 388–417 (2000) [doi:10.1038/sj.neo.7900118].
35. E. Keller et al., "Noninvasive measurement of regional cerebral blood flow and regional cerebral blood volume by near-infrared spectroscopy and indocyanine green dye dilution," NeuroImage **20**(2), 828–839 (2003) [doi:10.1016/S1053-8119(03)00315-X].
36. N. A. Clark, J. H. Lunacek, and G. B. Benedek, "A Study of Brownian Motion Using Light Scattering" (2015).
37. B. Chu, *Laser Light Scattering: Basic Principles and Practice*, Courier Corporation (2007).
38. H. C. Hulst and H. C. van de Hulst, *Light Scattering by Small Particles*, Courier Corporation (1981).

39. B. J. Berne and R. Pecora, *Dynamic Light Scattering: With Applications to Chemistry, Biology, and Physics*, Courier Corporation (2000).
40. G. G. Fuller et al., "The measurement of velocity gradients in laminar flow by homodyne light-scattering spectroscopy," J. Fluid Mech. **100**(03), 555 (1980) [doi:10.1017/S0022112080001280].
41. G. Maret and P. E. Wolf, "Multiple light scattering from disordered media. The effect of brownian motion of scatterers," Z. Physik B - Condensed Matter **65**(4), 409–413 (1987) [doi:10.1007/BF01303762].
42. M. J. Stephen, "Temporal fluctuations in wave propagation in random media," Phys. Rev. B **37**(1), 1–5 (1988) [doi:10.1103/PhysRevB.37.1].
43. D. J. Pine et al., "Diffusing-wave spectroscopy: dynamic light scattering in the multiple scattering limit," J. Phys. France **51**(18), 2101–2127 (1990) [doi:10.1051/jphys:0199000510180210100].
44. G. Maret and P. E. Wolf, "Multiple light scattering from disordered media. The effect of brownian motion of scatterers," Z. Physik B - Condensed Matter **65**(4), 409–413 (1987) [doi:10.1007/BF01303762].
45. H. Liu et al., "DISR: Deep Infrared Spectral Restoration Algorithm for Robot Sensing and Intelligent Visual Tracking Systems," in 2019 IEEE/RSJ International Conference on Intelligent Robots and Systems (IROS), pp. 8012–8017, IEEE, Macau, China (2019) [doi:10.1109/IROS40897.2019.8967891].
46. T. Liu et al., "Fast Blind Instrument Function Estimation Method for Industrial Infrared Spectrometers," IEEE Trans. Ind. Inf., 1–1 (2018) [doi:10.1109/TII.2018.2794449].
47. T. Liu et al., "Flexible FTIR Spectral Imaging Enhancement for Industrial Robot Infrared Vision Sensing," IEEE Trans. Ind. Inf. **16**(1), 544–554 (2020) [doi:10.1109/TII.2019.2934728].
48. I. Fredriksson, C. Fors, and J. Johansson, "Laser Doppler Flowmetry – A Theoretical Framework" (2012).
49. T. Vo-Dinh, *Biomedical Photonics Handbook: Biomedical Diagnostics*, CRC Press (2014).
50. D. Briers et al., "Laser speckle contrast imaging: theoretical and practical limitations," J. Biomed. Opt **18**(6), 066018 (2013) [doi:10.1117/1.JBO.18.6.066018].
51. R. Bi, J. Dong, and K. Lee, "Multi-channel deep tissue flowmetry based on temporal diffuse speckle contrast analysis," Opt. Express **21**(19), 22854 (2013) [doi:10.1364/OE.21.022854].
52. R. Bi, J. Dong, and K. Lee, "Deep tissue flowmetry based on diffuse speckle contrast analysis," Opt. Lett. **38**(9), 1401 (2013) [doi:10.1364/OL.38.001401].
53. D. A. Boas, "DIFFUSE PHOTON PROBES OF STRUCTURAL AND DYNAMICAL PROPERTIES OF TURBID MEDIA: THEORY AND BIOMEDICAL APPLICATIONS."
54. C. Cheung et al., "*In vivo* cerebrovascular measurement combining diffuse near-infrared absorption and correlation spectroscopies," Phys. Med. Biol. **46**(8), 2053–2065 (2001) [doi:10.1088/0031-9155/46/8/302].
55. D. A. Boas, L. E. Campbell, and A. G. Yodh, "Scattering and Imaging with Diffusing Temporal Field Correlations," Phys. Rev. Lett. **75**(9), 1855–1858 (1995) [doi:10.1103/PhysRevLett.75.1855].
56. D. A. Boas et al., "Diffusion of temporal field correlation with selected applications," presented at CIS Selected Papers: Coherence Domain Methods in Biomedical Optics, 9 February 1996, Saratov, Russia, 34–46 [doi:10.1117/12.231685].
57. D. A. Boas et al., "Establishing the diffuse correlation spectroscopy signal relationship with blood flow," Neurophoton **3**(3), 031412 (2016) [doi:10.1117/1.NPh.3.3.031412].
58. D. A. Boas, L. E. Campbell, and A. G. Yodh, "Scattering and Imaging with Diffusing Temporal Field Correlations," Phys. Rev. Lett. **75**(9), 1855–1858 (1995) [doi:10.1103/PhysRevLett.75.1855].

59. D. A. Boas and A. G. Yodh, "Spatially varying dynamical properties of turbid media probed with diffusing temporal light correlation," J. Opt. Soc. Am. A **14**(1), 192 (1997) [doi:10.1364/JOSAA.14.000192].
60. L. Kreiss et al., "Beneath the Surface: Revealing Deep-Tissue Blood Flow in Human Subjects with Massively Parallelized Diffuse Correlation Spectroscopy," arXiv:2403.03968, arXiv (2024).
61. W. Zhou et al., "Functional interferometric diffusing wave spectroscopy of the human brain," Sci. Adv. **7**(20), eabe0150 (2021) [doi:10.1126/sciadv.abe0150].
62. G. Yu et al., "Time-dependent blood flow and oxygenation in human skeletal muscles measured with noninvasive near-infrared diffuse optical spectroscopies," J. Biomed. Opt. **10**(2), 024027 (2005) [doi:10.1117/1.1884603].
63. M. Belau et al., "Noninvasive observation of skeletal muscle contraction using near-infrared time-resolved reflectance and diffusing-wave spectroscopy," J. Biomed. Opt. **15**(5), 057007 (2010) [doi:10.1117/1.3503398].
64. Y. Shang et al., "Effects of muscle fiber motion on diffuse correlation spectroscopy blood flow measurements during exercise," Biomed. Opt. Express **1**(2), 500 (2010) [doi:10.1364/BOE.1.000500].
65. G. Yu et al., "Intraoperative evaluation of revascularization effect on ischemic muscle hemodynamics using near-infrared diffuse optical spectroscopies," J. Biomed. Opt. **16**(2), 027004 (2011) [doi:10.1117/1.3533320].
66. N. Munk et al., "Noninvasively measuring the hemodynamic effects of massage on skeletal muscle: A novel hybrid near-infrared diffuse optical instrument," Journal of Bodywork and Movement Therapies **16**(1), 22–28 (2012) [doi:10.1016/j.jbmt.2011.01.018].
67. K. G. Guoqiang Yu, "Diffuse Correlation Spectroscopy (DCS) for Assessment of Tissue Blood Flow in Skeletal Muscle: Recent Progress," Anat Physiol **03**(02) (2013) [doi:10.4172/2161-0940.1000128].
68. V. Quaresima et al., "Diffuse correlation spectroscopy and frequency-domain near-infrared spectroscopy for measuring microvascular blood flow in dynamically exercising human muscles," Journal of Applied Physiology **127**(5), 1328–1337 (2019) [doi:10.1152/japplphysiol.00324.2019].
69. T. Durduran et al., "Diffuse optical measurement of blood flow in breast tumors," Opt. Lett. **30**(21), 2915 (2005) [doi:10.1364/OL.30.002915].
70. C. Zhou et al., "Diffuse optical monitoring of blood flow and oxygenation in human breast cancer during early stages of neoadjuvant chemotherapy," J. Biomed. Opt. **12**(5), 051903 (2007) [doi:10.1117/1.2798595].
71. H. S. Yazdi et al., "Mapping breast cancer blood flow index, composition, and metabolism in a human subject using combined diffuse optical spectroscopic imaging and diffuse correlation spectroscopy," J. Biomed. Opt **22**(4), 045003 (2017) [doi:10.1117/1.JBO.22.4.045003].
72. D. Grosenick et al., "Review of optical breast imaging and spectroscopy," J. Biomed. Opt **21**(9), 091311 (2016) [doi:10.1117/1.JBO.21.9.091311].
73. L. He et al., "Noncontact diffuse correlation tomography of human breast tumor," J. Biomed. Opt **20**(8), 086003 (2015) [doi:10.1117/1.JBO.20.8.086003].
74. R. Choe et al., "Optically Measured Microvascular Blood Flow Contrast of Malignant Breast Tumors," PLoS ONE **9**(6), J. Najbauer, Ed., e99683 (2014) [doi:10.1371/journal.pone.0099683].
75. S. H. Chung et al., "Macroscopic optical physiological parameters correlate with microscopic proliferation and vessel area breast cancer signatures," Breast Cancer Res **17**(1), 72 (2015) [doi:10.1186/s13058-015-0578-z].
76. C. Cheung et al., "*In vivo* cerebrovascular measurement combining diffuse near-infrared absorption and correlation spectroscopies," Phys. Med. Biol. **46**(8), 2053–2065 (2001) [doi:10.1088/0031-9155/46/8/302].

77. J. P. Culver et al., "Diffuse Optical Tomography of Cerebral Blood Flow, Oxygenation, and Metabolism in Rat during Focal Ischemia," J Cereb Blood Flow Metab **23**(8), 911–924 (2003) [doi:10.1097/01.WCB.0000076703.71231.BB].
78. J. Li et al., "Noninvasive detection of functional brain activity with near-infrared diffusing-wave spectroscopy," J. Biomed. Opt. **10**(4), 044002 (2005) [doi:10.1117/1.2007987].
79. C. Zhou et al., "Diffuse optical correlation tomography of cerebral blood flow during cortical spreading depression in rat brain," Opt. Express **14**(3), 1125 (2006) [doi:10.1364/OE.14.001125].
80. J. Li et al., "Transient functional blood flow change in the human brain measured noninvasively by diffusing-wave spectroscopy," Opt. Lett. **33**(19), 2233 (2008) [doi:10.1364/OL.33.002233].
81. E. M. Buckley et al., "Cerebral hemodynamics in preterm infants during positional intervention measured with diffuse correlation spectroscopy and transcranial Doppler ultrasound," Opt. Express **17**(15), 12571 (2009) [doi:10.1364/OE.17.012571].
82. M. N. Kim et al., "Noninvasive Measurement of Cerebral Blood Flow and Blood Oxygenation Using Near-Infrared and Diffuse Correlation Spectroscopies in Critically Brain-Injured Adults," Neurocrit Care **12**(2), 173–180 (2010) [doi:10.1007/s12028-009-9305-x].
83. C. Cheung et al., "*In vivo* cerebrovascular measurement combining diffuse near-infrared absorption and correlation spectroscopies," Phys. Med. Biol. **46**(8), 2053–2065 (2001) [doi:10.1088/0031-9155/46/8/302].
84. T. Durduran et al., "Diffuse optical measurement of blood flow, blood oxygenation, and metabolism in a human brain during sensorimotor cortex activation," Opt. Lett. **29**(15), 1766 (2004) [doi:10.1364/OL.29.001766].
85. E. M. Buckley et al., "Diffuse correlation spectroscopy for measurement of cerebral blood flow: future prospects," Neurophoton **1**(1), 011009 (2014) [doi:10.1117/1.NPh.1.1.011009].
86. G. Yu, "Diffuse Correlation Spectroscopy (DCS): A Diagnostic Tool for Assessing Tissue Blood Flow in Vascular-Related Diseases and Therapies," CMIR **8**(3), 194–210 (2012) [doi:10.2174/157340512803759875].
87. R. C. Mesquita et al., "Direct measurement of tissue blood flow and metabolism with diffuse optics," Phil. Trans. R. Soc. A. **369**(1955), 4390–4406 (2011) [doi:10.1098/rsta.2011.0232].
88. H. Ayaz et al., "Optical imaging and spectroscopy for the study of the human brain: status report," Neurophoton. **9**(S2) (2022) [doi:10.1117/1.NPh.9.S2.S24001].
89. T. Durduran et al., "Diffuse optics for tissue monitoring and tomography," Rep. Prog. Phys. **73**(7), 076701 (2010) [doi:10.1088/0034-4885/73/7/076701].
90. Y. Shang, T. Li, and G. Yu, "Clinical applications of near-infrared diffuse correlation spectroscopy and tomography for tissue blood flow monitoring and imaging," Physiol. Meas. **38**(4), R1–R26 (2017) [doi:10.1088/1361-6579/aa60b7].
91. S. Moka et al., "Frequency domain diffuse correlation spectroscopy: a new method for simultaneous estimation of static and dynamic tissue optical properties," in Multiscale Imaging and Spectroscopy III, K. C. Maitland, D. M. Roblyer, and P. J. Campagnola, Eds., p. 20, SPIE, San Francisco, United States (2022) [doi:10.1117/12.2610115].
92. B. J. Ackerson et al., "Correlation transfer - Application of radiative transfer solution methods to photon correlation problems," Journal of Thermophysics and Heat Transfer **6**(4), 577–588 (1992) [doi:10.2514/3.11537].
93. R. L. Dougherty et al., "Correlation transfer: Development and application," Journal of Quantitative Spectroscopy and Radiative Transfer **52**(6), 713–727 (1994) [doi:10.1016/0022-4073(94)90037-X].
94. D. A. Boas and A. G. Yodh, "Spatially varying dynamical properties of turbid media probed with diffusing temporal light correlation," J. Opt. Soc. Am. A **14**(1), 192 (1997) [doi:10.1364/JOSAA.14.000192].

95. M. S. Patterson et al., "Absorption spectroscopy in tissue-simulating materials: a theoretical and experimental study of photon paths," Appl. Opt. **34**(1), 22 (1995) [doi:10.1364/AO.34.000022].
96. D. A. Boas, "DIFFUSE PHOTON PROBES OF STRUCTURAL AND DYNAMICAL PROPERTIES OF TURBID MEDIA: THEORY AND BIOMEDICAL APPLICATIONS." (1996)
97. G. Maret and P. E. Wolf, "Multiple light scattering from disordered media. The effect of brownian motion of scatterers," Z. Physik B - Condensed Matter **65**(4), 409–413 (1987) [doi:10.1007/BF01303762].
98. A. Kienle and M. S. Patterson, "Determination of the optical properties of semi-infinite turbid media from frequency-domain reflectance close to the source," Phys. Med. Biol. **42**(9), 1801–1819 (1997) [doi:10.1088/0031-9155/42/9/011].
99. T. Durduran, "Noninvasive measurements of tissue hemodynamics with hybrid diffuse optical methods," Med. Phys. **31**(7), 2178–2178 (2004) [doi:10.1118/1.1763412].
100. S. A. Carp et al., "Validation of diffuse correlation spectroscopy measurements of rodent cerebral blood flow with simultaneous arterial spin labeling MRI; towards MRI-optical continuous cerebral metabolic monitoring," Biomed. Opt. Express **1**(2), 553 (2010) [doi:10.1364/BOE.1.000553].
101. C. Zhou et al., "Diffuse optical monitoring of hemodynamic changes in piglet brain with closed head injury," J. Biomed. Opt. **14**(3), 034015 (2009) [doi:10.1117/1.3146814].
102. R. M. Forti et al., "Non-invasive diffuse optical monitoring of cerebral physiology in an adult swine-model of impact traumatic brain injury," Biomed. Opt. Express **14**(6), 2432 (2023) [doi:10.1364/BOE.486363].
103. J. Li et al., "Noninvasive detection of functional brain activity with near-infrared diffusing-wave spectroscopy," J. Biomed. Opt. **10**(4), 044002 (2005) [doi:10.1117/1.2007987].
104. F. Jaillon et al., "Activity of the human visual cortex measured non-invasively by diffusing-wave spectroscopy," Opt. Express **15**(11), 6643 (2007) [doi:10.1364/OE.15.006643].
105. G. Dietsche et al., "Fiber-based multispeckle detection for time-resolved diffusing-wave spectroscopy: characterization and application to blood flow detection in deep tissue," Appl. Opt. **46**(35), 8506 (2007) [doi:10.1364/AO.46.008506].
106. J. Li et al., "Transient functional blood flow change in the human brain measured noninvasively by diffusing-wave spectroscopy," Opt. Lett. **33**(19), 2233 (2008) [doi:10.1364/OL.33.002233].
107. L. Koban et al., "Processing of emotional words measured simultaneously with steady-state visually evoked potentials and near-infrared diffusing-wave spectroscopy," BMC Neurosci **11**(1), 85 (2010) [doi:10.1186/1471-2202-11-85].
108. P. Zirak et al., "Effects of acetazolamide on the micro- and macro-vascular cerebral hemodynamics: a diffuse optical and transcranial doppler ultrasound study," Biomed. Opt. Express **1**(5), 1443 (2010) [doi:10.1364/BOE.1.001443].
109. B. L. Edlow et al., "The effects of healthy aging on cerebral hemodynamic responses to posture change," Physiol. Meas. **31**(4), 477–495 (2010) [doi:10.1088/0967-3334/31/4/002].
110. G. Yu et al., "Noninvasive Monitoring of Murine Tumor Blood Flow During and After Photodynamic Therapy Provides Early Assessment of Therapeutic Efficacy," Clinical Cancer Research **11**(9), 3543–3552 (2005) [doi:10.1158/1078-0432.CCR-04-2582].
111. U. Sunar et al., "Hemodynamic responses to antivascular therapy and ionizing radiation assessed by diffuse optical spectroscopies," Opt. Express **15**(23), 15507 (2007) [doi:10.1364/OE.15.015507].
112. Y. Shang et al., "Portable optical tissue flow oximeter based on diffuse correlation spectroscopy," Opt. Lett. **34**(22), 3556 (2009) [doi:10.1364/OL.34.003556].
113. U. Sunar et al., "Noninvasive diffuse optical measurement of blood flow and blood oxygenation for monitoring radiation therapy in patients with head and neck tumors: a pilot study," J. Biomed. Opt. **11**(6), 064021 (2006) [doi:10.1117/1.2397548].

114. G. Yu et al., "Real-time In Situ Monitoring of Human Prostate Photodynamic Therapy with Diffuse Light," Photochem Photobiol **82**(5), 1279 (2006) [doi:10.1562/2005-10-19-RA-721].
115. L. Dong et al., "Noninvasive diffuse optical monitoring of head and neck tumor blood flow and oxygenation during radiation delivery," Biomed. Opt. Express **3**(2), 259 (2012) [doi:10.1364/BOE.3.000259].
116. T. Durduran, "Noninvasive measurements of tissue hemodynamics with hybrid diffuse optical methods," Med. Phys. **31**(7), 2178–2178 (2004) [doi:10.1118/1.1763412].
117. A. Kienle et al., "Noninvasive determination of the optical properties of two-layered turbid media," Appl. Opt. **37**(4), 779 (1998) [doi:10.1364/AO.37.000779].
118. L. Gagnon et al., "Double-layer estimation of intra- and extracerebral hemoglobin concentration with a time-resolved system," J. Biomed. Opt. **13**(5), 054019 (2008) [doi:10.1117/1.2982524].
119. F. Lesage, L. Gagnon, and M. Dehaes, "Diffuse optical-MRI fusion and applications," presented at Biomedical Optics (BiOS) 2008, 7 February 2008, San Jose, CA, 68500C [doi:10.1117/12.766745].
120. L. Gagnon et al., "Investigation of diffuse correlation spectroscopy in multi-layered media including the human head," Opt. Express **16**(20), 15514 (2008) [doi:10.1364/OE.16.015514].
121. A. Kienle et al., "Noninvasive determination of the optical properties of two-layered turbid media," Appl. Opt. **37**(4), 779 (1998) [doi:10.1364/AO.37.000779].
122. A. Kienle and T. Glanzmann, "*In vivo* determination of the optical properties of muscle with time-resolved reflectance using a layered model," Phys. Med. Biol. **44**(11), 2689–2702 (1999) [doi:10.1088/0031-9155/44/11/301].
123. A. Kienle et al., "Investigation of two-layered turbid media with time-resolved reflectance," Appl. Opt. **37**(28), 6852 (1998) [doi:10.1364/AO.37.006852].
124. K. Verdecchia et al., "Assessment of a multi-layered diffuse correlation spectroscopy method for monitoring cerebral blood flow in adults," Biomed. Opt. Express **7**(9), 3659 (2016) [doi:10.1364/BOE.7.003659].
125. H. Zhao, E. Sathialingam, and E. M. Buckley, "Accuracy of diffuse correlation spectroscopy measurements of cerebral blood flow when using a three-layer analytical model," Biomed. Opt. Express **12**(11), 7149 (2021) [doi:10.1364/BOE.438303].
126. Q. Wang et al., "Quantification of blood flow index in diffuse correlation spectroscopy using a robust deep learning method," J. Biomed. Opt. **29**(01) (2024) [doi:10.1117/1.JBO.29.1.015004].
127. H. Zhao et al., "Comparison of diffuse correlation spectroscopy analytical models for measuring cerebral blood flow in adults," J. Biomed. Opt. **28**(12) (2023) [doi:10.1117/1.JBO.28.12.126005].
128. T. J. Farrell, M. S. Patterson, and B. Wilson, "A diffusion theory model of spatially resolved, steady-state diffuse reflectance for the noninvasive determination of tissue optical properties *in vivo*," Med. Phys. **19**(4), 879–888 (1992) [doi:10.1118/1.596777].
129. J. Sutin et al., "Time-domain diffuse correlation spectroscopy," Optica **3**(9), 1006 (2016) [doi:10.1364/OPTICA.3.001006].
130. A. G. Yodh, P. D. Kaplan, and D. J. Pine, "Pulsed diffusing-wave spectroscopy: High resolution through nonlinear optical gating," Phys. Rev. B **42**(7), 4744–4747 (1990) [doi:10.1103/PhysRevB.42.4744].
131. A. Kienle and M. S. Patterson, "Improved solutions of the steady-state and the time-resolved diffusion equations for reflectance from a semi-infinite turbid medium," J. Opt. Soc. Am. A **14**(1), 246 (1997) [doi:10.1364/JOSAA.14.000246].
132. S. Samaei et al., "Performance assessment of laser sources for time-domain diffuse correlation spectroscopy," Biomed. Opt. Express **12**(9), 5351 (2021) [doi:10.1364/BOE.432363].
133. J. Li et al., "Analytical models for time-domain diffuse correlation spectroscopy for multi-layer and heterogeneous turbid media," Biomed. Opt. Express **8**(12), 5518 (2017) [doi:10.1364/BOE.8.005518].

134. L. Dong et al., "Simultaneously Extracting Multiple Parameters via Fitting One Single Autocorrelation Function Curve in Diffuse Correlation Spectroscopy," IEEE Trans. Biomed. Eng. **60**(2), 361–368 (2013) [doi:10.1109/TBME.2012.2226885].
135. S. A. Carp et al., "Diffuse correlation spectroscopy measurements of blood flow using 1064 nm light," J. Biomed. Opt. **25**(09) (2020) [doi:10.1117/1.JBO.25.9.097003].
136. D. Mazumder et al., "Optimization of time domain diffuse correlation spectroscopy parameters for measuring brain blood flow," Neurophoton. **8**(03) (2021) [doi:10.1117/1.NPh.8.3.035005].
137. D. Irwin et al., "Influences of tissue absorption and scattering on diffuse correlation spectroscopy blood flow measurements," Biomed. Opt. Express **2**(7), 1969 (2011) [doi:10.1364/BOE.2.001969].
138. K. Schätzel, "Noise in Photon Correlation and Photon Structure Functions," Optica Acta: International Journal of Optics **30**(2), 155–166 (1983) [doi:10.1080/713821145].
139. D. E. Koppel, "Statistical accuracy in fluorescence correlation spectroscopy," Phys. Rev. A **10**(6), 1938–1945 (1974) [doi:10.1103/PhysRevA.10.1938].
140. X. Cheng et al., "Measuring neuronal activity with diffuse correlation spectroscopy: a theoretical investigation," Neurophoton. **8**(03) (2021) [doi:10.1117/1.NPh.8.3.035004].
141. E. J. Sie et al., "High-sensitivity multispeckle diffuse correlation spectroscopy," Neurophoton. **7**(03) (2020) [doi:10.1117/1.NPh.7.3.035010].
142. M. Helton et al., "Numerical approach to quantify depth-dependent blood flow changes in real-time using the diffusion equation with continuous-wave and time-domain diffuse correlation spectroscopy," Biomed. Opt. Express **14**(1), 367 (2023) [doi:10.1364/BOE.469419].
143. P. Zhang et al., "Approaches to denoise the diffuse optical signals for tissue blood flow measurement," Biomed. Opt. Express **9**(12), 6170 (2018) [doi:10.1364/BOE.9.006170].
144. S. A. Carp et al., "Combined multi-distance frequency domain and diffuse correlation spectroscopy system with simultaneous data acquisition and real-time analysis," Biomed. Opt. Express **8**(9), 3993 (2017) [doi:10.1364/BOE.8.003993].
145. J. Sunwoo et al., "Diffuse correlation spectroscopy blood flow monitoring for intraventricular hemorrhage vulnerability in extremely low gestational age newborns," Sci Rep **12**(1), 12798 (2022) [doi:10.1038/s41598-022-16499-3].
146. D. Tamborini et al., "Portable System for Time-Domain Diffuse Correlation Spectroscopy," IEEE Trans. Biomed. Eng. **66**(11), 3014–3025 (2019) [doi:10.1109/TBME.2019.2899762].
147. S. Moka et al., "Frequency domain diffuse correlation spectroscopy: a new method for simultaneous estimation of static and dynamic tissue optical properties," in Multiscale Imaging and Spectroscopy III, K. C. Maitland, D. M. Roblyer, and P. J. Campagnola, Eds., p. 20, SPIE, San Francisco, United States (2022) [doi:10.1117/12.2610115].
148. I. J. Bigio and S. Fantini, *Quantitative Biomedical Optics: Theory, Methods, and Applications*, Cambridge University Press (2016).
149. T. Bellini et al., "Effects of finite laser coherence in quasielastic multiple scattering," Phys. Rev. A **44**(8), 5215–5223 (1991) [doi:10.1103/PhysRevA.44.5215].
150. A. Biswas et al., "Fast diffuse correlation spectroscopy with a low-cost, fiber-less embedded diode laser," Biomed. Opt. Express **12**(11), 6686 (2021) [doi:10.1364/BOE.435136].
151. M. N. Kim et al., "Noninvasive Measurement of Cerebral Blood Flow and Blood Oxygenation Using Near-Infrared and Diffuse Correlation Spectroscopies in Critically Brain-Injured Adults," Neurocrit Care **12**(2), 173–180 (2010) [doi:10.1007/s12028-009-9305-x].
152. D. T. Delpy et al., "Estimation of optical pathlength through tissue from direct time of flight measurement," Phys. Med. Biol. **33**(12), 1433–1442 (1988) [doi:10.1088/0031-9155/33/12/008].
153. "A. N. S Institute, American National Standard for Safe Use of Lasers (Laser Institute of America, 2007)."

154. A. I. Zavriyev et al., "The role of diffuse correlation spectroscopy and frequency-domain near-infrared spectroscopy in monitoring cerebral hemodynamics during hypothermic circulatory arrests," JTCVS Techniques **7**, 161–177 (2021) [doi:10.1016/j.xjtc.2021.01.023].
155. S. Y. Lee et al., "Noninvasive optical assessment of resting-state cerebral blood flow in children with sickle cell disease" (2019).
156. K. C. Wu et al., "Enhancing diffuse correlation spectroscopy pulsatile cerebral blood flow signal with near-infrared spectroscopy photoplethysmography," Neurophoton. **10**(03) (2023) [doi:10.1117/1.NPh.10.3.035008].
157. W. Liu et al., "Fast and sensitive diffuse correlation spectroscopy with highly parallelized single photon detection," APL Photonics **6**(2), 026106 (2021) [doi:10.1063/5.0031225].
158. C. Huang et al., "Low-cost compact diffuse speckle contrast flowmeter using small laser diode and bare charge-coupled-device," J. Biomed. Opt **21**(8), 080501 (2016) [doi:10.1117/1.JBO.21.8.080501].
159. E. James and P. R. T. Munro, "Diffuse Correlation Spectroscopy: A Review of Recent Advances in Parallelisation and Depth Discrimination Techniques," Sensors **23**(23), 9338 (2023) [doi:10.3390/s23239338].
160. S. A. Carp et al., "Diffuse correlation spectroscopy measurements of blood flow using 1064 nm light," J. Biomed. Opt. **25**(09) (2020) [doi:10.1117/1.JBO.25.9.097003].
161. D. O'Connor, *Time-correlated single photon counting*, Academic Press (2012).
162. N. Ozana et al., "Functional Time Domain Diffuse Correlation Spectroscopy," Front. Neurosci. **16**, 932119 (2022) [doi:10.3389/fnins.2022.932119].
163. D. Tamborini et al., "Portable System for Time-Domain Diffuse Correlation Spectroscopy," IEEE Trans. Biomed. Eng. **66**(11), 3014–3025 (2019) [doi:10.1109/TBME.2019.2899762].
164. Y. Shang et al., "Diffuse optical monitoring of repeated cerebral ischemia in mice," Opt. Express **19**(21), 20301 (2011) [doi:10.1364/OE.19.020301].
165. Y. Lin et al., "Noncontact diffuse correlation spectroscopy for noninvasive deep tissue blood flow measurement," J. Biomed. Opt. **17**(1), 010502 (2012) [doi:10.1117/1.JBO.17.1.010502].
166. R. Cheng et al., "Noninvasive optical evaluation of spontaneous low frequency oscillations in cerebral hemodynamics," NeuroImage **62**(3), 1445–1454 (2012) [doi:10.1016/j.neuroimage.2012.05.069].
167. S. Han et al., "Non-Invasive Monitoring of Temporal and Spatial Blood Flow during Bone Graft Healing Using Diffuse Correlation Spectroscopy," PLoS ONE **10**(12), J. Cray, Ed., e0143891 (2015) [doi:10.1371/journal.pone.0143891].
168. C. J. Stapels et al., "A scalable correlator for multichannel diffuse correlation spectroscopy," presented at SPIE BiOS, 7 March 2016, San Francisco, California, United States, 969816 [doi:10.1117/12.2213114].
169. P. Farzam et al., "Pre-clinical longitudinal monitoring of hemodynamic response to anti-vascular chemotherapy by hybrid diffuse optics," Biomed. Opt. Express **8**(5), 2563 (2017) [doi:10.1364/BOE.8.002563].
170. E. Sathialingam et al., "Small separation diffuse correlation spectroscopy for measurement of cerebral blood flow in rodents," Biomed. Opt. Express **9**(11), 5719 (2018) [doi:10.1364/BOE.9.005719].
171. C.-S. Poon, F. Long, and U. Sunar, "Deep learning model for ultrafast quantification of blood flow in diffuse correlation spectroscopy," 10.
172. L. Cortese et al., "Recipes for diffuse correlation spectroscopy instrument design using commonly utilized hardware based on targets for signal-to-noise ratio and precision," Biomed. Opt. Express **12**(6), 3265 (2021) [doi:10.1364/BOE.423071].
173. K. R. Cowdrick et al., "Impaired cerebrovascular reactivity in pediatric sickle cell disease using diffuse correlation spectroscopy," Biomed. Opt. Express **14**(11), 5696 (2023) [doi:10.1364/BOE.499274].

174. M. Nakabayashi et al., "Deep-learning-based separation of shallow and deep layer blood flow rates in diffuse correlation spectroscopy," Biomed. Opt. Express **14**(10), 5358 (2023) [doi:10.1364/BOE.498693].
175. M. A. Wayne et al., "Massively parallel, real-time multispeckle diffuse correlation spectroscopy using a 500 × 500 SPAD camera," Biomed. Opt. Express **14**(2), 703 (2023) [doi:10.1364/BOE.473992].
176. S. Samaei et al., "Continuous-wave parallel interferometric near-infrared spectroscopy (CW πNIRS) with a fast two-dimensional camera," Biomed. Opt. Express **13**(11), 5753 (2022) [doi:10.1364/BOE.472643].
177. I. Freund, "Joseph W. Goodman: Speckle Phenomena in Optics: Theory and Applications: Roberts & Company (Englewood, Colorado), 2007," J Stat Phys **130**(2), 413–414 (2007) [doi:10.1007/s10955-007-9440-8].
178. C. Zhou et al., "Diffuse optical correlation tomography of cerebral blood flow during cortical spreading depression in rat brain," Opt. Express **14**(3), 1125 (2006) [doi:10.1364/OE.14.001125].
179. C. Zhou, "IN-VIVO OPTICAL IMAGING AND SPECTROSCOPY OF CEREBRAL HEMODYNAMICS."
180. J. D. Johansson et al., "A multipixel diffuse correlation spectroscopy system based on a single photon avalanche diode array," J. Biophotonics **12**(11) (2019) [doi:10.1002/jbio.201900091].
181. F. M. Della Rocca et al., "Field programmable gate array compression for large array multispeckle diffuse correlation spectroscopy," J. Biomed. Opt. **28**(05) (2023) [doi:10.1117/1.JBO.28.5.057001].
182. L. He et al., "Using optical fibers with different modes to improve the signal-to-noise ratio of diffuse correlation spectroscopy flow-oximeter measurements," J. Biomed. Opt **18**(3), 037001 (2013) [doi:10.1117/1.JBO.18.3.037001].
183. D. A. Boas, L. E. Campbell, and A. G. Yodh, "Scattering and Imaging with Diffusing Temporal Field Correlations," Phys. Rev. Lett. **75**(9), 1855–1858 (1995) [doi:10.1103/PhysRevLett.75.1855].
184. W. G. Lawrence et al., "A comparison of avalanche photodiode and photomultiplier tube detectors for flow cytometry," presented at Biomedical Optics (BiOS) 2008, 7 February 2008, San Jose, CA, 68590M [doi:10.1117/12.758958].
185. C. Niclass et al., "Design and characterization of a CMOS 3-D image sensor based on single photon avalanche diodes," IEEE J. Solid-State Circuits **40**(9), 1847–1854 (2005) [doi:10.1109/JSSC.2005.848173].
186. D. Stoppa et al., "A CMOS 3-D Imager Based on Single Photon Avalanche Diode," IEEE Trans. Circuits Syst. I **54**(1), 4–12 (2007) [doi:10.1109/TCSI.2006.888679].
187. D. Mosconi et al., "CMOS Single-Photon Avalanche Diode Array for Time-Resolved Fluorescence Detection," in 2006 Proceedings of the 32nd European Solid-State Circuits Conference, pp. 564–567, IEEE, Montreux (2006) [doi:10.1109/ESSCIR.2006.307487].
188. J. A. Richardson, L. A. Grant, and R. K. Henderson, "Low Dark Count Single-Photon Avalanche Diode Structure Compatible With Standard Nanometer Scale CMOS Technology," IEEE Photon. Technol. Lett. **21**(14), 1020–1022 (2009) [doi:10.1109/LPT.2009.2022059].
189. "https://cordis.europa.eu/project/id/029217."
190. D.-U. Li et al., "Real-time fluorescence lifetime imaging system with a 32 × 32 013μm CMOS low dark-count single-photon avalanche diode array," Opt. Express **18**(10), 10257 (2010) [doi:10.1364/OE.18.010257].
191. J. Richardson et al., "A 32x32 50ps Resolution 10 bit Time to Digital Converter Array in 130nm CMOS for Time Correlated Imaging."
192. C. Veerappan et al., "A 160×128 single-photon image sensor with on-pixel 55ps 10b time-to-digital converter," in 2011 IEEE International Solid-State Circuits Conference, pp. 312–314, IEEE, San Francisco, CA, USA (2011) [doi:10.1109/ISSCC.2011.5746333].

193. E. Le Francois et al., "Combining time of flight and photometric stereo imaging for 3D reconstruction of discontinuous scenes," Opt. Lett. **46**(15), 3612 (2021) [doi:10.1364/OL.424000].
194. D. Xiao et al., "Dynamic fluorescence lifetime sensing with CMOS single-photon avalanche diode arrays and deep learning processors," Biomed. Opt. Express **12**(6), 3450 (2021) [doi:10.1364/BOE.425663].
195. K. Morimoto et al., "Megapixel time-gated SPAD image sensor for 2D and 3D imaging applications," Optica **7**(4), 346 (2020) [doi:10.1364/OPTICA.386574].
196. R. K. Henderson et al., "A 192 x 128 Time Correlated Single Photon Counting Imager in 40nm CMOS Technology."
197. E. J. Sie et al., "High-sensitivity multispeckle diffuse correlation spectroscopy," Neurophoton. **7**(03) (2020) [doi:10.1117/1.NPh.7.3.035010].
198. S. Xu et al., "Transient Motion Classification Through Turbid Volumes via Parallelized Single-Photon Detection and Deep Contrastive Embedding," Front. Neurosci. **16**, 908770 (2022) [doi:10.3389/fnins.2022.908770].
199. S. Xu et al., "Imaging Dynamics Beneath Turbid Media via Parallelized Single-Photon Detection," Advanced Science **9**(24), 2201885 (2022) [doi:10.1002/advs.202201885].
200. W. Zhou et al., "Highly parallel, interferometric diffusing wave spectroscopy for monitoring cerebral blood flow dynamics," Optica **5**(5), 518 (2018) [doi:10.1364/OPTICA.5.000518].
201. X. Liu et al., "A Wearable Fiber-Free Optical Sensor for Continuous Monitoring of Neonatal Cerebral Blood Flow and Oxygenation."
202. M. B. Robinson et al., "Interferometric diffuse correlation spectroscopy improves measurements at long source–detector separation and low photon count rate," J. Biomed. Opt. **25**(09) (2020) [doi:10.1117/1.JBO.25.9.097004].
203. V. Parfentyeva et al., "Fast time-domain diffuse correlation spectroscopy with superconducting nanowire single-photon detector: system validation and in vivo results," Sci Rep **13**(1), 11982 (2023) [doi:10.1038/s41598-023-39281-5].
204. N. Ozana et al., "Time-gated diffuse correlation spectroscopy for functional imaging via 1064 nm pulse laser shaping and superconducting nanowire single photon sensing," in Optical Techniques in Neurosurgery, Neurophotonics, and Optogenetics **11629**, p. 116292F, SPIE (2021) [doi:10.1117/12.2578909].
205. C. Schuck et al., "Matrix of Integrated Superconducting Single-Photon Detectors With High Timing Resolution," IEEE Trans. Appl. Supercond. **23**(3), 2201007–2201007 (2013) [doi:10.1109/TASC.2013.2239346].
206. I. Esmaeil Zadeh et al., "Superconducting nanowire single-photon detectors: A perspective on evolution, state-of-the-art, future developments, and applications," Applied Physics Letters **118**(19), 190502 (2021) [doi:10.1063/5.0045990].
207. W. Liu et al., "Fast and sensitive diffuse correlation spectroscopy with highly parallelized single photon detection," APL Photonics **6**(2), 026106 (2021) [doi:10.1063/5.0031225].
208. F. M. Della Rocca et al., "Field programmable gate array compression for large array multispeckle diffuse correlation spectroscopy," J. Biomed. Opt. **28**(05) (2023) [doi:10.1117/1.JBO.28.5.057001].
209. M. B. Robinson et al., "Portable, high speed blood flow measurements enabled by long wavelength, interferometric diffuse correlation spectroscopy (LW-iDCS)," Sci Rep **13**(1), 8803 (2023) [doi:10.1038/s41598-023-36074-8].
210. M. Diop et al., "Calibration of diffuse correlation spectroscopy with a time-resolved near-infrared technique to yield absolute cerebral blood flow measurements," Biomed. Opt. Express **2**(7), 2068 (2011) [doi:10.1364/BOE.2.002068].
211. K. Schätzel, M. Drewel, and S. Stimac, "Photon Correlation Measurements at Large Lag Times: Improving Statistical Accuracy," Journal of Modern Optics **35**(4), 711–718 (1988) [doi:10.1080/09500348814550731].

212. K. Schatzel, "Noise on photon correlation data. I. Autocorrelation functions," Quantum Opt. **2**(4), 287–305 (1990) [doi:10.1088/0954-8998/2/4/002].
213. K. Schätzel, "Correlation techniques in dynamic light scattering," Appl. Phys. B **42**(4), 193–213 (1987) [doi:10.1007/BF00693937].
214. L. Cipelletti and D. A. Weitz, "Ultralow-angle dynamic light scattering with a charge coupled device camera based multispeckle, multitau correlator," Review of Scientific Instruments **70**(8), 3214–3221 (1999) [doi:10.1063/1.1149894].
215. D. Magatti and F. Ferri, "Fast multi-tau real-time software correlator for dynamic light scattering," Appl. Opt. **40**(24), 4011 (2001) [doi:10.1364/AO.40.004011].
216. D. Magatti and F. Ferri, "25 ns software correlator for photon and fluorescence correlation spectroscopy," Review of Scientific Instruments **74**(2), 1135–1144 (2003) [doi:10.1063/1.1525876].
217. J. Dong et al., "Diffuse correlation spectroscopy with a fast Fourier transform-based software autocorrelator," J. Biomed. Opt **17**(9), 0970041 (2012) [doi:10.1117/1.JBO.17.9.097004].
218. https://lsinstruments.ch/en/products/lsi-correlator.
219. https://www.becker-hickl.com/applications/dcs-diffuse-correlation/.
220. https://www.alvgmbh.de/Products/Correlators/Discontinued_Models_/ALV-5000_EPP/alv-5000_epp.html.
221. https://www.photon-force.com/.
222. F. Martelli et al., "There's plenty of light at the bottom: statistics of photon penetration depth in random media," Sci Rep **6**(1), 27057 (2016) [doi:10.1038/srep27057].
223. S. Samaei et al., "New hybrid time-domain device for diffuse correlation spectroscopy and near-infrared spectroscopy for brain hemodynamic assessment," in Diffuse Optical Spectroscopy and Imaging VIII, D. Contini, Y. Hoshi, and T. D. O'Sullivan, Eds., p. 7, SPIE, Online Only, Germany (2021) [doi:10.1117/12.2615209].
224. M. Pagliazzi et al., "Time domain diffuse correlation spectroscopy with a high coherence pulsed source: in vivo and phantom results," Biomed. Opt. Express **8**(11), 5311 (2017) [doi:10.1364/BOE.8.005311].
225. L. Colombo et al., "Effects of the instrument response function and the gate width in time-domain diffuse correlation spectroscopy: model and validations," Neurophoton. **6**(03), 1 (2019) [doi:10.1117/1.NPh.6.3.035001].
226. L. Colombo et al., "Coherent fluctuations in time-domain diffuse optics," APL Photonics **5**(7), 071301 (2020) [doi:10.1063/5.0011838].
227. K. R. Cowdrick et al., "Agreement in cerebrovascular reactivity assessed with diffuse correlation spectroscopy across experimental paradigms improves with short separation regression," Neurophoton. **10**(02) (2023) [doi:10.1117/1.NPh.10.2.025002].
228. A. Von Lühmann et al., "Using the General Linear Model to Improve Performance in fNIRS Single Trial Analysis and Classification: A Perspective," Front. Hum. Neurosci. **14**, 30 (2020) [doi:10.3389/fnhum.2020.00030].
229. X. Cheng et al., "Time domain diffuse correlation spectroscopy: modeling the effects of laser coherence length and instrument response function," Opt. Lett. **43**(12), 2756 (2018) [doi:10.1364/OL.43.002756].
230. M. Pagliazzi et al., "In vivo time-gated diffuse correlation spectroscopy at quasi-null source-detector separation," Opt. Lett. **43**(11), 2450 (2018) [doi:10.1364/OL.43.002450].
231. L. Colombo et al., "In vivo time-domain diffuse correlation spectroscopy above the water absorption peak," Opt. Lett. **45**(13), 3377 (2020) [doi:10.1364/OL.392355].
232. S. Samaei et al., "Time-domain diffuse correlation spectroscopy (TD-DCS) for noninvasive, depth-dependent blood flow quantification in human tissue in vivo," Sci Rep **11**(1), 1817 (2021) [doi:10.1038/s41598-021-81448-5].

233. D. Milej et al., "Direct assessment of extracerebral signal contamination on optical measurements of cerebral blood flow, oxygenation, and metabolism," Neurophoton. **7**(04) (2020) [doi:10.1117/1.NPh.7.4.045002].
234. J. Wu et al., "Two-layer analytical model for estimation of layer thickness and flow using Diffuse Correlation Spectroscopy," PLoS ONE **17**(9), A. Dalla Mora, Ed., e0274258 (2022) [doi:10.1371/journal.pone.0274258].
235. R. M. Forti et al., "Optimizing a two-layer method for hybrid diffuse correlation spectroscopy and frequency-domain diffuse optical spectroscopy cerebral measurements in adults," Neurophoton. **10**(02) (2023) [doi:10.1117/1.NPh.10.2.025008].
236. W. B. Baker et al., "Pressure modulation algorithm to separate cerebral hemodynamic signals from extracerebral artifacts," Neurophoton **2**(3), 035004 (2015) [doi:10.1117/1.NPh.2.3.035004].
237. W. B. Baker et al., "Modified Beer-Lambert law for blood flow," Biomed. Opt. Express **5**(11), 4053 (2014) [doi:10.1364/BOE.5.004053].
238. O. Kholiqov et al., "Time-of-flight resolved light field fluctuations reveal deep human tissue physiology," Nat Commun **11**(1), 391 (2020) [doi:10.1038/s41467-019-14228-5].
239. Y. Shang and G. Yu, "A $N$ th-order linear algorithm for extracting diffuse correlation spectroscopy blood flow indices in heterogeneous tissues," Appl. Phys. Lett. **105**(13), 133702 (2014) [doi:10.1063/1.4896992].
240. Y. Shang et al., "Extraction of diffuse correlation spectroscopy flow index by integration of $N$ th-order linear model with Monte Carlo simulation," Appl. Phys. Lett. **104**(19), 193703 (2014) [doi:10.1063/1.4876216].
241. V. N. Vapnik, "An overview of statistical learning theory," IEEE Trans. Neural Netw. **10**(5), 988–999 (1999) [doi:10.1109/72.788640].
242. P. Zhang et al., "Approaches to denoise the diffuse optical signals for tissue blood flow measurement," Biomed. Opt. Express **9**(12), 6170 (2018) [doi:10.1364/BOE.9.006170].
243. R. Dechter, "Learning while searching in constraint-satisfaction-problems," in Proceedings of the Fifth AAAI National Conference on Artificial Intelligence, pp. 178–183, AAAI Press, Philadelphia, Pennsylvania (1986).
244. W. Ma et al., "Deep learning for the design of photonic structures," Nat. Photonics **15**(2), 77–90 (2021) [doi:10.1038/s41566-020-0685-y].
245. A. C. Mater and M. L. Coote, "Deep Learning in Chemistry," J. Chem. Inf. Model. **59**(6), 2545–2559 (2019) [doi:10.1021/acs.jcim.9b00266].
246. T. Ching et al., "Opportunities and obstacles for deep learning in biology and medicine," J. R. Soc. Interface. **15**(141), 20170387 (2018) [doi:10.1098/rsif.2017.0387].
247. Y. Zhang et al., "An Investigation of Deep Learning Models for EEG-Based Emotion Recognition," Front. Neurosci. **14**, 622759 (2020) [doi:10.3389/fnins.2020.622759].
248. X. Liu et al., "Deep learning in ECG diagnosis: A review," Knowledge-Based Systems **227**, 107187 (2021) [doi:10.1016/j.knosys.2021.107187].
249. P. Zhang et al., "Signal Processing for Diffuse Correlation Spectroscopy with Recurrent Neural Network of Deep Learning," in 2019 IEEE Fifth International Conference on Big Data Computing Service and Applications (BigDataService), pp. 328–332, IEEE, Newark, CA, USA (2019) [doi:10.1109/BigDataService.2019.00058].
250. C.-S. Poon, F. Long, and U. Sunar, "Deep learning model for ultrafast quantification of blood flow in diffuse correlation spectroscopy," Biomed. Opt. Express **11**(10), 5557 (2020) [doi:10.1364/BOE.402508].
251. Z. Li et al., "Quantification of blood flow index in diffuse correlation spectroscopy using long short-term memory architecture," Biomed. Opt. Express **12**(7), 4131 (2021) [doi:10.1364/BOE.423777].

252. J. Feng et al., "Cerebral blood flow monitoring using a ConvGRU model based on diffuse correlation spectroscopy," Infrared Physics & Technology **129**, 104541 (2023) [doi:10.1016/j.infrared.2022.104541].
253. P. Zhang et al., "Signal Processing for Diffuse Correlation Spectroscopy with Recurrent Neural Network of Deep Learning," in 2019 IEEE Fifth International Conference on Big Data Computing Service and Applications (BigDataService), pp. 328–332, IEEE, Newark, CA, USA (2019) [doi:10.1109/BigDataService.2019.00058].
254. Z. Li et al., "Quantification of blood flow index in diffuse correlation spectroscopy using long short-term memory architecture," Biomed. Opt. Express **12**(7), 4131 (2021) [doi:10.1364/BOE.423777].
255. M. Diop et al., "Comparison of time-resolved and continuous-wave near-infrared techniques for measuring cerebral blood flow in piglets," JBO **15**(5), 057004, SPIE (2010) [doi:10.1117/1.3488626].
256. M. Diop et al., "Time-resolved near-infrared technique for bedside monitoring of absolute cerebral blood flow," in Advanced Biomedical and Clinical Diagnostic Systems VIII **7555**, pp. 115–123, SPIE (2010) [doi:10.1117/12.842521].
257. D. Milej et al., "Characterizing dynamic cerebral vascular reactivity using a hybrid system combining time-resolved near-infrared and diffuse correlation spectroscopy," Biomed. Opt. Express **11**(8), 4571 (2020) [doi:10.1364/BOE.392113].
258. C. S. Roy and C. S. Sherrington, "On the Regulation of the Blood-supply of the Brain," The Journal of Physiology **11**(1–2), 85–158 (1890) [doi:10.1113/jphysiol.1890.sp000321].
259. H. C. Lou, L. Edvinsson, and E. T. MacKenzie, "The concept of coupling blood flow to brain function: Revision required?," Ann Neurol. **22**(3), 289–297 (1987) [doi:10.1002/ana.410220302].
260. U. Dirnagl et al., "Coupling of cerebral blood flow to neuronal activation: role of adenosine and nitric oxide," American Journal of Physiology-Heart and Circulatory Physiology **267**(1), H296–H301 (1994) [doi:10.1152/ajpheart.1994.267.1.H296].
261. D. Malonek and A. Grinvald, "Interactions Between Electrical Activity and Cortical Microcirculation Revealed by Imaging Spectroscopy: Implications for Functional Brain Mapping," Science **272**(5261), 551–554 (1996) [doi:10.1126/science.272.5261.551].
262. J. B. Mandeville et al., "Evidence of a Cerebrovascular Postarteriole Windkessel with Delayed Compliance," J Cereb Blood Flow Metab **19**(6), 679–689 (1999) [doi:10.1097/00004647-199906000-00012].
263. S. A. Carp, M. B. Robinson, and M. A. Franceschini, "Diffuse correlation spectroscopy: current status and future outlook," Neurophoton. **10**(01) (2023) [doi:10.1117/1.NPh.10.1.013509].
264. D. A. Boas and A. G. Yodh, "Spatially varying dynamical properties of turbid media probed with diffusing temporal light correlation," J. Opt. Soc. Am. A **14**(1), 192 (1997) [doi:10.1364/JOSAA.14.000192].
265. C. Cheung et al., "*In vivo* cerebrovascular measurement combining diffuse near-infrared absorption and correlation spectroscopies," Phys. Med. Biol. **46**(8), 2053–2065 (2001) [doi:10.1088/0031-9155/46/8/302].
266. C. Menon et al., "An Integrated Approach to Measuring Tumor Oxygen Status Using Human Melanoma Xenografts as a Model 1."
267. A. Marrero et al., "Aminolevulinic Acid-Photodynamic Therapy Combined with Topically Applied Vascular Disrupting Agent Vadimezan Leads to Enhanced Antitumor Responses," Photochemistry and Photobiology **87**(4), 910–919 (2011) [doi:10.1111/j.1751-1097.2011.00943.x].
268. G. Yu et al., "Noninvasive Monitoring of Murine Tumor Blood Flow During and After Photodynamic Therapy Provides Early Assessment of Therapeutic Efficacy," Clinical Cancer Research **11**(9), 3543–3552 (2005) [doi:10.1158/1078-0432.CCR-04-2582].

269. T. M. Busch et al., "Fluence rate-dependent intratumor heterogeneity in physiologic and cytotoxic responses to Photofrin photodynamic therapy," Photochem Photobiol Sci **8**(12), 1683–1693 (2009) [doi:10.1039/b9pp00004f].
270. J. P. Culver et al., "Diffuse Optical Tomography of Cerebral Blood Flow, Oxygenation, and Metabolism in Rat during Focal Ischemia," J Cereb Blood Flow Metab **23**(8), 911–924 (2003) [doi:10.1097/01.WCB.0000076703.71231.BB].
271. M. Diop et al., "Calibration of diffuse correlation spectroscopy with a time-resolved near-infrared technique to yield absolute cerebral blood flow measurements," Biomed. Opt. Express **2**(7), 2068 (2011) [doi:10.1364/BOE.2.002068].
272. R. C. Mesquita et al., "Optical Monitoring and Detection of Spinal Cord Ischemia," PLoS ONE **8**(12), J. A. Coles, Ed., e83370 (2013) [doi:10.1371/journal.pone.0083370].
273. K. Verdecchia et al., "Quantifying the cerebral metabolic rate of oxygen by combining diffuse correlation spectroscopy and time-resolved near-infrared spectroscopy," J. Biomed. Opt **18**(2), 027007 (2013) [doi:10.1117/1.JBO.18.2.027007].
274. S. Han et al., "Non-invasive diffuse correlation tomography reveals spatial and temporal blood flow differences in murine bone grafting approaches," Biomed. Opt. Express **7**(9), 3262 (2016) [doi:10.1364/BOE.7.003262].
275. A. Ruesch et al., "Estimating intracranial pressure using pulsatile cerebral blood flow measured with diffuse correlation spectroscopy," Biomed. Opt. Express **11**(3), 1462 (2020) [doi:10.1364/BOE.386612].
276. M. A. Franceschini et al., "The effect of different anesthetics on neurovascular coupling," NeuroImage **51**(4), 1367–1377 (2010) [doi:10.1016/j.neuroimage.2010.03.060].
277. C. Huang et al., "Speckle contrast diffuse correlation tomography of cerebral blood flow in perinatal disease model of neonatal piglets," Journal of Biophotonics **14**(4), e202000366 (2021) [doi:10.1002/jbio.202000366].
278. A. Villringer and B. Chance, "Non-invasive optical spectroscopy and imaging of human brain function," Trends in Neurosciences **20**(10), 435–442, Elsevier (1997) [doi:10.1016/S0166-2236(97)01132-6].
279. R. M. Danen et al., "Regional Imager for Low-Resolution Functional Imaging of the Brain with Diffusing Near-Infrared Light," Photochemistry and Photobiology **67**(1), 33–40 (1998) [doi:10.1111/j.1751-1097.1998.tb05162.x].
280. E. O. Ohuma et al., "National, regional, and global estimates of preterm birth in 2020, with trends from 2010: a systematic analysis," The Lancet **402**(10409), 1261–1271, Elsevier (2023) [doi:10.1016/S0140-6736(23)00878-4].
281. "Preterm birth," <https://www.who.int/news-room/fact-sheets/detail/preterm-birth> (accessed 21 August 2023).
282. U. Kiechl-Kohlendorfer et al., "Adverse neurodevelopmental outcome in preterm infants: risk factor profiles for different gestational ages," Acta Paediatrica **98**(5), 792–796 (2009) [doi:10.1111/j.1651-2227.2009.01219.x].
283. N. Roche-Labarbe et al., "Noninvasive optical measures of CBV, StO2, CBF index, and rCMRO2 in human premature neonates' brains in the first six weeks of life," Hum. Brain Mapp. **31**(3), 341–352 (2010) [doi:10.1002/hbm.20868].
284. P.-Y. Lin et al., "Reduced cerebral blood flow and oxygen metabolism in extremely preterm neonates with low-grade germinal matrix- intraventricular hemorrhage," Sci Rep **6**(1), 25903 (2016) [doi:10.1038/srep25903].
285. A. Rajaram et al., "Assessing cerebral blood flow, oxygenation and cytochrome c oxidase stability in preterm infants during the first 3 days after birth," Sci Rep **12**(1), 181 (2022) [doi:10.1038/s41598-021-03830-7].
286. T. Durduran et al., "Optical measurement of cerebral hemodynamics and oxygen metabolism in neonates with congenital heart defects," J. Biomed. Opt. **15**(3), 037004 (2010) [doi:10.1117/1.3425884].

287. E. M. Buckley et al., "Early postoperative changes in cerebral oxygen metabolism following neonatal cardiac surgery: Effects of surgical duration," The Journal of Thoracic and Cardiovascular Surgery **145**(1), 196-205.e1 (2013) [doi:10.1016/j.jtcvs.2012.09.057].
288. K. Shaw et al., "The use of novel diffuse optical spectroscopies for improved neuromonitoring during neonatal cardiac surgery requiring antegrade cerebral perfusion," Front Pediatr **11**, 1125985 (2023) [doi:10.3389/fped.2023.1125985].
289. M. Dehaes et al., "Cerebral Oxygen Metabolism in Neonatal Hypoxic Ischemic Encephalopathy during and after Therapeutic Hypothermia," J Cereb Blood Flow Metab **34**(1), 87–94 (2014) [doi:10.1038/jcbfm.2013.165].
290. J. Sutin et al., "Association of cerebral metabolic rate following therapeutic hypothermia with 18-month neurodevelopmental outcomes after neonatal hypoxic ischemic encephalopathy," eBioMedicine **94**, Elsevier (2023) [doi:10.1016/j.ebiom.2023.104673].
291. I. M. Inocencio et al., "Cerebral haemodynamic response to somatosensory stimulation in preterm lambs is enhanced following sildenafil and inhaled nitric oxide administration," Frontiers in Physiology **14** (2023).
292. N. Roche-Labarbe et al., "Near-Infrared Spectroscopy Assessment of Cerebral Oxygen Metabolism in the Developing Premature Brain," J Cereb Blood Flow Metab **32**(3), 481–488 (2012) [doi:10.1038/jcbfm.2011.145].
293. G. Côté-Corriveau et al., "Associations between neurological examination at term-equivalent age and cerebral hemodynamics and oxygen metabolism in infants born preterm," Frontiers in Neuroscience **17** (2023).
294. P.-Y. Lin et al., "Regional and Hemispheric Asymmetries of Cerebral Hemodynamic and Oxygen Metabolism in Newborns," Cerebral Cortex **23**(2), 339–348 (2013) [doi:10.1093/cercor/bhs023].
295. V. Dumont et al., "Somatosensory prediction in the premature neonate brain," Developmental Cognitive Neuroscience **57**, 101148 (2022) [doi:10.1016/j.dcn.2022.101148].
296. D. R. Busch et al., "Cerebral Blood Flow Response to Hypercapnia in Children with Obstructive Sleep Apnea Syndrome," Sleep **39**(1), 209–216 (2016) [doi:10.5665/sleep.5350].
297. M. Nourhashemi et al., "Preictal neuronal and vascular activity precedes the onset of childhood absence seizure: direct current potential shifts and their correlation with hemodynamic activity," NPh **10**(2), 025005, SPIE (2023) [doi:10.1117/1.NPh.10.2.025005].
298. S. Y. Lee et al., "Quantifying the Cerebral Hemometabolic Response to Blood Transfusion in Pediatric Sickle Cell Disease With Diffuse Optical Spectroscopies," Front. Neurol. **13**, 869117 (2022) [doi:10.3389/fneur.2022.869117].
299. A. Rajaram et al., "Perfusion and Metabolic Neuromonitoring during Ventricular Taps in Infants with Post-Hemorrhagic Ventricular Dilatation," Brain Sciences **10**(7), 452 (2020) [doi:10.3390/brainsci10070452].
300. B. N. Pasley and R. D. Freeman, "Neurovascular coupling," Scholarpedia **3**(3), 5340 (2008) [doi:10.4249/scholarpedia.5340].
301. T. Durduran et al., "Diffuse optical measurement of blood flow, blood oxygenation, and metabolism in a human brain during sensorimotor cortex activation," Opt. Lett. **29**(15), 1766 (2004) [doi:10.1364/OL.29.001766].
302. G. M. Tellis, R. C. Mesquita, and A. G. Yodh, "Use of Diffuse Correlation Spectroscopy To Measure Brain Blood Flow Differences During Speaking and Nonspeaking Tasks for Fluent Speakers and Persons Who Stutter."
303. R. C. Mesquita et al., "Blood flow and oxygenation changes due to low-frequency repetitive transcranial magnetic stimulation of the cerebral cortex," J. Biomed. Opt **18**(6), 067006 (2013) [doi:10.1117/1.JBO.18.6.067006].
304. K. R. Cowdrick et al., "Agreement in cerebrovascular reactivity assessed with diffuse correlation spectroscopy across experimental paradigms improves with short separation regression," NPh **10**(2), 025002, SPIE (2023) [doi:10.1117/1.NPh.10.2.025002].

305. C. Udina et al., "Dual-task related frontal cerebral blood flow changes in older adults with mild cognitive impairment: A functional diffuse correlation spectroscopy study," Front. Aging Neurosci. **14**, 958656 (2022) [doi:10.3389/fnagi.2022.958656].
306. L. Shoemaker et al., "Using depth-enhanced diffuse correlation spectroscopy and near-infrared spectroscopy to isolate cerebral hemodynamics during transient hypotension," NPh **10**(2), 025013, SPIE (2023) [doi:10.1117/1.NPh.10.2.025013].
307. T. W. Johnson et al., "Cerebral Blood Flow Hemispheric Asymmetry in Comatose Adults Receiving Extracorporeal Membrane Oxygenation," Front. Neurosci. **16**, 858404 (2022) [doi:10.3389/fnins.2022.858404].
308. Y. Shang et al., "Cerebral monitoring during carotid endarterectomy using near-infrared diffuse optical spectroscopies and electroencephalogram," Phys. Med. Biol. **56**(10), 3015 (2011) [doi:10.1088/0031-9155/56/10/008].
309. K. Kaya et al., "Intraoperative Cerebral Hemodynamic Monitoring during Carotid Endarterectomy via Diffuse Correlation Spectroscopy and Near-Infrared Spectroscopy," Brain Sciences **12**(8), 1025 (2022) [doi:10.3390/brainsci12081025].
310. R. C. Mesquita et al., "Diffuse optical characterization of an exercising patient group with peripheral artery disease," J. Biomed. Opt **18**(5), 057007 (2013) [doi:10.1117/1.JBO.18.5.057007].
311. A. Lafontant et al., "Comparison of optical measurements of critical closing pressure acquired before and during induced ventricular arrhythmia in adults," Neurophoton. **9**(03) (2022) [doi:10.1117/1.NPh.9.3.035004].
312. T. Durduran et al., "Transcranial optical monitoring of cerebrovascular hemodynamics in acute stroke patients," Opt. Express **17**(5), 3884 (2009) [doi:10.1364/OE.17.003884].
313. C. G. Favilla et al., "Optical Bedside Monitoring of Cerebral Blood Flow in Acute Ischemic Stroke Patients During Head-of-Bed Manipulation," Stroke **45**(5), 1269–1274 (2014) [doi:10.1161/STROKEAHA.113.004116].
314. K. C. Wu et al., "Validation of diffuse correlation spectroscopy measures of critical closing pressure against transcranial Doppler ultrasound in stroke patients," JBO **26**(3), 036008, SPIE (2021) [doi:10.1117/1.JBO.26.3.036008].
315. E. Sathialingam, "Microvascular cerebral blood flow response to intrathecal nicardipine is associated with delayed cerebral ischemia," Frontiers in Neurology.
316. P. Zirak et al., "Transcranial diffuse optical monitoring of microvascular cerebral hemodynamics after thrombolysis in ischemic stroke," J. Biomed. Opt **19**(1), 018002 (2014) [doi:10.1117/1.JBO.19.1.018002].
317. M. N. Kim et al., "Continuous Optical Monitoring of Cerebral Hemodynamics During Head-of-Bed Manipulation in Brain-Injured Adults," Neurocrit Care **20**(3), 443–453 (2014) [doi:10.1007/s12028-013-9849-7].
318. G. Yu et al., "Time-dependent blood flow and oxygenation in human skeletal muscles measured with noninvasive near-infrared diffuse optical spectroscopies," J. Biomed. Opt. **10**(2), 024027 (2005) [doi:10.1117/1.1884603].
319. G. Yu et al., "Validation of diffuse correlation spectroscopy for muscle blood flow with concurrent arterial spin labeled perfusion MRI," Opt. Express **15**(3), 1064 (2007) [doi:10.1364/OE.15.001064].
320. Y. Shang et al., "Noninvasive optical characterization of muscle blood flow, oxygenation, and metabolism in women with fibromyalgia," Arthritis Res Ther **14**(6), R236 (2012) [doi:10.1186/ar4079].
321. Y. Matsuda et al., "Evaluation of Local Skeletal Muscle Blood Flow in Manipulative Therapy by Diffuse Correlation Spectroscopy," Front. Bioeng. Biotechnol. **9**, 800051 (2022) [doi:10.3389/fbioe.2021.800051].

322. C.-G. Bangalore-Yogananda et al., "Concurrent measurement of skeletal muscle blood flow during exercise with diffuse correlation spectroscopy and Doppler ultrasound," Biomed. Opt. Express **9**(1), 131 (2018) [doi:10.1364/BOE.9.000131].
323. B. Henry et al., "Hybrid diffuse optical techniques for continuous hemodynamic measurement in gastrocnemius during plantar flexion exercise," J. Biomed. Opt **20**(12), 125006 (2015) [doi:10.1117/1.JBO.20.12.125006].
324. T. Durduran et al., "Diffuse optical measurement of blood flow in breast tumors," Opt. Lett. **30**(21), 2915 (2005) [doi:10.1364/OL.30.002915].
325. R. Choe et al., "Optically Measured Microvascular Blood Flow Contrast of Malignant Breast Tumors," PLoS ONE **9**(6), J. Najbauer, Ed., e99683 (2014) [doi:10.1371/journal.pone.0099683].
326. U. Sunar et al., "Monitoring photobleaching and hemodynamic responses to HPPH-mediated photodynamic therapy of head and neck cancer: a case report," Opt. Express **18**(14), 14969 (2010) [doi:10.1364/OE.18.014969].
327. C. Zhou et al., "Diffuse optical monitoring of blood flow and oxygenation in human breast cancer during early stages of neoadjuvant chemotherapy," J. Biomed. Opt. **12**(5), 051903 (2007) [doi:10.1117/1.2798595].
328. S. H. Chung et al., "Macroscopic optical physiological parameters correlate with microscopic proliferation and vessel area breast cancer signatures," Breast Cancer Res **17**(1), 72 (2015) [doi:10.1186/s13058-015-0578-z].
329. L. Dong et al., "Noninvasive diffuse optical monitoring of head and neck tumor blood flow and oxygenation during radiation delivery," Biomed. Opt. Express **3**(2), 259 (2012) [doi:10.1364/BOE.3.000259].
330. R. Choe and T. Durduran, "Diffuse Optical Monitoring of the Neoadjuvant Breast Cancer Therapy," IEEE J. Select. Topics Quantum Electron. **18**(4), 1367–1386 (2012) [doi:10.1109/JSTQE.2011.2177963].
331. G. Yu, "Near-infrared diffuse correlation spectroscopy in cancer diagnosis and therapy monitoring," J. Biomed. Opt. **17**(1), 010901 (2012) [doi:10.1117/1.JBO.17.1.010901].
332. T. Durduran and A. G. Yodh, "Diffuse correlation spectroscopy for non-invasive, micro-vascular cerebral blood flow measurement," NeuroImage **85**, 51–63 (2014) [doi:10.1016/j.neuroimage.2013.06.017].
333. T. Li et al., "Simultaneous measurement of deep tissue blood flow and oxygenation using noncontact diffuse correlation spectroscopy flow-oximeter," Sci Rep **3**(1), 1358 (2013) [doi:10.1038/srep01358].
334. L. Dong et al., "Diffuse optical measurements of head and neck tumor hemodynamics for early prediction of chemoradiation therapy outcomes," J. Biomed. Opt **21**(8), 085004 (2016) [doi:10.1117/1.JBO.21.8.085004].
335. C.-S. Poon et al., "First-in-clinical application of a time-gated diffuse correlation spectroscopy system at 1064 nm using superconducting nanowire single photon detectors in a neuro intensive care unit," Biomed. Opt. Express **13**(3), 1344 (2022) [doi:10.1364/BOE.448135].
336. S. Tagliabue et al., "Comparison of cerebral metabolic rate of oxygen, blood flow, and bispectral index under general anesthesia," Neurophoton. **10**(01) (2023) [doi:10.1117/1.NPh.10.1.015006].
337. C. Lindner et al., "Diffuse Optical Characterization of the Healthy Human Thyroid Tissue and Two Pathological Case Studies," PLOS ONE (2016).
338. "A. N. S. Institute, American National Standard for safe Use of Lasers ( Laser Institute of Aemerica, 2007)."
339. C. Bruschini et al., "Single-photon avalanche diode imagers in biophotonics: review and outlook," Light Sci Appl **8**(1), 87 (2019) [doi:10.1038/s41377-019-0191-5].
340. E. James and S. Powell, "Fourier domain diffuse correlation spectroscopy with heterodyne holographic detection," Biomed. Opt. Express **11**(11), 6755 (2020) [doi:10.1364/BOE.400525].

341. J. Xu et al., "Interferometric speckle visibility spectroscopy (ISVS) for human cerebral blood flow monitoring," APL Photonics **5**(12), 126102 (2020) [doi:10.1063/5.0021988].
342. W. Zhou et al., "Multi-exposure interferometric diffusing wave spectroscopy," Opt. Lett. **46**(18), 4498 (2021) [doi:10.1364/OL.427746].
343. D. Borycki, O. Kholiqov, and V. J. Srinivasan, "Reflectance-mode interferometric near-infrared spectroscopy quantifies brain absorption, scattering, and blood flow index in vivo," Opt. Lett. **42**(3), 591 (2017) [doi:10.1364/OL.42.000591].
344. O. Kholiqov et al., "Scanning interferometric near-infrared spectroscopy," Opt. Lett. **47**(1), 110 (2022) [doi:10.1364/OL.443533].
345. W. Zhou, M. Zhao, and V. J. Srinivasan, "Interferometric diffuse optics: recent advances and future outlook," Neurophoton. **10**(01) (2022) [doi:10.1117/1.NPh.10.1.013502].
346. W. Zhou et al., "Highly parallel, interferometric diffusing wave spectroscopy for monitoring cerebral blood flow dynamics," Optica **5**(5), 518 (2018) [doi:10.1364/OPTICA.5.000518].
347. N. Ozana et al., "Superconducting nanowire single-photon sensing of cerebral blood flow," Neurophoton. **8**(03) (2021) [doi:10.1117/1.NPh.8.3.035006].
348. F. Villa et al., "CMOS Imager With 1024 SPADs and TDCs for Single-Photon Timing and 3-D Time-of-Flight," IEEE J. Select. Topics Quantum Electron. **20**(6), 364–373 (2014) [doi:10.1109/JSTQE.2014.2342197].
349. F. Villa et al., "SPAD Smart Pixel for Time-of-Flight and Time-Correlated Single-Photon Counting Measurements," IEEE Photonics J. **4**(3), 795–804 (2012) [doi:10.1109/JPHOT.2012.2198459].
350. M. Gersbach et al., "A Time-Resolved, Low-Noise Single-Photon Image Sensor Fabricated in Deep-Submicron CMOS Technology," IEEE J. Solid-State Circuits **47**(6), 1394–1407 (2012) [doi:10.1109/JSSC.2012.2188466].
351. S. Jahromi et al., "A single chip laser radar receiver with a 9×9 SPAD detector array and a 10-channel TDC," in ESSCIRC Conference 2015 - 41st European Solid-State Circuits Conference (ESSCIRC), pp. 364–367, IEEE, Graz, Austria (2015) [doi:10.1109/ESSCIRC.2015.7313903].
352. E. Charbon, "Single-photon imaging in complementary metal oxide semiconductor processes," Phil. Trans. R. Soc. A. **372**(2012), 20130100 (2014) [doi:10.1098/rsta.2013.0100].
353. Y. Wang et al., "Low-Hardware Consumption, Resolution-Configurable Gray Code Oscillator Time-to-Digital Converters Implemented in 16 nm, 20 nm, and 28 nm FPGAs," IEEE Trans. Ind. Electron. **70**(4), 4256–4266 (2023) [doi:10.1109/TIE.2022.3174299].
354. W. Xie et al., "128-Channel High-Linearity Resolution-Adjustable Time-to-Digital Converters for LiDAR Applications: Software Predictions and Hardware Implementations," IEEE Trans. Ind. Electron. **69**(4), 4264–4274 (2022) [doi:10.1109/TIE.2021.3076708].
355. W. Xie, H. Chen, and D. D.-U. Li, "Efficient Time-to-Digital Converters in 20 nm FPGAs With Wave Union Methods," IEEE Trans. Ind. Electron. **69**(1), 1021–1031 (2022) [doi:10.1109/TIE.2021.3053905].
356. H. Chen and D. D.-U. Li, "Multichannel, Low Nonlinearity Time-to-Digital Converters Based on 20 and 28 nm FPGAs," IEEE Trans. Ind. Electron. **66**(4), 3265–3274 (2019) [doi:10.1109/TIE.2018.2842787].
357. H. Chen, Y. Zhang, and D. D.-U. Li, "A Low Nonlinearity, Missing-Code Free Time-to-Digital Converter Based on 28-nm FPGAs With Embedded Bin-Width Calibrations," IEEE Trans. Instrum. Meas. **66**(7), 1912–1921 (2017) [doi:10.1109/TIM.2017.2663498].
358. Y. Wang et al., "A two-stage interpolation time-to-digital converter implemented in 20 nm and 28 nm FGPAs."
359. M. Barth et al., "Correlation of Clinical Outcome with Pressure-, Oxygen-, and Flow-Related Indices of Cerebrovascular Reactivity in Patients Following Aneurysmal SAH," Neurocrit Care **12**(2), 234–243 (2010) [doi:10.1007/s12028-009-9287-8].

360. C. Caprara and C. Grimm, "From oxygen to erythropoietin: Relevance of hypoxia for retinal development, health and disease," Progress in Retinal and Eye Research **31**(1), 89–119 (2012) [doi:10.1016/j.preteyeres.2011.11.003].
361. V. Edul, A. Dubin, and C. Ince, "The Microcirculation as a Therapeutic Target in the Treatment of Sepsis and Shock," Semin Respir Crit Care Med **32**(05), 558–568 (2011) [doi:10.1055/s-0031-1287864].
362. P. Schober and L. A. Schwarte, "From system to organ to cell: oxygenation and perfusion measurement in anesthesia and critical care," J Clin Monit Comput **26**(4), 255–265 (2012) [doi:10.1007/s10877-012-9350-4].
363. S. M. White et al., "Longitudinal *In Vivo* Imaging to Assess Blood Flow and Oxygenation in Implantable Engineered Tissues," Tissue Engineering Part C: Methods **18**(9), 697–709 (2012) [doi:10.1089/ten.tec.2011.0744].
364. U. Wolf et al., "Localized irregularities in hemoglobin flow and oxygenation in calf muscle in patients with peripheral vascular disease detected with near-infrared spectrophotometry," Journal of Vascular Surgery **37**(5), 1017–1026 (2003) [doi:10.1067/mva.2003.214].
365. C. Cheung et al., "*In vivo* cerebrovascular measurement combining diffuse near-infrared absorption and correlation spectroscopies," Phys. Med. Biol. **46**(8), 2053–2065 (2001) [doi:10.1088/0031-9155/46/8/302].
366. G. Yu et al., "Noninvasive Monitoring of Murine Tumor Blood Flow During and After Photodynamic Therapy Provides Early Assessment of Therapeutic Efficacy," Clinical Cancer Research **11**(9), 3543–3552 (2005) [doi:10.1158/1078-0432.CCR-04-2582].
367. D. A. Boas et al., "Establishing the diffuse correlation spectroscopy signal relationship with blood flow," Neurophoton **3**(3), 031412 (2016) [doi:10.1117/1.NPh.3.3.031412].
368. L. Gagnon et al., "Investigation of diffuse correlation spectroscopy in multi-layered media including the human head," Opt. Express **16**(20), 15514 (2008) [doi:10.1364/OE.16.015514].
369. J. Feng et al., "Cerebral blood flow monitoring using a ConvGRU model based on diffuse correlation spectroscopy," Infrared Physics & Technology **129**, 104541 (2023) [doi:10.1016/j.infrared.2022.104541].
370. J. W. Goodman, "Statistical Properties of Laser Speckle Patterns," in Laser Speckle and Related Phenomena, J. C. Dainty, Ed., pp. 9–75, Springer, Berlin, Heidelberg (1975) [doi:10.1007/978-3-662-43205-1_2].
371. R. Bi et al., "Optical methods for blood perfusion measurement—theoretical comparison among four different modalities," J. Opt. Soc. Am. A **32**(5), 860 (2015) [doi:10.1364/JOSAA.32.000860].
372. J. Liu et al., "Quantitatively assessing flow velocity by the slope of the inverse square of the contrast values versus camera exposure time," Opt. Express **22**(16), 19327 (2014) [doi:10.1364/OE.22.019327].
373. J. Liu et al., "Quantitative model of diffuse speckle contrast analysis for flow measurement," J. Biomed. Opt **22**(7), 076016 (2017) [doi:10.1117/1.JBO.22.7.076016].
374. J. Liu et al., "Simultaneously extracting multiple parameters via multi-distance and multi-exposure diffuse speckle contrast analysis," Biomed. Opt. Express **8**(10), 4537 (2017) [doi:10.1364/BOE.8.004537].
375. C. Yeo, H. Kim, and C. Song, "Cerebral Blood Flow Monitoring by Diffuse Speckle Contrast Analysis during MCAO Surgery in the Rat," Current Optics and Photonics **1**(5), 433–439 (2017) [doi:10.3807/COPP.2017.1.5.433].
376. C. Yeo et al., "Avian embryo monitoring during incubation using multi-channel diffuse speckle contrast analysis," Biomed. Opt. Express **7**(1), 93 (2016) [doi:10.1364/BOE.7.000093].
377. V. Rajan et al., "Review of methodological developments in laser Doppler flowmetry," Lasers Med Sci **24**(2), 269–283 (2009) [doi:10.1007/s10103-007-0524-0].
378. A. Humeau et al., "Laser Doppler perfusion monitoring and imaging: novel approaches," Med Bio Eng Comput **45**(5), 421 (2007) [doi:10.1007/s11517-007-0170-5].

379. C. P. Valdes et al., "Speckle contrast optical spectroscopy, a non-invasive, diffuse optical method for measuring microvascular blood flow in tissue," Biomed. Opt. Express **5**(8), 2769 (2014) [doi:10.1364/BOE.5.002769].
380. B. Kim et al., "Measuring human cerebral blood flow and brain function with fiber-based speckle contrast optical spectroscopy system," Commun Biol **6**(1), 844 (2023) [doi:10.1038/s42003-023-05211-4].
381. S. A. Carp, M. B. Robinson, and M. A. Franceschini, "Diffuse correlation spectroscopy: current status and future outlook," Neurophoton. **10**(01) (2023) [doi:10.1117/1.NPh.10.1.013509].
382. L. Wang, "CML - Monte Carlo modeling of light transport in multi-la.yered. tissues," Computer Methods and Programs in Biomedicine (1995).
383. C. Zhu and Q. Liu, "Review of Monte Carlo modeling of light transport in tissues," J. Biomed. Opt **18**(5), 050902 (2013) [doi:10.1117/1.JBO.18.5.050902].
384. S. M. Hernandez and L. Pollonini, "NIRSplot: A Tool for Quality Assessment of fNIRS Scans," in Biophotonics Congress: Biomedical Optics 2020 (Translational, Microscopy, OCT, OTS, BRAIN), p. BM2C.5, Optica Publishing Group, Washington, DC (2020) [doi:10.1364/BRAIN.2020.BM2C.5].
385. Y. Zhao et al., "NIRS-ICA: A MATLAB Toolbox for Independent Component Analysis Applied in fNIRS Studies," Front. Neuroinform. **15**, 683735 (2021) [doi:10.3389/fninf.2021.683735].
386. S. L. Jacques, "Time resolved propagation of ultrashort laser pulses within turbid tissues," Appl. Opt. **28**(12), 2223 (1989) [doi:10.1364/AO.28.002223].
387. Q. Fang and D. A. Boas, "Monte Carlo Simulation of Photon Migration in 3D Turbid Media Accelerated by Graphics Processing Units," Opt. Express **17**(22), 20178 (2009) [doi:10.1364/OE.17.020178].
388. Q. Fang, "Mesh-based Monte Carlo method using fast ray-tracing in Plücker coordinates," Biomed. Opt. Express **1**(1), 165 (2010) [doi:10.1364/BOE.1.000165].
389. M. M. Wu, R. Horstmeyer, and S. A. Carp, "scatterBrains: an open database of human head models and companion optode locations for realistic Monte Carlo photon simulations," J. Biomed. Opt. **28**(10) (2023) [doi:10.1117/1.JBO.28.10.100501].
390. L. Wang and S. L. Jacques, "Monte Carlo Modeling of Light Transport in Multi-layered Tissues in Standard C." Comput. Methods Programs Biomed., 47(2) 131-146 (1995).